# Superconductivity at 253 K in lanthanum-yttrium ternary hydrides

Dmitrii V. Semenok,[1,&,*] Ivan A. Troyan,[2,&] Alexander G. Kvashnin,[1,&,*] Anna G. Ivanova,[2] Michael Hanfland,[3] Andrey V. Sadakov,[4] Oleg A. Sobolevskiy,[4] Kirill S. Pervakov,[4] Alexander G. Gavriliuk,[2,6] Igor S. Lyubutin,[2] Konstantin Glazyrin[9], Nico Giordano[9], Denis Karimov,[2] Alexander Vasiliev,[2,5] Ryosuke Akashi,[7] Vladimir M. Pudalov[4,8] and Artem R. Oganov[1,*]

[1] Skolkovo Institute of Science and Technology, Skolkovo Innovation Center, 3 Nobel Street, Moscow 121205, Russia
[2] Shubnikov Institute of Crystallography, Federal Scientific Research Center Crystallography and Photonics, Russian Academy of Sciences, 59 Leninsky Prospekt, Moscow 119333, Russia
[3] ID15B High Pressure Diffraction Beamline, ESRF, BP220, Grenoble 38043, France
[4] P. N. Lebedev Physical Institute, Russian Academy of Sciences, Moscow 119991, Russia
[5] National Research Center Kurchatov Institute, Moscow, Russia
[6] Institute for Nuclear Research, Russian Academy of Sciences, Moscow 117312, Russia
[7] University of Tokyo, 7-3-1 Hongo, Bunkyo, Tokyo 113-8654, Japan
[8] National Research University Higher School of Economics, Moscow 101000, Russia
[9] Deutsches Elektronen-Synchrotron, D-22607 Hamburg, Germany

# Abstract

Polyhydrides offer intriguing perspectives as high-temperature superconductors. Here we report the high-pressure synthesis of a series of lanthanum–yttrium ternary hydrides: cubic hexahydride (La,Y)$H_6$ with a critical temperature $T_C = 237 \pm 5$ K and decahydrides (La,Y)$H_{10}$ with a maximum $T_C \sim 253$ K and an extrapolated upper critical magnetic field $B_{C2}(0)$ up to 135 T at 183 GPa. This is one of the first examples of ternary high-$T_C$ superconducting hydrides. Our experiments show that a part of the atoms in the structures of recently discovered $Im\bar{3}m$-$YH_6$ and $Fm\bar{3}m$-$LaH_{10}$ can be replaced with lanthanum (~70%) and yttrium (~25%), respectively, with a formation of unique ternary superhydrides containing incorporated La@$H_{24}$ and Y@$H_{32}$ which are specific for $Im\bar{3}m$-$LaH_6$ and $Fm\bar{3}m$-$YH_{10}$. Ternary La–Y hydrides were obtained at pressures of 170–196 GPa via the laser heating of $P6_3/mmc$ lanthanum–yttrium alloys in the ammonia borane medium at temperatures above 2000 K. A novel tetragonal (La,Y)$H_4$ was discovered as an impurity phase in synthesized cubic (La,Y)$H_6$. The current–voltage measurements show that the critical current density $J_C$ in (La,Y)$H_{10}$ may exceed 2500 A/mm$^2$ at 4.2 K, which is comparable with that for commercial superconducting wires such as NbTi, $Nb_3Sn$. Hydrides that are unstable in a pure form may nevertheless be stabilized at relatively low pressures in solid solutions with superhydrides having the same structure.

**Highlights**

- Stabilization of “impossible” $LaH_6$ and $YH_{10}$ in solid solutions with $Im\bar{3}m$-$YH_6$ and $Fm\bar{3}m$-$LaH_{10}$
- Superconductivity in cubic (La,Y)$H_{10}$ at 253 K and 183 GPa
- Extrapolated upper critical magnetic field for (La,Y)$H_{10}$ is up to 135 T
- Extrapolated critical current density in (La,Y)$H_{10}$ is above 2500 A/mm$^2$ at 4.2 K.

# Introduction

Metallic hydrogen is expected to display remarkable superconducting properties stemming from its high Debye temperature and strong electron-phonon coupling [1–4]. Recent series of successful experimental syntheses of high-temperature superconductors $LaH_{10}$, $YH_6$ and $YH_9$ [5–9] containing the metallic H sublattice motivated us to attempt to raise the critical temperature of the La–H and Y–H systems by combining the elements into ternary La–Y hydrides.

Several binary hydrides of yttrium and lanthanum have been discovered in the last few years. In the work of Li et al. (2015),[10] $Im\bar{3}m$-$YH_6$ was predicted to be stable at pressures over 110 GPa, with the superconducting transition temperature, found via the numerical solution of the Migdal–Eliashberg equations, in the range from 251 to 264 K at 120 GPa. Recently, we have synthesized $YH_6$ and shown[8] that it has a much lower critical temperature, about 224 K at 166 GPa, which contradicts the theoretical results mentioned above. In 2017, in a detailed theoretical study of yttrium and lanthanum hydrides, Liu et al.[11] have found that lanthanum hexahydride $Im\bar{3}m$-$LaH_6$, isostructural to $YH_6$, less symmetrical $R\bar{3}m$-$LaH_6$, and $Fm\bar{3}m$-$YH_{10}$ are thermodynamically unstable and lie above the convex hull, which is a surface of the formation energy as a function of the chemical composition, at pressures below 250 GPa. We wondered whether these unstable compounds could be stabilized in ternary alloys with a similar structure.

Ternary La–Y hydrides have been theoretically studied by our group[12] and by Kostrzewa et al.[13] Using the neural network, we have predicted that the combination of elements from the "lability belt",[14] such us Mg, Ca, Ba, Sc, Y, La, and Th, forms the main group of high-$T_C$ ternary hydride superconductors, and the La–Y–H is one of the most prospective systems. Kostrzewa et al., analyzing the dependence of the logarithmically averaged frequency $\omega_{log}$ on the mass of atoms, have found that the La–Y–H compounds may have $T_C$ up to 274 K at 190 GPa. Thus, the theoretical analysis shows that the La–Y–H system is very attractive for achieving high critical temperatures of superconductivity.

In this work, we experimentally studied the superconducting properties of cubic hexahydrides $La_xY_{1-x}H_6$ and decahydrides $La_xY_{1-x}H_{10}$ obtained via the laser heating of yttrium–lanthanum alloys with ammonia borane (AB, $NH_3BH_3$). Samples were compressed to 165–196 GPa in symmetrical diamond anvil cells and heated by a series of laser pulses of millisecond duration. The superconducting properties of the yttrium–lanthanum polyhydrides were investigated using the electrical transport measurements in different current modes and external magnetic fields.

# Results and Discussion

## *Structural search*

The study of yttrium hexahydride, synthesized at moderate pressures,[8] has demonstrated that its critical temperature of superconductivity is significantly lower than has been theoretically predicted.[15] Is it possible to improve the parameters of the superconducting state by introducing another metal into the Y–H system? We noticed that lanthanum could form $Im\bar{3}m$-$LaH_6$, which lies only slightly above the area of thermodynamic stability (i.e. convex hull, Figure 1) at 150–200 GPa.[11] This cubic $LaH_6$ can be, at least kinetically, stabilized by an isostructural environment of $YH_6$, which is of great interest in terms of high-temperature superconductivity because of the structural analogy with yttrium hexahydride.

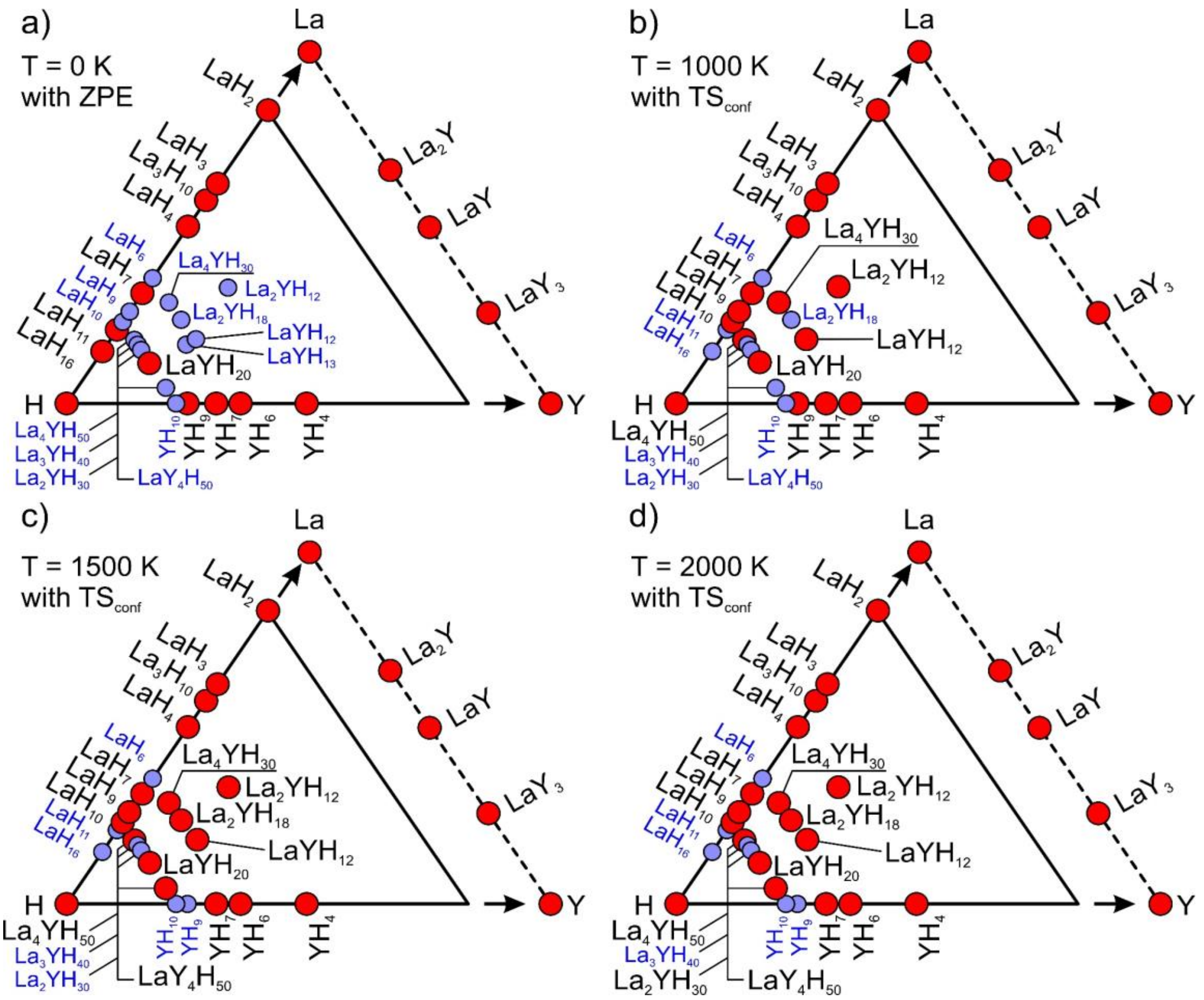

Figure 1. Ternary convex hulls of the La–Y–H system at pressure 200 GPa and temperatures (a) 0 K, (b) 1000 K, (c) 1500 K, and (d) 2000 K, calculated with a contribution of the configurational entropy ($TS_{conf}$) and the zero-point energy (ZPE). Invisible Z-axis corresponds to the enthalpy of formation and directed perpendicular to the plane of the figure. Stable and metastable phases are shown in red and blue, respectively.

An evolutionary search using USPEX[16–18] for thermodynamically stable phases in the La–Y–H system (Figure 1) shows that at 200 GPa and 0 K none of the ternary hydrides lies on the 3D convex hull except $LaYH_{20}$. Below in this paragraph we denote the symmetry of compounds without distinguishing Y and La. Surprisingly, $P6_3/mmc$-$LaYH_{20}$ is stable at 0 K (Figure 1a), whereas the cubic modification $Fm\bar{3}m$-$LaYH_{20}$ is 50 meV/atom above the convex hull. An increase in the temperature to 1000 K leads to the stabilization of tetragonal $I4/mmm$-$La_2YH_{12}$ and cubic hexahydrides $La_4YH_{30}$ and $LaYH_{12}$. These phases are metastable at 0 K, with the distances from the convex hull equal to 82, 38, and 53 meV/atom, respectively. Moreover, $La_4YH_{50}$, which belongs to the $La_{1-x}Y_xH_{10}$ structural type ($x = 0.2$), becomes stable at 1000 K. These structures can be considered as solid solutions of the La and Y atoms in the metallic sublattice of $Fm\bar{3}m$-$LaH_{10}$. Other phases that belong to the $La_{1-x}Y_xH_{10}$ group ($La_3YH_{40}$, $La_2YH_{30}$, and $LaY_4H_{50}$) are 5, 8, and 13 meV/atom above the convex hull at 1000 K (Figure 1b). Further increase in the temperature to 1500 K leads to the stabilization of $LaY_4H_{50}$ and $La_2YH_{18}$ (Figure 1c), whereas an even higher temperature of 2000 K stabilizes the $La_2YH_{30}$ phase (Figure 1d). Above 1800 K, $Fm\bar{3}m$-$LaYH_{20}$ also becomes stable (at 2000 K, $P6_3/mmc$-$LaYH_{20}$ is 5 meV/atom above the convex hull). Under all the studied pressure-temperature conditions, pure $Im\bar{3}m$-$LaH_6$ and $Fm\bar{3}m$-$YH_{10}$ were unstable. Thus, high-temperature laser synthesis may lead to stabilization and "freezing" of a wide range of ternary La–Y polyhydrides.

## *High-pressure synthesis*

To synthesize promising superconducting ternary yttrium–lanthanum hydrides, we prepared a series of La–Y alloys: $La_4Y$, $La_3Y$, $La_2Y$, LaY, and $LaY_4$. The yttrium (>99.99%) and lanthanum (>99.99%) metal pieces were mixed with the selected molar ratio and a total weight of 1g, pressurized and several times melted in an argon arc with sample rotation for homogenization of the melt. Subsequent X-ray and electron diffraction, X-ray fluorescence (XRF) and energy-dispersive X-ray (EDX) analyses showed that the hexagonal phases *P*$6_3$*/mmc*-$La_xY_{1-x}$ are the main product of the sintering (Supporting Information Figure S10-S11, S14-S15).

For loading high-pressure diamond anvil cells (DACs), we took the material from the homogeneous region of the alloy surface with the desired La:Y ratio determined by the EDX and XRF. To measure the synchrotron X-ray diffraction (XRD) and the critical temperature of superconductivity of synthesized lanthanum–yttrium hydrides, we used seven DACs M1-2, M2_S, SL1, SL1_S, SL3 and SL3_S, with a 50 µm culet beveled to 300 µm at 8.5°, equipped with four ~200 nm thick Ta electrodes with ~80 nm gold plating (Figure 3c) that were sputtered onto the piston diamond. Composite gaskets consisting of a tungsten ring and a $CaF_2$/epoxy mixture were used to isolate the electrical leads. Lanthanum–yttrium pieces with a thickness of ~1–2 µm were sandwiched between the electrodes and ammonia borane $NH_3BH_3$ (AB) in the gasket hole with a diameter of 20 µm and a thickness of 10–12 µm. Laser heating of samples above 2000 K at pressures of 170–196 GPa by several 100 µs pulses led to the formation of ternary lanthanum–yttrium hydrides whose structure was analyzed using the X-ray diffraction. Because each laser heating and cooling cycle resulted in some pressure change in the DACs, the pressure in each experiment was determined separately. The detailed description of the DACs is presented in Supporting Information.

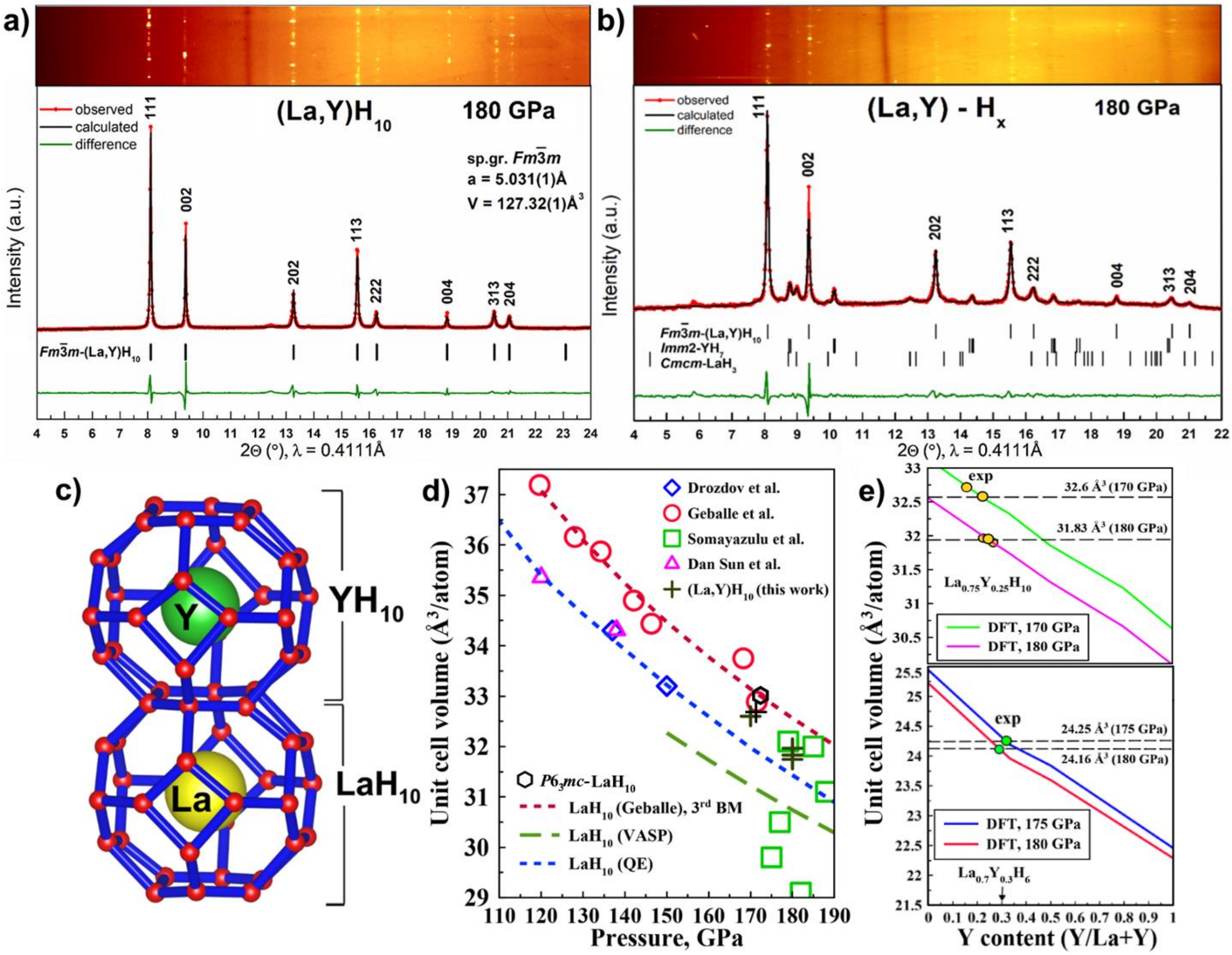

Figure 2. X-ray diffraction study of La-Y hydrides. Experimental diffraction patterns and Le Bail refinements of the cell parameters of (a) *Fm*$\bar{3}$*m*-(La,Y)$H_{10}$ (DAC SL1) and (b) *Imm*2-$YH_7$ and *Cmcm*-$LaH_3$ (DAC M1). The experimental data, fit and residues are shown by red, black and green lines, respectively. (c) Fragment of crystal structure of (La,Y)$H_{10}$ where Y and La are neighbors (for illustrative purposes). (d) Pressure–unit cell volume diagram for *fcc* $LaH_{10}$: circles, squares, rhombuses, triangles, and crosses show the experimental data, lines depict the theoretical calculations. (e) Estimates of the Y content in (La,Y)$H_{10}$ and (La,Y)$H_6$ obtained using the experimental unit cell volumes.

In five out of seven diamond anvil cells prepared for the XRD studies, the main observed phase was $Fm\bar{3}m$-(La,Y)$H_{10}$ (Figure 2a,b, Supporting Information Figures S17–S20) with a cell volume of ~31.8 Å$^3$ at 180 GPa, which is about 0.5–0.83 Å$^3$/f.u. lower than the volume of $Fm\bar{3}m$-$LaH_{10}$ found by Geballe et al.[5] (Figure 2d, Supporting Information Table S4). The decrease in the unit cell volume is due to the replacement of large La atoms with Y, which has a smaller size. In the discovered ternary cubic La–Y polyhydride, the substitution of the La atoms with Y resulted in the formation of the Y@$H_{32}$ inclusion with a locally distorted H-cage specific for $Fm\bar{3}m$-$YH_{10}$ (Figure 2c).[11]

There are large differences in the parameters of the unit cell of $LaH_{10}$ reported by different research groups.[11,5–7,19] In the first experiment by Geballe et al.,[5] the unit cell volume of $LaH_{10}$ at about 170 GPa exceeds 33 Å$^3$, which is significantly larger than 31.2 Å$^3$ calculated theoretically using VASP (Figure 1d, red circles). This prompted the authors of Ref. [5] to attribute the undefined stoichiometry $LaH_{9–12}$ to the discovered compound.[5]

In the subsequent studies by Drozdov et al.[6] and Dan Sun et al.[20], the experimental cell volume values of the synthesized $LaH_{10}$ ($V$ = 33.2 Å$^3$ at 150 GPa) are smaller by about 1.2 Å$^3$/f.u. than those observed by Geballe et al.[5] (Figure 2d). However, they are larger than those predicted using VASP,[21–24] and are in close agreement with the equations of state (EoS) calculated using the Quantum ESPRESSO[25,26] pseudopotentials (Figure 2d, "QE", blue line). In both cases the Perdew–Burke–Ernzerhof functional[27] in the generalized gradient approximation was used. The experimental data of Dan Sun et al.[20] were obtained in the low pressure range of 120–150 GPa, where distortion of the ideal cubic structure of $Fm\bar{3}m$-$LaH_{10}$ (→$C2/m$) occurs[20] and, at the same time, there is a possibility of losing some hydrogen from the crystal lattice.

Given the substantial uncertainty in both theoretical and experimental equations of states, and taking into account that substitution of La with Y in $LaH_{10}$ should lead to a decrease in the unit cell volume, we performed additional experiment with pure La squeezed with AB in DAC SL3_S at 171 GPa. After laser heating of the sample, we detected formation of $Fm\bar{3}m$-$LaH_{10}$ with the unit cell volume 32.82 Å$^3$ (Supporting Figure S21) and an impurity of hexagonal ($P6_3mc$) $LaH_{10}$ (cell volume 33.05 Å$^3$, Supporting Figure S22) which was also detected in previous experiments. [6] Thus, taking into account that our experimental result for V($LaH_{10}$) is close to the data of Ref. [5] we chose the 3rd order Birch–Murnaghan interpolation[28,29] of the experimental EoS found by Geballe et al.[5] as a reference (Figure 2d, red dotted line). Using this EoS, we determined the composition of synthesized (La,Y)$H_{10}$ as $La_3YH_{40}$ or $La_{0.75}Y_{0.25}H_{10}$ (Figure 2e), in which the La:Y ratio is close to that of the loaded sample in DACs M2, SL1 and SL1_S. Indeed, the results of the X-ray structural analysis show that the cleanest samples of (La,Y)$H_{10}$ were obtained in DACs SL1, SL1_S and M2, whereas in DACs M1 (loaded with LaY) and SL3 (loaded with $LaY_4$) we detected a notable amount of previously described[8] impurities: pseudocubic *Imm*2-$YH_7$ and, probably, *Cmcm*-$LaH_3$ (Supporting Information Figures S18-S19).

For DAC M2_S, loaded with $La_2Y$/AB, the X-ray diffraction patterns (Figure 3a,b) were identical to the one previously observed for $Im\bar{3}m$-$YH_6$.[8] However, the obtained unit cell volumes 48.5-48.3 Å$^3$ (Z = 2, 175-180 GPa) are in sharp contrast with that of $YH_6$ (< 46 Å$^3$, Figure 3e) in the same pressure interval. Given that the volume of the hypothetical $Im\bar{3}m$-$LaH_6$ is above 50 Å$^3$ (Z=2), the resulting product (La,Y)$H_6$ can be described as $La_2YH_{18}$ (more accurately, $La_{0.7}Y_{0.3}H_6$, Figures 3d, 2e) with a structure of $Im\bar{3}m$-$Y_3H_{18}$ where approximately every two out of three yttrium atoms are replaced by lanthanum. At the center of DAC M2_S, at 180 GPa we also detected a tetragonal phase impurity similar to previously detected *I4/mmm*-$YH_4$ [8], but with a unit cell volume larger by 1.4 Å$^3$/f.u. at this pressure. This enables to assign this compound the composition $La_2YH_{12}$ (Figure 3e). The significant difference between the results of synthesis in DACs M1-2, Sl1, SL3 and M2_S is probably due to the lack of hydrogen in the DAC M2_S, which is indicated by the presence of impurities of lower tetrahydrides (Figure 3).

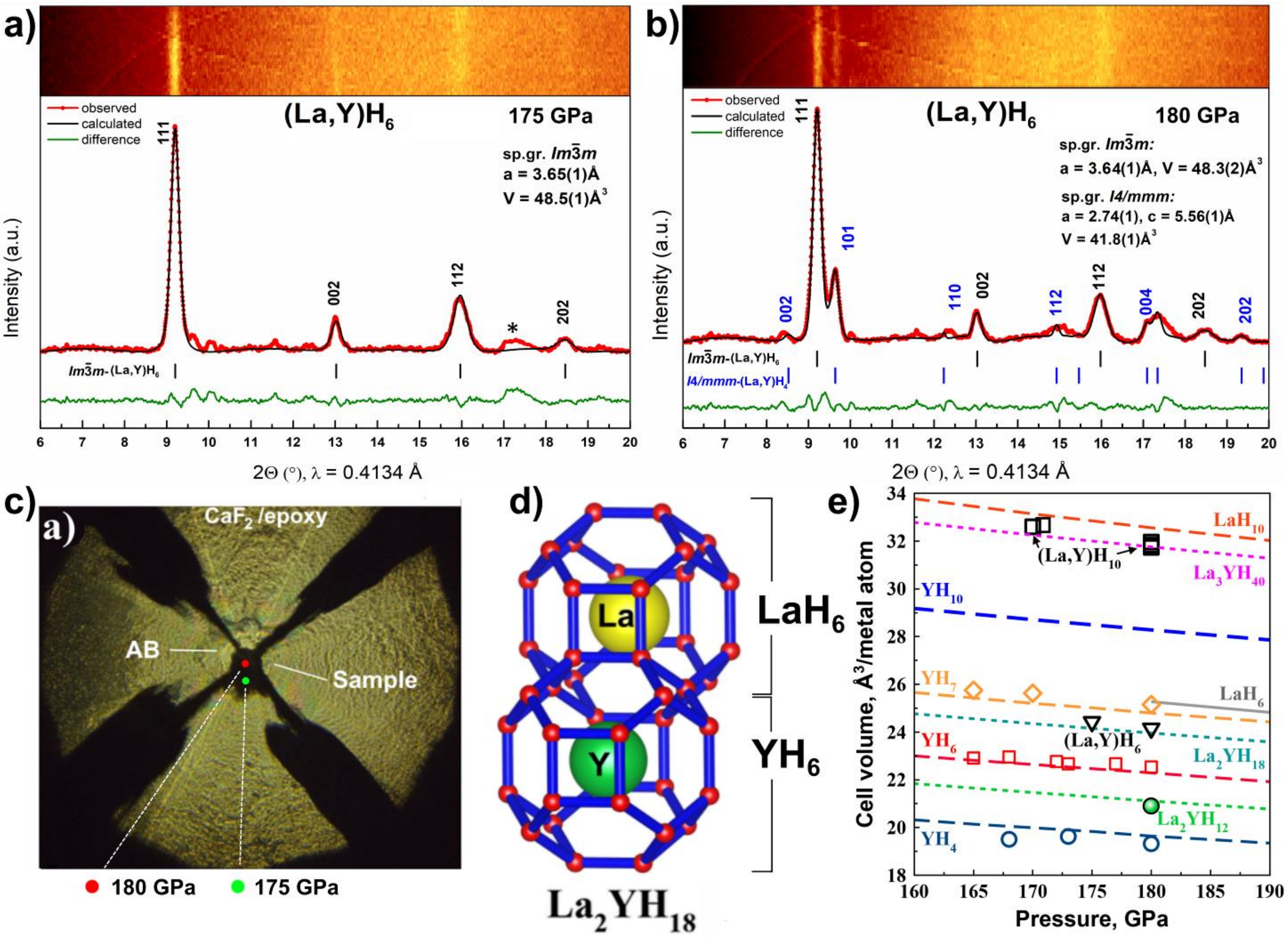

Figure 3. X-ray diffraction study of La-Y hydrides in DAC M2_S. (a, b) Experimental diffraction patterns and the Le Bail refinements of the cell parameters of $Im\bar{3}m$-(La,Y)$H_6$ and tetragonal *I4/mmm*-(La,Y)$H_4$ at 175 and 180 GPa. Experimental data, fit, and residues are shown in red, black, and green, respectively. (c) Optical microscopy of the loaded DAC: sample, $NH_3BH_3$ medium, and four Ta/Au electrodes. (d) Fragment of crystal structures of (La,Y)$H_6$ where Y and La are neighbors (for illustrative purposes). (e) Pressure–unit cell volume diagram of the studied La–Y–H phases.

### *Superconductivity*

The superconducting properties of the obtained lanthanum–yttrium hydrides were studied by measuring temperature dependence of the electrical resistance of the samples in various current modes using the four-probe method, with and without an external magnetic field (Figure 4). The $Fm\bar{3}m$-(La,Y)$H_{10}$ polyhydrides, synthesized from $La_2Y$ and $La_4Y$, have similar properties and exhibit relatively wide (15–17 K) superconducting transition with $T_C$ = 245–253 K at 183–199 GPa, with the resistance drop to 0.1 mΩ (Figure 4a,b). In several cells we detected an additional shelf in the $R(T)$ dependence at ~237 ± 5 K, possibly due to the presence of $Im\bar{3}m$-(La,Y)$H_6$ formed with the lack of hydrogen. The admixture $YH_6$, which forms sometimes, corresponds to the transition at 224–226 K (Supporting Information Figures S26 and S28).

Measurements in external magnetic fields of 0–16 T (Figure 4c,d, Supporting Information Figures S26-S27) show an almost linear dependence of $T_C(B)$ with the slope $dB_{C2}/dT \approx -0.76$ T/K near 230–250 K (Figure 4f). The extrapolated upper critical magnetic field $\mu_0 H_{c2}(0)$ is 90–135 T, which is lower than that of $YH_6$, even though $T_C$, the density of electronic states $N(E_F)$, and $N_H(E_F)$ for (La,Y)$H_{10}$ are notably higher (see below) than similar parameters for $YH_6$. This additionally indicates a possible anomaly in the mechanism of superconductivity in $Im\bar{3}m$-$YH_6$.[8]

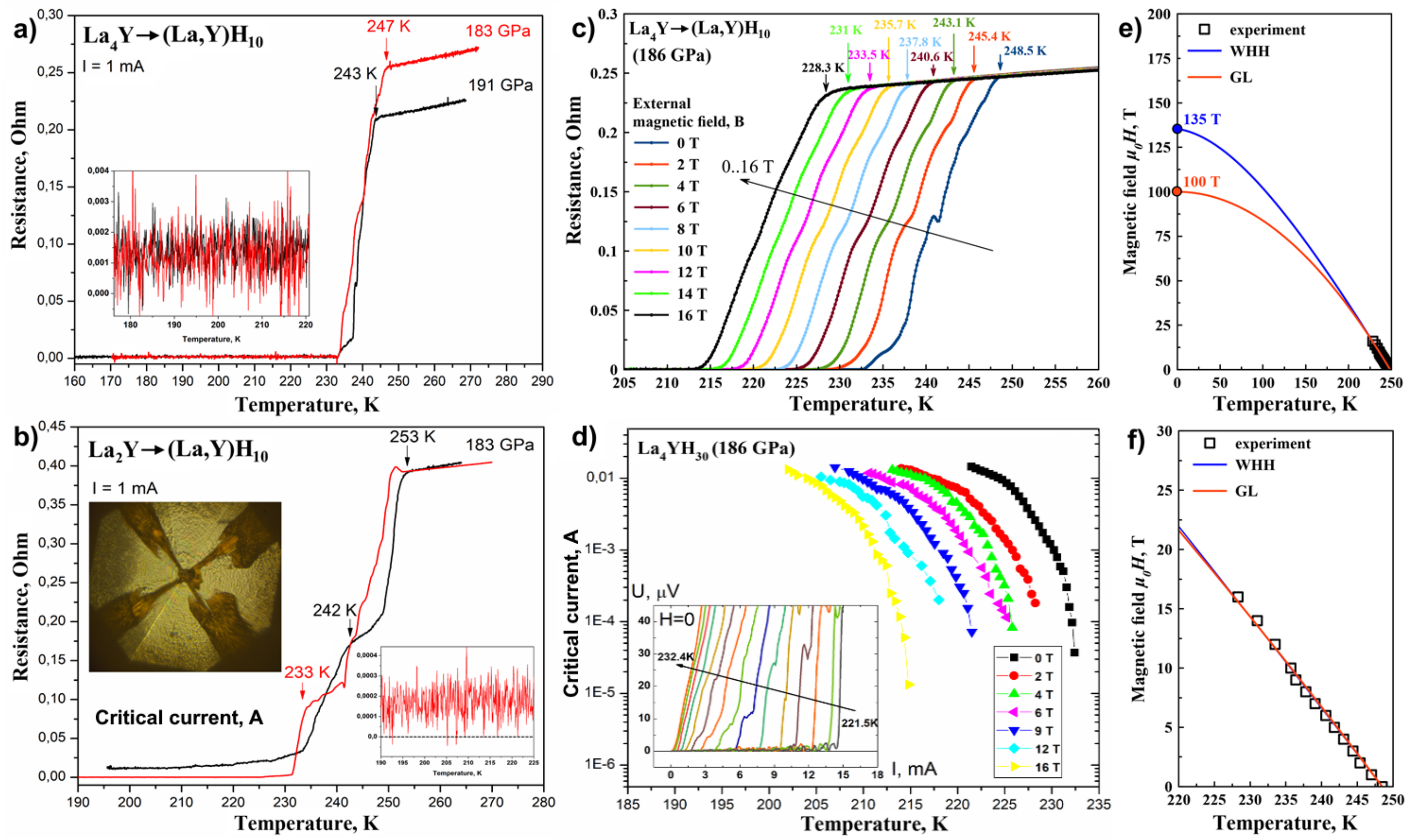

Figure 4. Superconducting transitions in $Fm\bar{3}m$-(La,Y)$H_{10}$: (a) temperature dependence of the electrical resistance for the sample obtained from $La_4Y$. Inset: residual resistance after cooling below $T_C$. (b) Temperature dependence of the resistance for the sample obtained from $La_2Y$. Insets: residual resistance after cooling below $T_C$ and a photo of the DAC culet with electrodes. (c) Dependence of the electrical resistance of (La,Y)$H_{10}$ on the external magnetic field (0–16 T) at 186 GPa and 0.1 mA current. The critical temperatures were determined at the onset of the resistance drop. (d) Dependence of the critical current on temperature and external magnetic field (0–16 T). The critical currents were measured near $T_C$. Inset: current-voltage characteristic near a SC transition. (e) Extrapolation of the upper critical magnetic field using the Werthamer–Helfand–Hohenberg theory[30] (WHH) and the Ginzburg–Landau[31] theory (GL). (f) Dependence of the critical temperature $T_C$ on the applied magnetic field.

We also investigated the pressure dependence of $T_C$ for the (La,Y)$H_{10}$ samples obtained from $La_2Y$ and $La_4Y$ alloys (Supporting Information Figure S27). For the first sample, when the pressure in the DAC decreases from 196 to 183 GPa, the critical temperature of the corresponding superhydride increases from 244.5 K to 253 K with a gradient $dT_C/dP$ = –0.65 K/GPa, whereas for the second sample, the measured gradient was –0.13 K/GPa (Supporting Information Figure S27a). The observed increase in the critical temperature is probably due to the loss of dynamic stability of the $Fm\bar{3}m$-$LaH_{10}$ superlattice and growth of the electron–phonon coupling (EPC) coefficient.

One of the distinguishing features of superconductors is the existence of an upper limit of current density $J_C$ at which superconductivity disappears and the material acquires nonzero electrical resistance. The critical currents and voltage–current ($U$–$I$) characteristics for the (La,Y)$H_{10}$ sample obtained from $La_4Y$ alloy were investigated in the range from $10^{-5}$ to $10^{-2}$ A in external magnetic fields at 186 GPa (Figure 4d). The critical current density was estimated on the basis of the fact that the sample size cannot exceed the size of the culet (50 μm), and the thickness of the sample is smaller than that of the gasket before the cell is loaded (~10 μm). Zero-field cooling (ZFC) shows that the critical current density in (La,Y)$H_{10}$ is over $2 \times 10^6$ A/m$^2$ at 230 K.

Analysis of the pinning force ($F_p = B \cdot I_C$) dependence on magnetic field (Supporting Figure S28a) shows that according to Dew-Hughes [32] the pinning in (La,Y)$H_{10}$ in the first approximation can be described as "dl-pinning". This allows us to extrapolate $I_C$(T) data to low temperatures within the single vortex model $J_C = J_{c0}(1 - T/T_C)^{5/2}(1 + T/T_C)^{-1/2}$. The extrapolation shows that at 4.2 K the critical current $I_C$ in the sample may reach 6 Ampers and the critical current density $J_C$ may exceed 12 kA/mm$^2$. The extrapolation using the Ginzburg–Landau model[31] ($J_C = J_{C0}(1 - T/T_C)^{3/2}$) gives much lower values at 4.2 K: the critical current $I_C$ in the sample can reach 1.25 A and the critical current density $J_C$ can exceed 2500 A/mm$^2$. This is comparable to the parameters of NbTi and YBCO,[33] and slightly higher than those in recently studied $Im\bar{3}m$-$YH_6$.[8]

The first-principles analysis of the superconducting properties of the La–Y–H phases was performed within the harmonic approximation for phonons. We calculated the parameters of the electron–phonon coupling and the superconducting state for a series of hexahydrides ($Im\bar{3}m$-$LaH_6$, cubic $La_4YH_{30}$, $La_2YH_{18}$, $Pm\bar{3}m$-$LaYH_{12}$) and for a decahydride — pseudocubic $R\bar{3}m$-$LaYH_{20}$ (obtained from $Fm\bar{3}m$-$La_2H_{20}$ by the La→Y replacement) at 180 GPa using the tetrahedron method of integration over the Brillouin zone[34] (Supporting Information Tables S5-6). To simplify the analysis, here we discuss in detail only the results of the calculations for high-symmetry decahydride $R\bar{3}m$-$LaYH_{20}$ with a regular arrangement of the lanthanum and yttrium atoms.

At 180 GPa, the EPC coefficient λ of $Pm\bar{3}m$-$LaYH_{12}$ reaches 2.82, $\omega_{log}$ is 847 K, and the critical temperature $T_C$ calculated using the Migdal–Eliashberg approach[35,36] is 223–241 K within the common range of the Coulomb pseudopotential $\mu^*$ = 0.15–0.1. A decrease in the concentration of yttrium in (La,Y)$H_6$ leads to an increase in $T_C$ of $La_2YH_{18}$ and $La_4YH_{30}$ to 265–270 K ($\mu^*$ = 0.1). Keeping in mind that the H sublattice in (La,Y)$H_6$ has almost the same structure as in $YH_6$, we can expect a negative anharmonic contribution $\Delta T_C$ ~ 25–30 K, as in pure yttrium hexahydride.[8] These estimates are in satisfactory agreement with the experimentally observed critical temperature of 237 ± 5 K. Calculations within the Migdal–Eliashberg (ME) theory show that the expected upper critical magnetic field $\mu_0 H_{C2}(0)$ of (La,Y)$H_6$ is ~70–80 T and the superconducting gap is around 60 meV (Supporting Information Table S6). We believe that the enhanced superconducting properties of (La,Y)$H_6$ compared to pure $YH_6$ are related to the presence of distorted $Im\bar{3}m$-$LaH_6$ in the crystal structure, which results in an increase in the electron–phonon interaction coefficient from 2.24 in $YH_6$ (harmonic approach[8]) to 2.41–2.82 in (La,Y)$H_6$.

A recent study[8] has shown that the density functional theory for superconductors (SCDFT), which incorporates the pair-breaking Coulomb repulsion and the retardation effect without empirical parameters such as $\mu^*$, revealed an anomaly of $YH_6$: its $T_C^{SCDFT}$ = 160 K significantly deviates from the experimental value of 224 K at 166 GPa. A similar phenomenon is also observed for $LaH_6$ and $LaYH_{12}$. Solving the SCDFT gap equation with the harmonic $\alpha^2F(\omega)$ (Supporting Information eq S1, Figure S29) at 180 GPa yields $T_C$ = 176 K for $LaH_6$ and 191 K for $LaYH_{12}$ with a standard error bar of ~2.5% originating from the random sampling step in solving the equation.[37,38] These absolute values of $T_C$ are substantially lower than the experimental data (~237 K), nevertheless the yttrium concentration dependence was reproduced: $\alpha^2F(\omega)$ for $LaH_6$ has a strong contribution in the 10–30 THz frequency interval (Supporting Information Figure S29f), which leads to $\lambda(LaH_6) > \lambda(YH_6)$ and $T_C(LaH_6) > T_C(YH_6)$. Moreover, in a surprising synergy, the superconducting properties of ternary (La,Y)$H_6$ are more pronounced both in theory and experiment than those of $LaH_6$ and $YH_6$: $T_C(LaYH_{12}) > T_C(LaH_6) > T_C(YH_6)$. Apart from the qualitative trend, underestimation of $T_C$ of (La,Y)$H_6$ when using the SCDFT indicates an anomalously large impact of the Coulomb repulsion and implies something beyond the conventional phonon-mediated superconductivity that boosts the critical temperature to the experimentally observed value. This possible anomaly would be worth future revisiting with improvement of the pairing interaction in the SCDFT [39].

Our calculations within the Migdal–Eliashberg (ME)[35,36] and the Bardeen–Cooper–Schrieffer[40,41] theories show that decahydride $R\bar{3}m$-$LaYH_{20}$ has the EPC coefficient of 3.87 and $\omega_{log}$ = 868 K, which are comparable with parameters of $LaH_{10}$,[42] and expected $T_C^{ME}$ = 281–300 K within the common range of the Coulomb pseudopotential $\mu^*$ = 0.15–0.1. Because the structure of $R\bar{3}m$-$LaYH_{20}$ is similar to that of $LaH_{10}$, we can expect the negative anharmonic contribution $\Delta T_C$ of ~25–35 K[19] to reduce $T_C$ to about 265 K. The calculations within the Migdal–Eliashberg theory show that for (La,Y)$H_{10}$ the expected upper critical magnetic field $\mu_0 H_{C2}(0)$ is 100 T, the superconducting gap is around 70 meV, and the coherence length $\xi_{BCS} = 0.5\sqrt{h/\pi e H_{C2}}$ is close to 16 Å (Supporting Information Table S6). In contrast to the $Im\bar{3}m$-$XH_6$ (X = La, Y) structures, the SCDFT calculations yield $T_C$ = 252 K for $R\bar{3}m$-$LaYH_{20}$ at 180 GPa, which is close to both the ME and experimental results. Experimentally found $T_C$ = 253 K and extrapolated $\mu_0 H_{C2}(0)$ = 100–135 T of (La,Y)$H_{10}$ are in close agreement with the theoretically calculated values, which means that superconductivity in La–Y–H ternary decahydrides may be described well in the framework of the classical Bardeen–Cooper–Schrieffer and Migdal–Eliashberg theories.

It is interesting to trace changes in the density of the electronic states of the La–Y–H phases that may shed light on their superconducting properties (Supporting Information Figures S30–S34). At 180 GPa, pure $Im\bar{3}m$-$YH_6$ has a total density of electronic states at the Fermi level $N(E_F)$ = 0.69 states/eV/f.u., where 0.22 states/eV/f.u. come from hydrogen, which is much lower than in $Fm\bar{3}m$-$LaH_{10}$ (0.36 states/eV/f.u.). The substitution of yttrium by lanthanum in $YH_6$ leads to a shift of the Fermi level and a significant increase in the density of electronic states. At La:Y ratio of 1:1, $N(E_F)$ reaches 0.81 states/eV/metal atom, where 0.25 states/eV/metal atom come from the H sublattice. For pure $Im\bar{3}m$-$LaH_6$, the density of states at $E_F$ is 1.56 states/eV/f.u. (0.61 states/eV/f.u. come from H), which is much higher than for $YH_6$. The relative density $N_H(E_F)/N(E_F)$ on the H sublattice increases in the series $YH_6$ (32%) < $LaH_6$ (39%) < $LaH_{10}$ (43%). Thus, the introduction of the La atoms into the $Im\bar{3}m$ structure of yttrium hexahydride may lead to a notable improvement of $T_C$ by increasing the electron density of states on the hydrogen sublattice.

## Conclusions

In this research, the novel high-$T_C$ ternary superconducting hydrides $Im\bar{3}m$-(La,Y)$H_6$ and $Fm\bar{3}m$-(La,Y)$H_{10}$ were experimentally discovered together with *I4/mmm*-(La,Y)$H_4$ at pressures of 170–196 GPa. Using the La–Y alloys (in ratio 1:1, 2:1, 3:1, 4:1, and 1:4) as a source for high-pressure synthesis, we replaced about 25% of the lanthanum atoms in the structure of $LaH_{10}$ with yttrium. Moreover, we found that about 70% of the yttrium atoms in $YH_6$ can be replaced by La without decomposition of the $Im\bar{3}m$ sodalite-like structure of the hexahydride. In other words, superhydrides that do not exist at 180 GPa can nevertheless be stabilized as a solid solution: inclusions of Y@$H_{32}$ with the local H environment specific for $Fm\bar{3}m$-$YH_{10}$ can be synthesized in the $Fm\bar{3}m$-$LaH_{10}$ superlattice, whereas $Im\bar{3}m$-$LaH_6$ can be stabilized by introducing only 30% of yttrium.

At 183 GPa, the obtained $Fm\bar{3}m$-(La,Y)$H_{10}$ decahydrides demonstrate outstanding $T_C$ of up to 253 K, the superconducting gap of ~70 meV, and the extrapolated upper critical magnetic field of 100–135 T in close agreement with the calculations within the BCS and ME theories. The estimated critical current density (≥2500 A/mm$^2$) at 4.2 K in lanthanum–yttrium decahydride may exceed that of YBCO, NbTi, and $YH_6$.

The measured critical temperature of $Im\bar{3}m$-(La,Y)$H_6$ is 237 ± 5 K, which is higher than $T_C$ of pure $YH_6$. Calculations show that the improvement in $T_C$ is due to pseudocubic $LaH_6$, stabilized by the superlattice of $Im\bar{3}m$-$YH_6$. An anomalously large impact of the Coulomb repulsion was found in the lanthanum and lanthanum–yttrium hexahydrides within the SCDFT approach. The parameter-free SCDFT calculations for cubic $LaH_6$ and $LaYH_{12}$ yield substantially underestimated $T_C$ of 176 and 191 K, respectively, which implies the importance of effects missing in the conventional Migdal–Eliashberg theory.

We believe, the performed experiments will have a strong impact on subsequent studies of ternary metal–hydrogen systems, opening promising ways to stabilize "impossible" compounds in the form of a solid solution with known superhydrides at relatively low pressures.

# Author Contributions

[&] These authors contributed equally to this work.
I.A.T, D.V.S., M.H., K.G., N.G., A.G.I. and A.G.I. performed the experiments. I.A.T., A.V.S. and O.A.S. performed the superconductivity measurements. D.V.S., A.G.K. R.A. and A.R.O. prepared the theoretical analysis. D.V.S., A.G.K., and A.G.I. contributed to the interpretation of the results. D.V.S., A.G.K., and A.R.O. wrote the manuscript. A.V.S., O.A.S., and V.M.P. performed the magnetotransport experiments in high magnetic fields and participated in the data processing and discussions. D.K., A.V. and K.S.P. prepared the La-Y alloys. R. A. performed SCDFT calculations. All the authors provided critical feedback and helped shape the research, analysis, and manuscript.

## Acknowledgments

The work was performed on the ESRF (Grenoble, France) station ID15B, on SPring-8 (Sayo, Japan) station BL10XU (project No. 2019B1476), and at P02.2 station of PETRA III, DESY (Hamburg, Germany). The high-pressure experiments were supported by the Ministry of Science and Higher Education of the Russian Federation within the state assignment of the FSRC Crystallography and Photonics of RAS and by the Russian Science Foundation (Project No. 19-12-00414). A.G.K. thanks the RFBR foundation, № 19-03-00100, and F.A.C.I.E. foundation, grant UMNIK № 13408GU/2018, for the financial support of this work. A.R.O. thanks the Russian Science Foundation (grant 19-72-30043). The reported study was funded by the RFBR, project 20-32-90099. A.R.O and D.V.S. thank the Ministry of Science and Higher Education agreement No. 075-15-2020-808. The authors express their gratitude to the staff of the station BL10XU of SPring-8 synchrotron research facility, especially to Dr. Saori Kawaguchi (JASRI), for the tremendous assistance in the use of the station's equipment before and after the experiment. A.G.G. acknowledges the use of the facilities of the Center for Collective Use "Accelerator Center for Neutron Research of the Structure of Substance and Nuclear Medicine" of the INR RAS for high-pressure cell preparation. The research used resources of the LPI Shared Facility Center. A.V.S., O.A.S., K.S.P. and V.M.P. acknowledge support of the state assignment of the Ministry of Science and Higher Education of the Russian Federation (Project No. 0023-2019-0005). K.S.P. thanks the Russian Foundation for Basic Research (project 19-02-00888). We thank Igor Grishin (Skoltech) for proofreading of the manuscript.

SUPPORTING INFORMATION

# Superconductivity at 253 K in lanthanum-yttrium ternary hydrides

Dmitrii V. Semenok,[1,&,*] Ivan A. Troyan,[2,&] Alexander G. Kvashnin,[1,&,*] Anna G. Ivanova,[2] Michael Hanfland,[3] Andrey V. Sadakov,[4] Oleg A. Sobolevskiy,[4] Kirill S. Pervakov,[4] Alexander G. Gavriliuk,[2,6] Igor S. Lyubutin,[2] Konstantin Glazyrin,[9] Nico Giordano,[9] Denis Karimov,[2] Alexander Vasiliev,[2,5] Ryosuke Akashi,[7] Vladimir M. Pudalov[4,8] and Artem R. Oganov[1,*]

[1] Skolkovo Institute of Science and Technology, Skolkovo Innovation Center, 3 Nobel Street, Moscow 121205, Russia
[2] Shubnikov Institute of Crystallography, Federal Scientific Research Center Crystallography and Photonics, Russian Academy of Sciences, 59 Leninsky Prospekt, Moscow 119333, Russia
[3] ID15B High Pressure Diffraction Beamline, ESRF, BP220, Grenoble 38043, France
[4] P. N. Lebedev Physical Institute, Russian Academy of Sciences, Moscow 119991, Russia
[5] National Research Center Kurchatov Institute, Moscow, Russia
[6] Institute for Nuclear Research, Russian Academy of Sciences, Moscow 117312, Russia
[7] University of Tokyo, 7-3-1 Hongo, Bunkyo, Tokyo 113-8654, Japan
[8] National Research University Higher School of Economics, Moscow 101000, Russia
[9] Deutsches Elektronen-Synchrotron, D-22607 Hamburg, Germany

# Contents

# Methods

*Experiment*

To perform the X-ray diffraction study, seven diamond anvil cells (DACs M1, M2, M2_S, SL1, SL1_S, SL3 and SL3_S) were loaded. The diameter of the working surface of the diamond anvils was 280 μm beveled at an angle of 8.5° to a culet of 50 μm. The X-ray diffraction patterns of samples in the DACs M1-2, SL1 and SL3 were recorded at the ID15B synchrotron beamline at the European Synchrotron Radiation Facility (Grenoble, France) using a focused (5 × 5 μm) monochromatic X-ray beam with a wavelength of 0.4111 Å and Mar555 detector. The exposure time was 60 s. $CeO_2$ standard was used for the distance calibration. The X-ray diffraction data were analyzed and integrated using Dioptas software package (version 0.5).[1] The full profile analysis of the diffraction patterns and the calculation of the unit cell parameters were performed in JANA2006 program[2] using the Le Bail method.[3] The pressure in the DACs was determined via the Raman signal of diamond.[4]

The X-ray diffraction patterns of the $La_xY_{1-x}H_6$ sample (DAC M2_S) were recorded at the BL10XU beamline (SPring-8, Japan) using monochromatic synchrotron radiation and an imaging plate detector at room temperature.[5,6] The X-ray beam with a wavelength of 0.413 Å was focused in a 3 × 8 μm spot with a polymer refractive lens (SU-8, produced by ANKA).

The powder XRD experiments with DACs SL1_S and SL3_S were carried out at beamline P02.2 of PETRA III, DESY (Hamburg) [7]. The X-ray wavelength was λ=0.2906 Å, and the beam size was 2×2 $\mu m^2$ at full width at half maximum (FWHM). The calibration was performed using $CeO_2$ standard. For data collection, a fast flat panel detector XRD1621 from Perkin Elmer (2048 pixels × 2048 pixels with 200×200 $\mu m^2$ pixel size) was used.

**Table S1.** Experimental parameters of the DACs used for the X-ray diffraction studies. Data in the second column corresponds to the pressure of the laser-assisted synthesis.

| #cell | Pressure, GPa | Sample size, μm | Composition |
|---|---|---|---|
| M1 | 180 | 30 | $LaY/BH_3NH_3$ |
| M2 | 180 | 32 | $La_2Y/BH_3NH_3$ |
| M2_S | 175-180 | 32 | $La_2Y/BH_3NH_3$ |
| SL1 | 180 | 29 | $La_4Y/BH_3NH_3$ |
| SL1_S | 171 | 30 | $La_3Y/BH_3NH_3$ |
| SL3 | 170 | 29 | $Y_4La/BH_3NH_3$ |
| SL3_S | 171 | 35 | $La/BH_3NH_3$ |

The initial La–Y alloys used as the precursors for the high-pressure synthesis of ternary hydrides were fused and then characterized by XRD, energy-dispersive X-ray Fluorescence (EDXRF) analysis, and scanning/transmission electron microscopy (SEM/TEM). To prepare $La_2Y$ alloy, pure metals of La and Y (99.9%, CHEMCRAFT Ltd.) were crushed, washed in dilute HCl and acetone to remove impurities, and dried in a glove box. The components were weighed and mixed in a specified ratio. Heating was carried out resistively. The melt was kept in tantalum crucibles at a temperature of 1900 K in an inert atmosphere (helium) for one hour and quenched at an initial rate of 200 K/min.

The XRD patterns of the La-Y alloys were recorded using powder X-ray diffractometer Rigaku MiniFlex 600 (Rigaku, Japan) with $CuK_\alpha$ radiation (40 kV, 15 mA, Ni–$K_\beta$-filter) in the 2θ angle range from 20° to 115° with a scanning step of 0.02º and a rate of 0.7º/min. The phases were identified in the PXDRL program (Rigaku, Japan) using ICDD PDF-2 datasets (release 2017). The unit cell parameters were refined in JANA2006 program[2] using the Le Bail method.[3] Elemental analysis of the initial La-Y alloys was performed using energy-dispersive EDXRF in μ-XRF system ORBIS PC (EDAX, USA).

EDXRF spectra were collected under vacuum from 2 mm areas of the samples irradiated Rh x-ray tube at 30 kV and 100 μA during 30 sec.

The morphology and composition of the initial LaY alloy, used as a precursor for $La_2YH_x$ hydride (DAC M2_S), were studied in a scanning electron microscope (SEM) Scios (ThermoFisher Scientific, USA) equipped with an energy-dispersive X-ray (EDX) spectrometer equipped with an Octane Elite silicon drift detector (SDD) with $Si_3N_4$ window (EDAX, USA). The acceleration voltage was 30 kV. For SEM and EDS studies, the chunk of the alloy was polished on one side by diamond powder with sequentially reduced particle sizes. The specimen for scanning/transmission electron microscopy (S/TEM) was prepared in a focus ion beam (FIB)/SEM microscope Scios (ThermoFisher Scientific, USA) by a standard lift-out FIB technique at an accelerating voltage of $Ga^+$ ions of 30 kV at initial and 5 kV at final step. Microstructural and elemental analyses were performed in an Osiris TEM/STEM (ThermoFisher Scientific, USA) equipped with an EDX Super-X SDD spectrometer (Bruker, USA), a high-angle annular dark field (HAADF) electron detector (Fischione, USA) and a 2048x2018 Gatan CCD (Gatan, USA). Image processing was performed using a Digital Micrograph (Gatan, USA) and TIA (ThermoFisher Scientific, USA) software. Simulations of the ED patterns and images were produced using Stadelmann's EMS software package [8].

Magnetotransport measurements were performed on samples with at least two hydride phases, therefore the voltage contacts in the Van der Pauw method might have been connected to a low-$T_C$ phase. As a result, the superconducting transition in the main phase (La,Y)$H_{10}$ can be observed as an upward feature of the resistance–temperature curves because of the shunting effect in fine-grained samples.

In Figure S24, the plot of the normalized volume pinning force $F_p/F_p^{max}$ versus the reduced field $h=H/H_{c2}$ is drawn on the basis of interpolation of the $Jc$(T,B) data. Experimental data is fitted by Dew-Hughes (DH) model [9] for surface type normal pinning centers $f \sim h^p(1-h)^q$. For this model parameters are p=0.5, q=2, $h_{max}$=0.2, which is close to our fit. Depinning critical current for this type of pinning can be described within the single vortices model, where vortices are pinned on randomly distributed weak pinning centers via spatial fluctuations of the charge carrier mean free path, or in other words "dl-pinning".

*Theory*

Computational predictions of thermodynamic stability of the La–Y–H phases at 200 GPa were carried out using the variable-composition evolutionary algorithm USPEX.[10–12] The first generation consisting of 120 structures was produced using the random symmetric[12] and random topology[13] generators, whereas all subsequent generations contained 20% of random structures and 80% of those created using heredity, softmutation, and transmutation operators. The evolutionary searches were combined with structure relaxations using the density functional theory (DFT)[14,15] within the Perdew–Burke–Ernzerhof functional (generalized gradient approximation)[16] and the projector augmented wave method[17,18] as implemented in the VASP code.[19–21] The kinetic energy cutoff for plane waves was 600 eV. The Brillouin zone was sampled using Γ-centered $k$-points meshes with a resolution of $2\pi \times 0.05$ Å$^{-1}$. This methodology is similar to those used in our previous works.[22,23]

The equations of state of the discovered phases were calculated using the density functional theory (DFT)[14,15] within the generalized gradient approximation (the Perdew–Burke–Ernzerhof functional)[16] and the projector augmented wave method[17,18] as implemented in the VASP code.[19–21] The plane wave kinetic energy cutoff was set to 600 eV and the Brillouin zone was sampled using Γ-centered $k$-points meshes with a resolution of $2\pi \times 0.05$ Å$^{-1}$. We also calculated the phonon densities of states of the studied materials using the finite displacements method (VASP and PHONOPY[24,25]).

The calculations of the critical temperature of superconductivity $T_C$ were carried out using Quantum ESPRESSO (QE) package.[26,27] The phonon frequencies and electron–phonon coupling (EPC) coefficients were computed using the density functional perturbation theory,[28] employing the plane-wave pseudopotential method and the Perdew–Burke–Ernzerhof exchange–correlation functional.[16] In our ab initio calculations of the electron–phonon coupling (EPC) coefficient λ, the first Brillouin zone was sampled using a 3×3×3 or 4×4×4 $q$-points mesh and a denser 16×16×16 $k$-points mesh for the La–Y–H phases. $T_C$ was calculated by solving the Eliashberg equations[29] using the iterative self-consistent method for the imaginary part of the order parameter $\Delta(T, \omega)$ (superconducting gap) and the renormalization wave function $Z(T, \omega)$.[30] More approximate estimates of $T_C$ were made using the Allen–Dynes formula.[31]

Finally, we also solved the gap equation in the density functional theory for superconductors (SCDFT)[32,33] for evaluating $T_C$ of $Im\bar{3}m$-$LaH_6$, $LaYH_{12}$ and $R\bar{3}m$-$LaYH_{20}$ nonempirically

$$\Delta_{nk}(T) = -Z_{nk}(T)\Delta_{nk}(T) - \frac{1}{2}\sum K_{nkn'k'}(T)\frac{\tanh \beta\xi_{n'k'}}{\xi_{n'k'}}\Delta_{n'k'}(T) \qquad \text{(S1)}$$

The temperature (as $\beta = 1/T$) dependence of the order parameter $\Delta_{nk}$ indicates $T_C$. Labels $n$, $n'$, $k$, and $k'$ denote the Kohn–Sham band and crystal wave number indexes, respectively. $\xi_{nk}$ is the energy eigenvalue of state $nk$ measured from the Fermi level, as calculated using the standard Kohn–Sham equation for the normal state. $Z_{nk}(T)$ and $K_{nkn'k'}(T)$ represent the electron–phonon and electron–electron Coulomb interaction effects, the formulas for which have been constructed so that the self energy corrections, almost the same as those in the Eliashberg equations with the Migdal approximation,[29,34–36] are included (see Supplemental materials of Kruglov et al.[37] for details, which is based on Refs. [32,33,38,39]). We calculated the screened electron-electron Coulomb interaction within the random phase approximation[40], electronic density of states (DOS) of the normal state was used for solving Eq. (S1). We generated dense $\xi_{nk}$ data points entering Eq. (S1) around $E_F$ by a linear interpolation from the values on a dense k-point mesh.

.

**Table S2.** Detailed conditions for calculating $T_C$ of $YH_6$ within the SCDFT approach.

| Crystal structure setting | | $LaH_6$ | $LaYH_{12}$ | $LaYH_{20}$ |
|---|---|---|---|---|
| Unit cell | | $La_2H_{12}$ (cubic) | $La_2Y_2H_{24}$ | $LaYH_{20}$ |
| Charge density | *k* | 12×12×12 equal mesh | 8×8×8 equal mesh | 6×12×12 equal mesh |
| | Interpolation | 1st order Hermite Gaussian[40] with width = 0.020 Ry | | |
| Dielectric matrix ε | *k* for bands crossing $E_F$ | 15×15×15 equal mesh | 9×9×9 equal mesh | 9×15×15 equal mesh |
| | *k* for other bands | 5×5×5 equal mesh | 3×3×3 equal mesh | 3×5×5 equal mesh |
| | Number of unoccupied bands† | 98 | 184 | 103 |
| | Interpolation | Tetrahedron with the Rath–Freeman treatment[41] | | |
| DOS for phononic kernels | *k* | 19×19×19 equal mesh | 11×11×11 equal mesh | 9×19×19 equal mesh |
| | Interpolation | Tetrahedron with the Blöchl correction[17] | | |
| SCDFT gap function | Number of unoccupied bands† | 57 | 99 | 58 |
| | *k* for the electronic kernel | 5×5×5 equal mesh | 3×3×3 equal mesh | 3×5×5 equal mesh |
| | *k* for the KS energies | 19×19×19 equal mesh | 11×11×11 equal mesh | 9×19×19 equal mesh |
| | Sampling points for bands crossing $E_F$ | 6000 | | |
| | Sampling points for the other bands | 150 | | |
| | Sampling error in $T_C$ (%) | ~3.1 | ~2.4 | ~2.8 |

†States up to $E_F$ + 70 eV were taken into account.

# Cell parameters

Table S3. Crystal data of the constructed pseudocubic and pseudotetragonal lanthanum–yttrium hydrides at 180 GPa (relaxation was performed in VASP).

| Phase | Volume, $Å^3$/atom | Lattice parameters | Coordinates (*x*/*a*; *y*/*b*; *z*/*c*) |
|---|---|---|---|
| $Pm\bar{3}m$-$LaYH_{12}$ | 3.36 | $a$ = 3.612 Å | La1 La 0.00000 0.00000 0.00000<br>Y1 Y 0.50000 0.50000 0.50000<br>H1 H 0.24305 0.50000 0.00000 |
| *P*-1-$La_2YH_{12}$ (pseudotetragonal) | 4.133 | $a$ = 3.435 Å<br>$b$ = 5.111 Å<br>$c$ = 7.451 Å<br>α = 80.687°<br>β = 92.133 °<br>γ = 74.543 ° | La1 La -0.42465 0.24845 -0.16863<br>La2 La -0.25375 -0.24640 0.49572<br>Y1 Y -0.07610 0.25504 0.16479<br>H1 H 0.16533 0.07090 -0.04767<br>H2 H -0.32758 -0.42804 -0.04862<br>H3 H -0.15160 0.06545 -0.37662<br>H4 H 0.33730 -0.41983 -0.37902<br>H5 H -0.48523 0.08474 0.27809<br>H6 H 0.01538 -0.43181 0.28108<br>H7 H -0.46465 -0.13052 -0.08381<br>H8 H 0.03337 0.37466 -0.07958<br>H9 H 0.20744 -0.13170 -0.40758<br>H10 H -0.29244 0.37867 -0.41536<br>H11 H -0.12415 -0.11314 0.24587<br>H12 H 0.37927 0.37296 0.24863 |
| *P*-1-$La_2YH_{18}$ (pseudocubic) | 3.42 | $a$ = 3.146 Å<br>$b$ = 6.024 Å<br>$c$ = 7.916 Å<br>α = 101.981°<br>β = 97.620 °<br>γ = 79.960 ° | La1 La 0.25000 0.25000 0.00000<br>La2 La 0.41700 0.41700 -0.33300<br>Y1 Y -0.08300 -0.08300 -0.33300<br>H1 H 0.16700 -0.08300 -0.08300<br>H2 H -0.33300 0.41700 -0.08300<br>H3 H -0.16700 -0.41700 -0.41700<br>H4 H 0.33300 0.08300 -0.41700<br>H5 H 0.50000 0.25000 0.25000<br>H6 H 0.00000 -0.25000 0.25000<br>H7 H -0.20800 0.04200 -0.08300<br>H8 H 0.29200 -0.45800 -0.08300<br>H9 H 0.45800 -0.29200 -0.41700<br>H10 H -0.04200 0.20800 -0.41700<br>H11 H 0.12500 0.37500 0.25000<br>H12 H -0.37500 -0.12500 0.25000<br>H13 H 0.45800 -0.04200 -0.16700<br>H14 H -0.04200 0.45800 -0.16700<br>H15 H 0.12500 -0.37500 0.50000<br>H16 H -0.37500 0.12500 0.50000<br>H17 H -0.20800 0.29200 0.16700<br>H18 H 0.29200 -0.20800 0.16700 |
| *Fmmm*-$La_4YH_{30}$ (pseudocubic) | 3.34 | $a$ = 3.602 Å<br>$b$ = 5.093 Å<br>$c$ = 25.47 Å<br>α = 90 °<br>β = 90 °<br>γ = 90 ° | La1 La 0.00000 0.00000 0.40000<br>La2 La 0.00000 0.00000 -0.20000<br>Y1 Y 0.00000 0.00000 0.00000<br>H1 H 0.00000 -0.12500 -0.12500<br>H2 H 0.00000 0.37500 -0.22500<br>H3 H 0.00000 -0.12500 -0.32500<br>H4 H 0.00000 0.37500 -0.42500<br>H5 H 0.00000 0.37500 -0.02500<br>H6 H 0.25000 0.25000 0.35000<br>H7 H 0.25000 0.25000 0.25000<br>H8 H 0.25000 0.25000 -0.45000 |
| $R\bar{3}m$-$LaYH_{20}$<br>(pseudocubic) | 2.68 | $a = b$ = 3.4800 Å<br>$c$ = 16.874 Å<br>α = 90 º<br>β = 90 º<br>γ = 120 º | La1 La 0.00000 0.00000 0.50000<br>Y1 Y 0.00000 0.00000 0.00000<br>H1 H -0.16147 0.16147 0.23081<br>H2 H -0.49432 0.49432 0.06083<br>H3 H 0.00000 0.00000 0.19121<br>H4 H 0.00000 0.00000 0.31037<br>H5 H 0.00000 0.00000 0.12520<br>H6 H 0.00000 0.00000 -0.37618 |
| $P\bar{3}m1$-$La_2YH_{30}$<br>(pseudocubic) | 2.72 | $a = b$ = 3.4974 Å<br>$c$ = 8.5006 Å<br>α = 90 º<br>β = 90 º<br>γ = 120 º | La1 La 0.33333 0.66667 -0.16957<br>Y1 Y 0.00000 0.00000 0.50000<br>H1 H -0.16073 0.16073 -0.03807<br>H2 H -0.49354 0.49354 -0.38311<br>H3 H 0.17105 -0.17105 0.29759<br>H4 H 0.00000 0.00000 0.12642<br>H5 H 0.33333 0.66667 0.22024<br>H6 H 0.33333 0.66667 0.45284<br>H7 H 0.33333 0.66667 0.08853<br>H8 H 0.33333 0.66667 -0.41170<br>H9 H 0.00000 0.00000 0.25079 |
| *Cmmm*-$La_3YH_{40}$<br>(pseudocubic) | 2.74 | $a$ = 7.034 Å<br>$b$ = 9.879 Å<br>$c$ = 3.469 Å<br>α = 90 º<br>β = 90 º<br>γ = 90 º | La1 La 0.00000 0.00000 0.00000<br>La2 La 0.25000 0.25000 0.50000<br>Y1 Y 0.50000 0.00000 0.00000<br>H1 H -0.37770 0.31165 0.00000<br>H2 H 0.37338 0.06336 0.50000<br>H3 H -0.13146 0.06280 0.50000<br>H4 H 0.12029 0.31526 0.00000<br>H5 H -0.24496 0.44300 -0.26036<br>H6 H 0.00000 -0.30994 0.23903<br>H7 H 0.00000 0.19260 0.24256<br>H8 H -0.25420 0.37167 0.00000<br>H9 H 0.00000 -0.37421 0.50000<br>H10 H 0.00000 0.12471 0.50000 |
| $R\bar{3}m$-$La_4YH_{50}$<br>(pseudocubic) | 2.75 | $a = b$ = 3.5165 Å<br>$c$ = 42.51186 Å<br>α = 90 º<br>β = 90 º<br>γ = 120 º | La1 La 0.00000 0.00000 -0.39887<br>La2 La 0.00000 0.00000 0.20046<br>Y1 Y 0.00000 0.00000 0.00000<br>H1 H -0.49388 0.49388 0.22636<br>H2 H 0.17218 -0.17218 0.15887<br>H3 H -0.16154 0.16154 0.09087<br>H4 H -0.49449 0.49449 0.02278<br>H5 H -0.16175 0.16175 0.29276<br>H6 H 0.00000 0.00000 -0.07435<br>H7 H 0.00000 0.00000 -0.47498<br>H8 H 0.00000 0.00000 0.12377<br>H9 H 0.00000 0.00000 -0.27921<br>H10 H 0.00000 0.00000 0.32344<br>H11 H 0.00000 0.00000 -0.45008<br>H12 H 0.00000 0.00000 0.14921<br>H13 H 0.00000 0.00000 -0.25067<br>H14 H 0.00000 0.00000 0.35082<br>H15 H 0.00000 0.00000 -0.04985 |

# Structural characterization of the initial La–Y alloys

The SEM images and the EDX analysis results of the La-Y alloy obtained from several areas within the mechanically polished area of the specimen are shown in Figs. S1(c,d) and Figs. S2(a,e). SEM image (Fig. S1(c)) demonstrates the presence of laths, which are associated with the features found in light microscopy (LM) images (see Fig. S1(a,b)). The EDX spectra from the area highlighted by the red rectangle in Fig. S1(c) is presented in Fig. S4(d). It turned out that gross La:Y elemental ratio in this area is 57:43 at%. The SEM image and the EDS elemental map acquired at higher magnification are shown in Figs. S2(a,c). It can be concluded unambiguously that the laths observed in LM images are Y-dendrites. These dendrites could be formed as a result of the peritectic reaction during La-Y alloy solidification.

The EDX microanalysis results of the dendrite and matrix (see points 1&2 in Fig. S2(a,c)) demonstrate the absence of La in the dendrite crystal and 35:65 at% Y:La elemental ratio in the alloy.

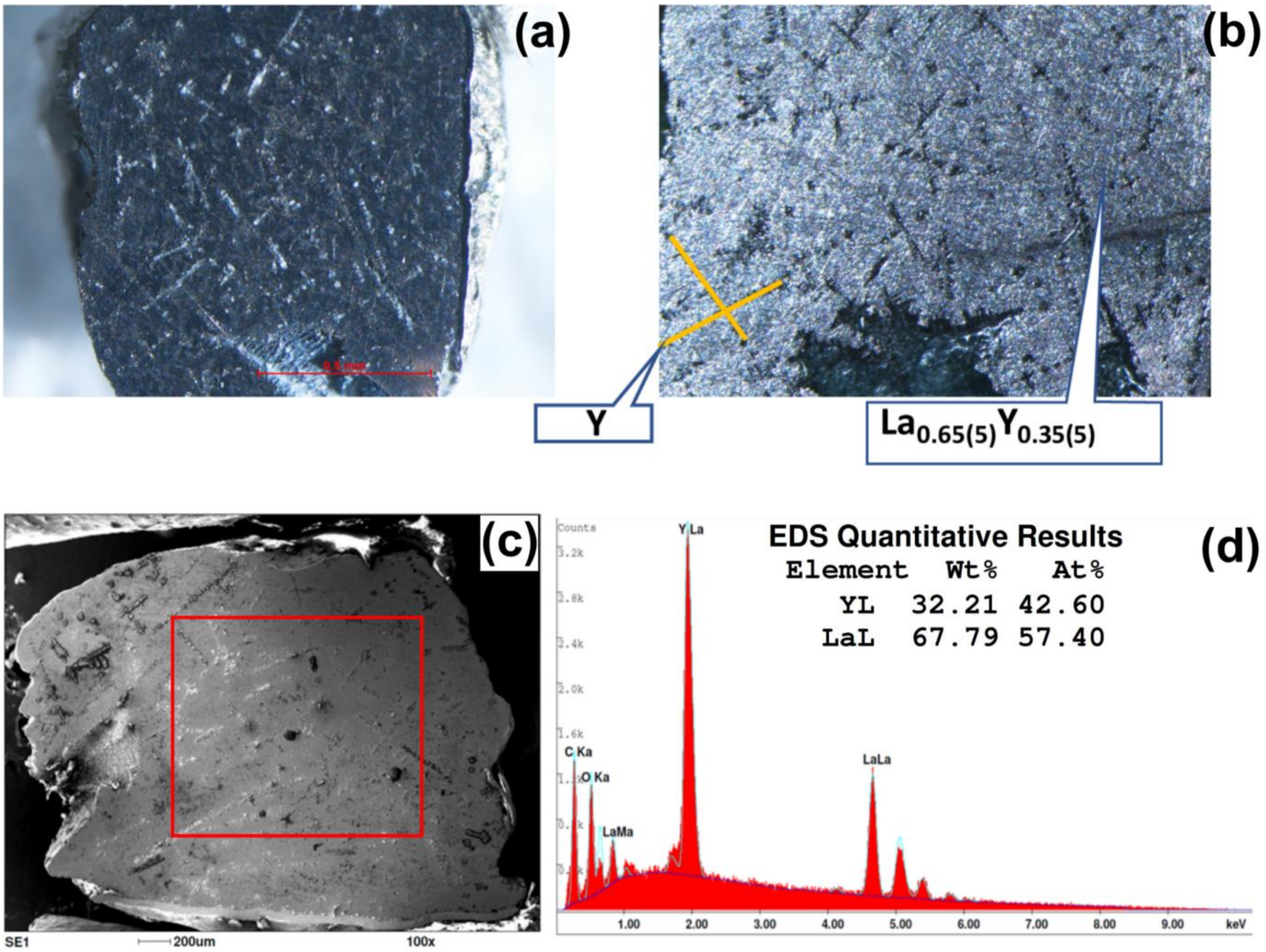

**Figure S1**. Light microscopy images of the initial $La_2Y$ alloy (a,c) at 0 GPa, SEM image (d), and EDX spectra from the area highlighted by the red rectangle in the SEM image.

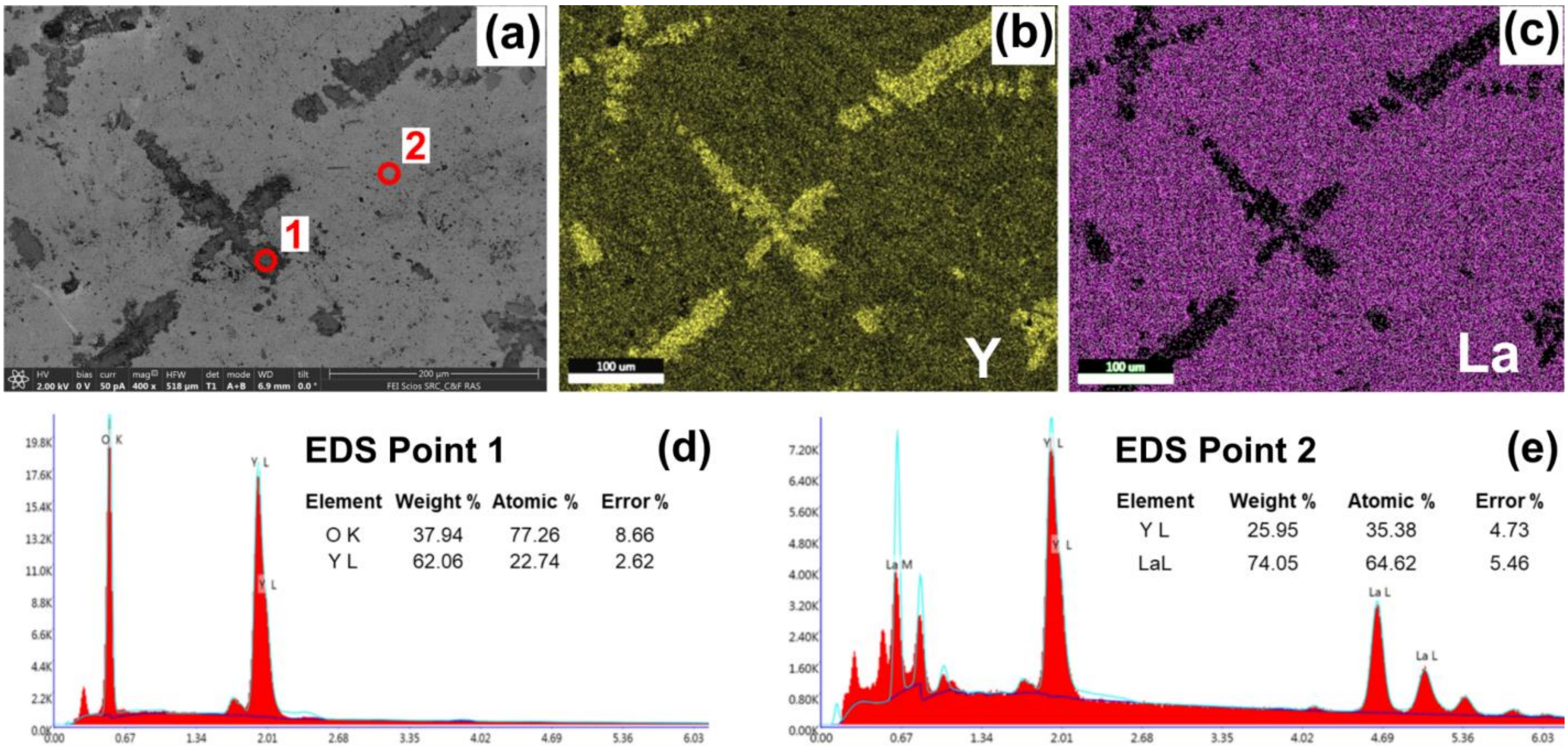

**Figure S2**. SEM image of the initial $La_2Y$ alloy at 0 GPa (a), elemental distribution maps of Y (b) and La (c), and analysis of EDX spectra collected from point 1 (d) and point 2 (e).

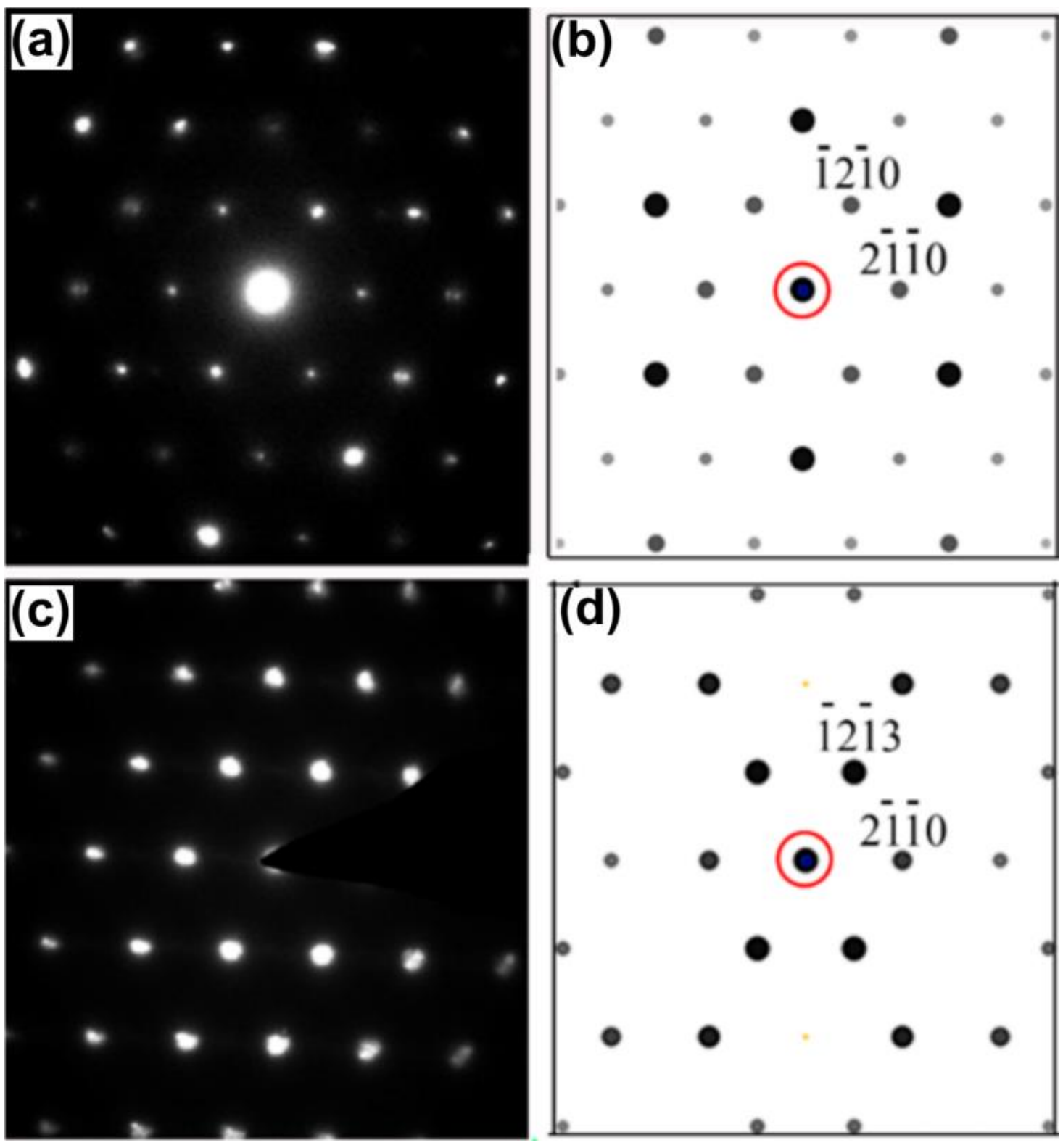

**Figure S3.** The selected area electron diffraction (SAED) from the transparent part of the $La_2Y$ alloy specimen (a, c) and the corresponding simulated selected area diffraction patterns (SADP) (b, c).

The electron diffraction (ED) study of the sample was performed on the next step. The selected area ED (SAED) pattern from the transparent area of the specimen (Fig. S3(a)) exhibited nearly uniform orientation with the distinct six-fold symmetry. That kind of symmetry pointed to hexagonal crystal lattice, and there are two choices between LaY compounds with *hcp* crystal structure: 1) $La_{0.5}Y_{0.5}$ [42] with unit cell dimensions as $a$ = 0.369(5) nm, $c$ = 1.185(2) nm, and 2) $La_{0.307}Y_{0.693}$ [43] with unit cell $a$ = 0.3702 nm and $c$ = 0.5844 nm. In both cases space group (S.G.) is $P6_3/mmc$. The lattice parameter $a$, obtained from our experiment, was slightly larger: $a$ = 0.375 nm. In order to resolve between two crystal lattices, the tilting experiment was performed. The SADP of one of the closest low index diffraction patterns to the zone axis shown in Fig.

S3(a) was approximately 16° away. After comparison of simulated SADP for two crystals with different unit cell parameters, we found that the best match between the observed and simulated diffraction patterns is for the crystal with the unit cell parameter *c* = 1.18 nm, which is slightly smaller than of LaY alloy described above. The results of SADP simulations are presented in Figs. S3(b, d). The estimation of angles between reflections and distances indicated that zone axis shown in Fig. 3(a, c) are [0001] and [11-23], respectively. Thus, the investigated LaY compound adopted *hcp* crystal structure with a parameter slightly larger than described in Ref. [42].

High-resolution TEM image of the alloy in [0001] zone axis is presented in Fig. S4. The bright and dark long period contrast variations looking as moiré fringes observed in different areas could arise from defects like dislocations in the specimen. The HAADF STEM mages of the sample demonstrate uniform contrast and we assume uniform distribution of the elements within the specimen.

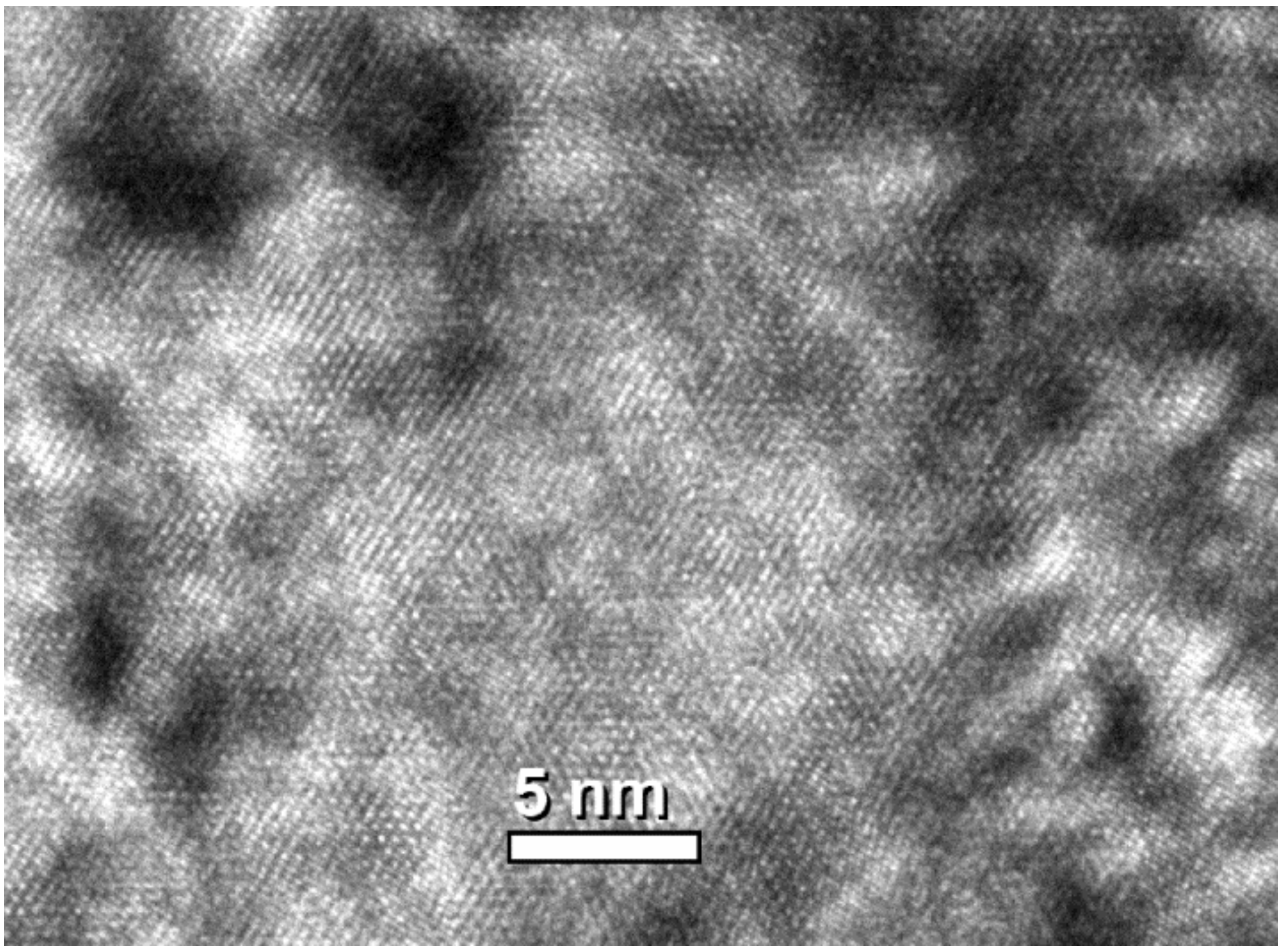

**Figure S4.** High resolution TEM image of the initial $La_2Y$ alloy in [0001] zone axis.

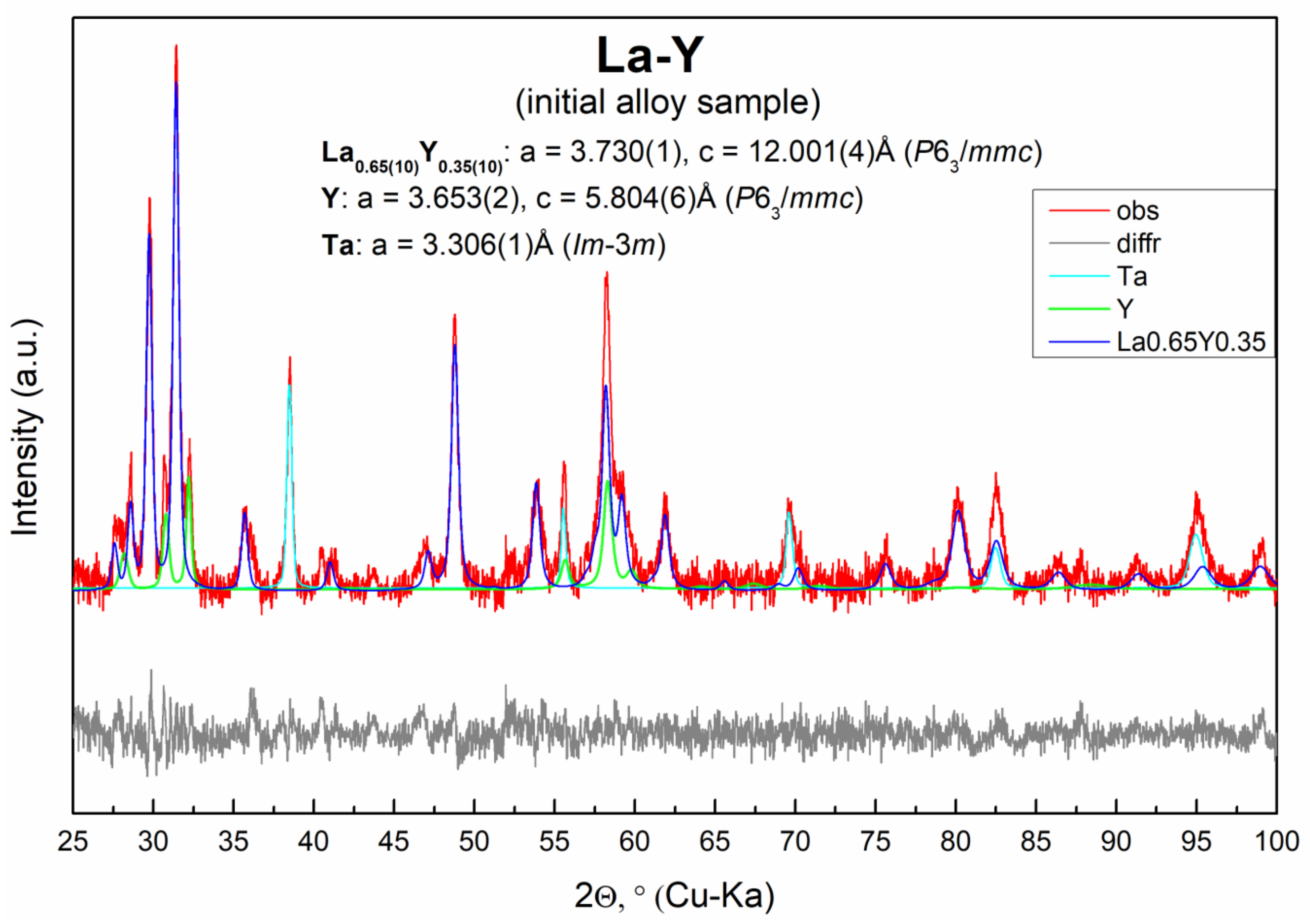

**Figure S5.** Powder X-ray diffraction analysis of the initial $La_2Y$ alloy at 0 GPa. The experimental data are shown in red; peak deconvolution of the fitted XRD profile — in blue ($La_{0.65}Y_{0.35}$), green (Y), and azure (Ta); residues are in gray. Tantalum is an impurity from the crucible material on the rim of the La-Y ingot.

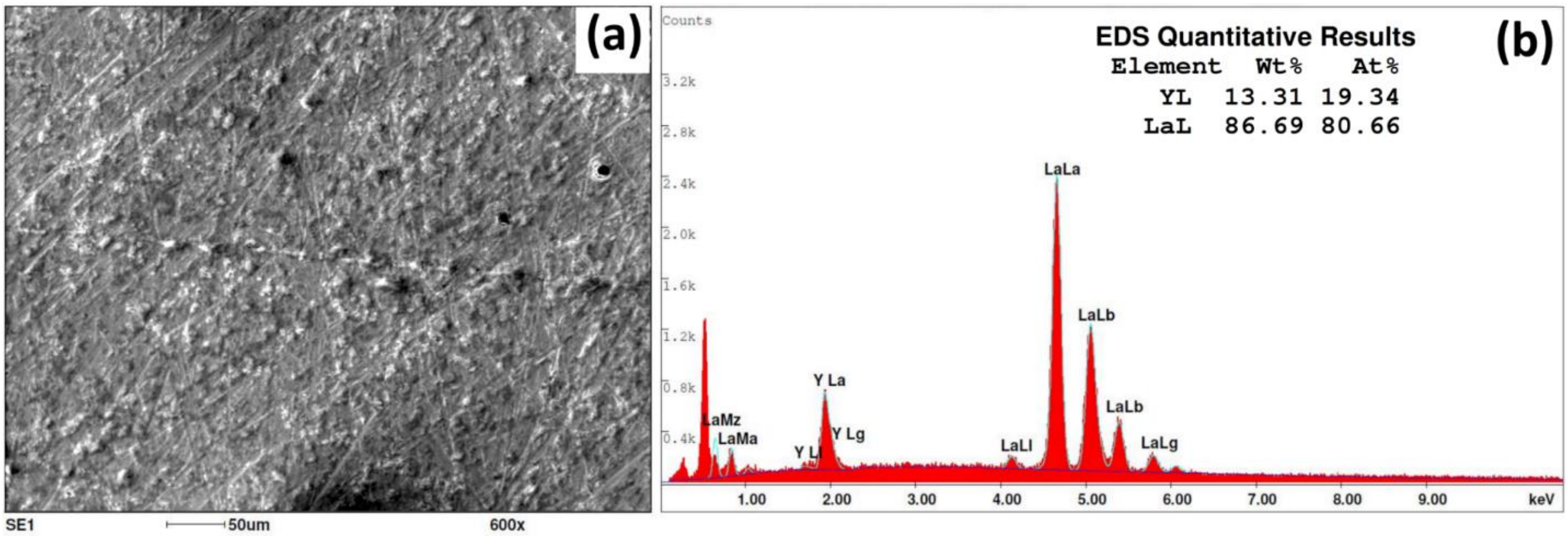

**Figure S6.** SEM image and EDX analysis of the whole shown area of the initial $La_4Y$ alloy at 0 GPa. The sample is uniform.

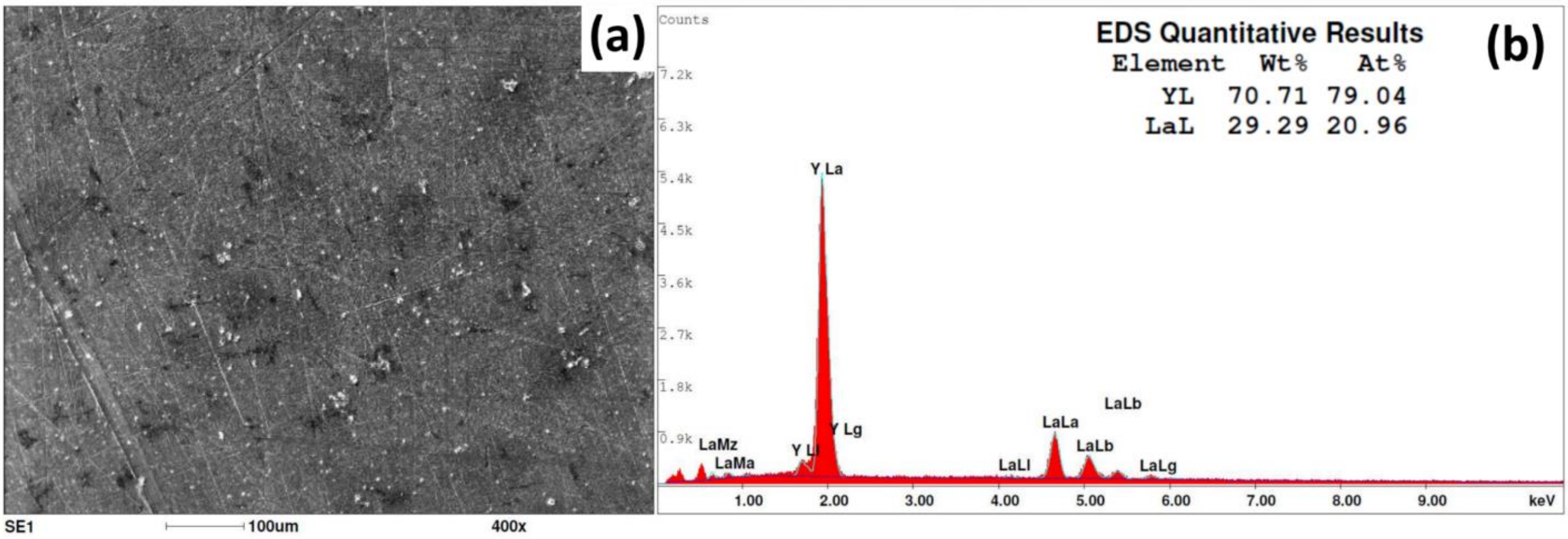

**Figure S7.** SEM image and EDX analysis of the whole shown area of the initial $LaY_4$ alloy at 0 GPa. The sample is uniform (see below).

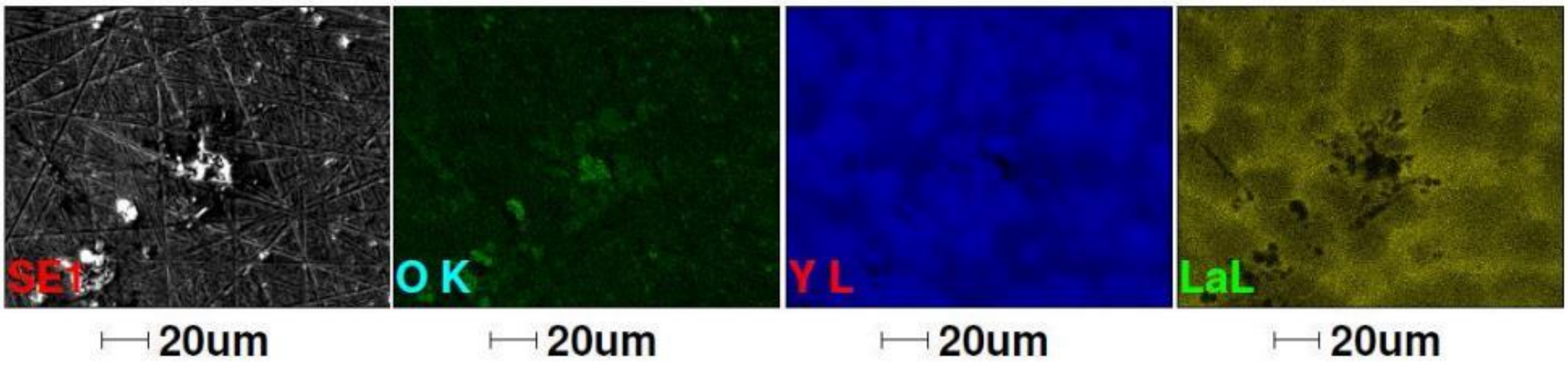

**Figure S8.** SEM image and elemental distribution maps in the initial $LaY_4$ alloy in a region with non-uniform distribution of Y and La. The size of regions with the nonstoichiometric La–Y composition is about 15 μm.

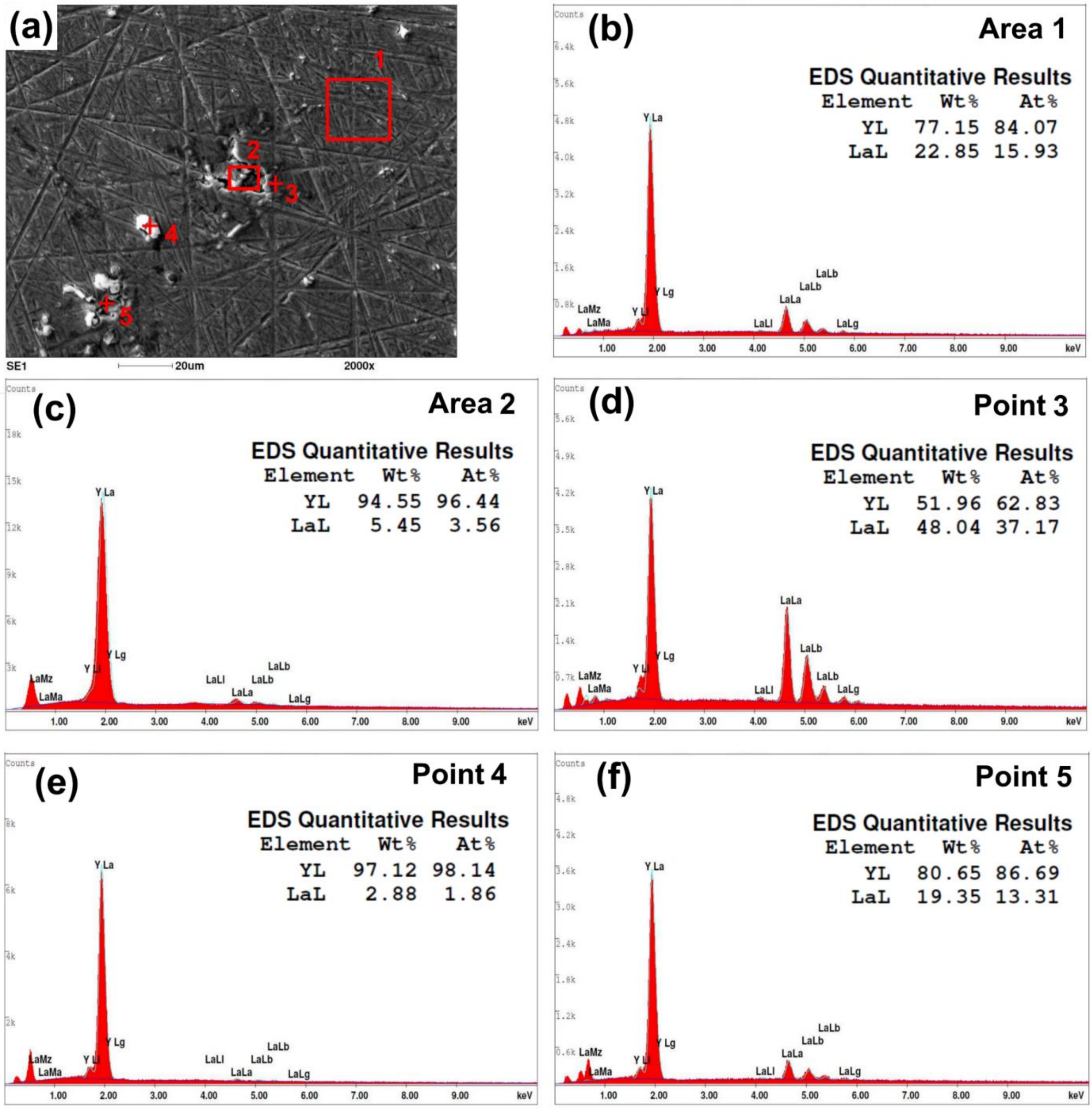

**Figure S9.** SEM image of $LaY_4$ alloy in a region with non-uniform distribution of Y and La (a), EDX spectra were recorded at the areas and points indicated on the SEM image (b-f).

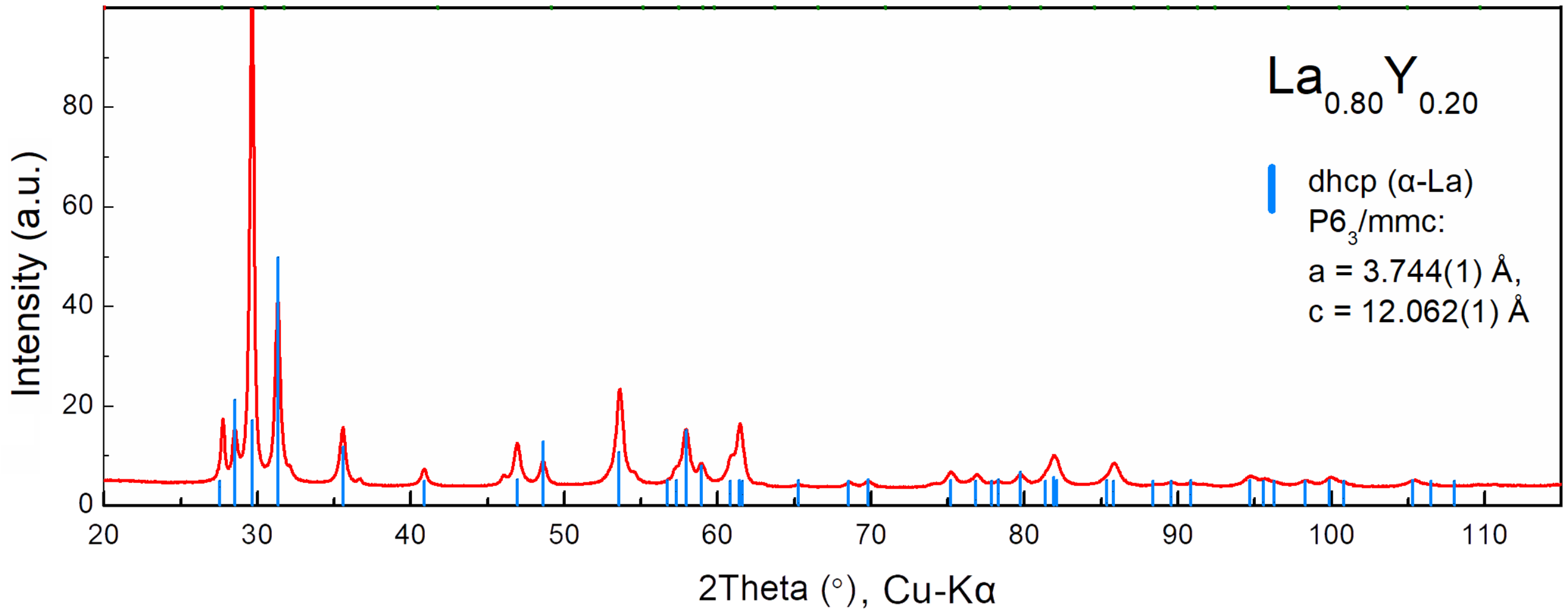

**Figure S10.** Powder X-ray diffraction analysis of the initial $La_4Y$ alloy at 0 GPa. The experimental data are shown in red. The positions of the Bragg reflections for the α-La phase with refined parameters (space group *P*$6_3$/*mmc*, *a* = 3.744 Å, *c* = 12.062 Å) are marked in blue.

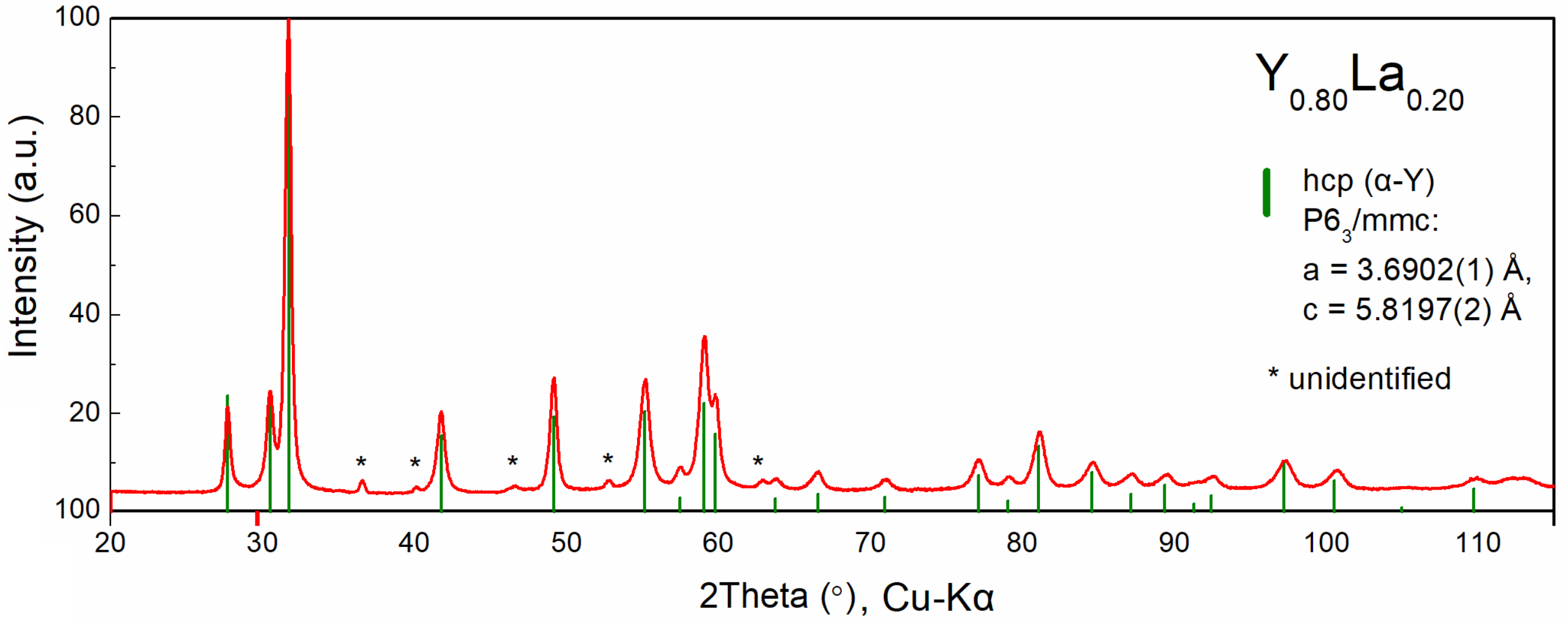

**Figure S11.** Powder X-ray diffraction analysis of the initial $LaY_4$ alloy at 0 GPa. The experimental data are shown in red. The positions of the Bragg reflections for the α-Y phase with refined parameters (space group *P*$6_3$/*mmc*, *a* = 3.6902 Å, *c* = 5.8197 Å) are marked in green. Asterisks indicate unidentified reflections.

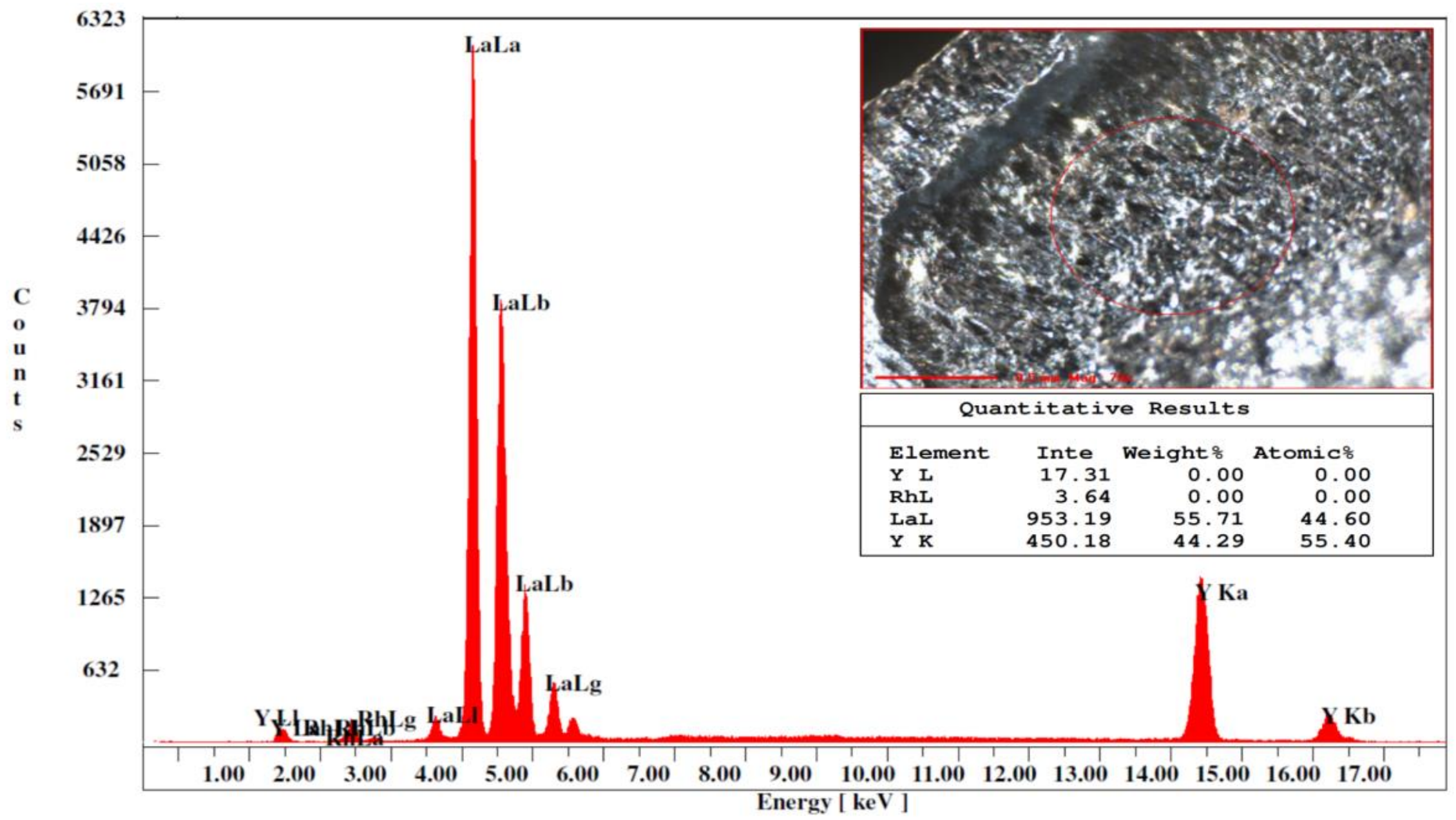

**Figure S12**. XRF analysis of the initial LaY (1:1) alloy at 0 GPa.

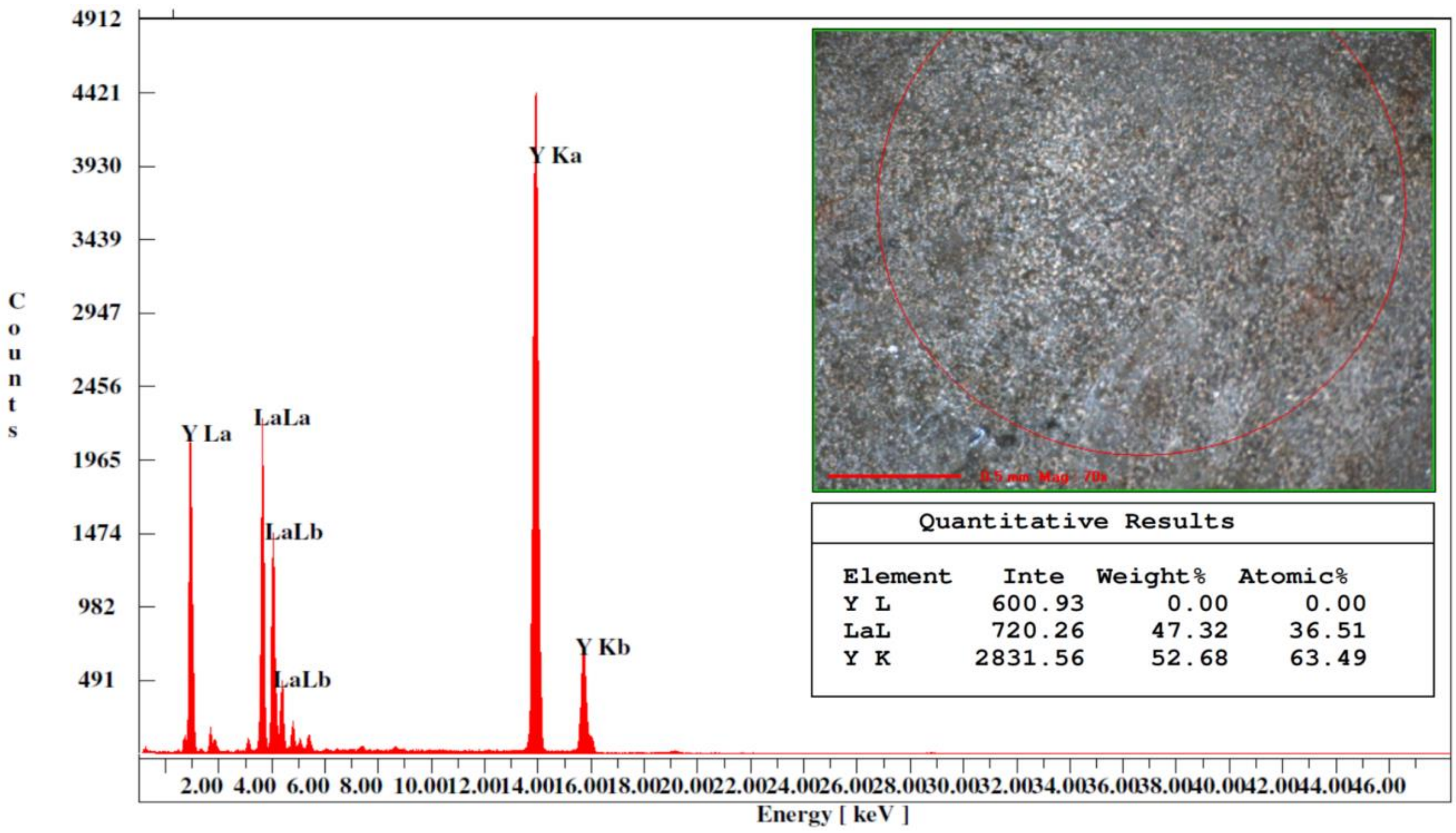

**Figure S13.** XRF analysis at a different sample (compare Figure S12) of the initial LaY (1:1) alloy at 0 GPa.

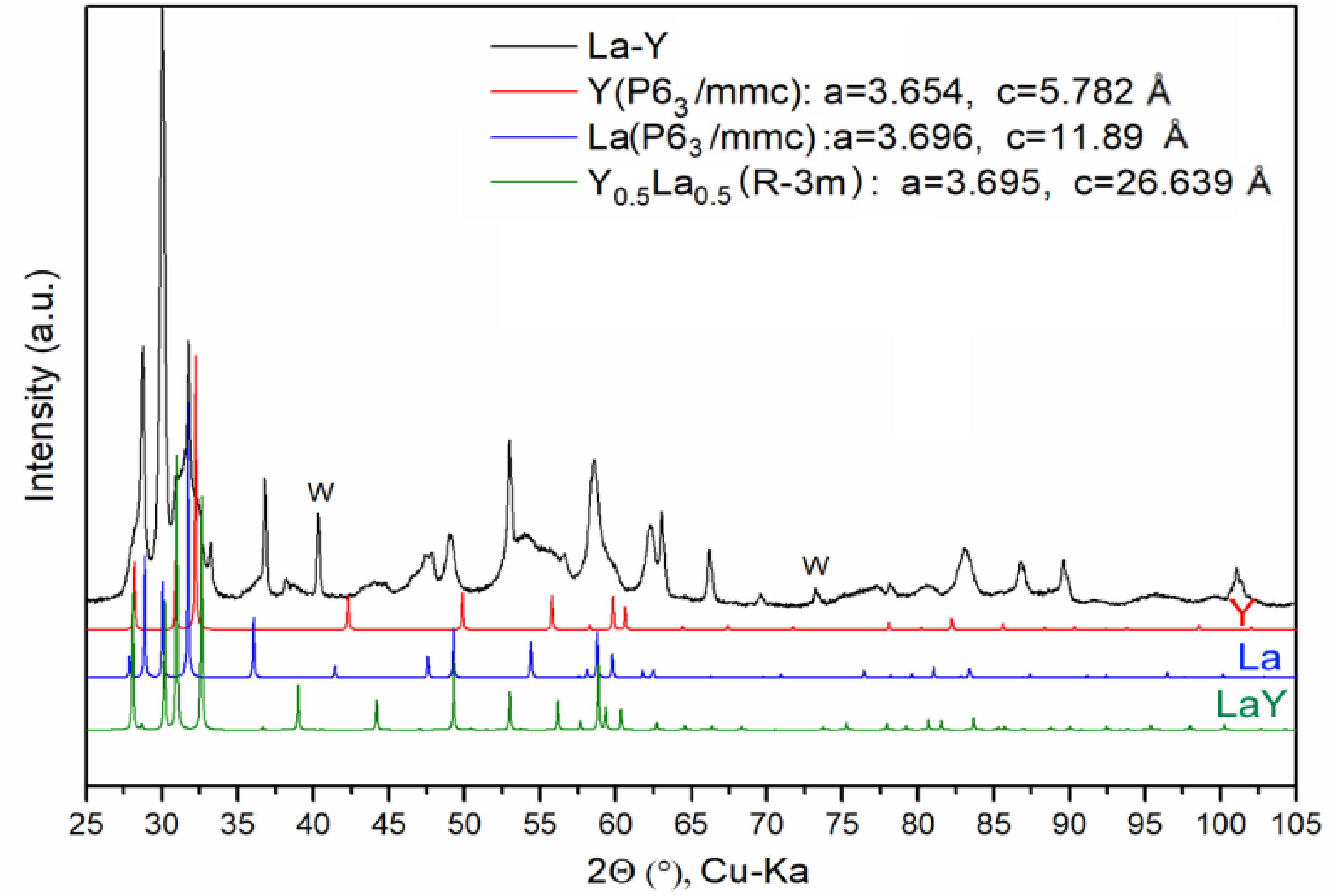

**Figure S14.** Powder X-ray diffraction analysis of the initial LaY alloy at 0 GPa. The positions of the Bragg reflections for *hcp*-Y, *hcp*-La, and $R\bar{3}m$-LaY (more matching phase ICSD# 642079 [44]) are shown in red, blue, and green, respectively.

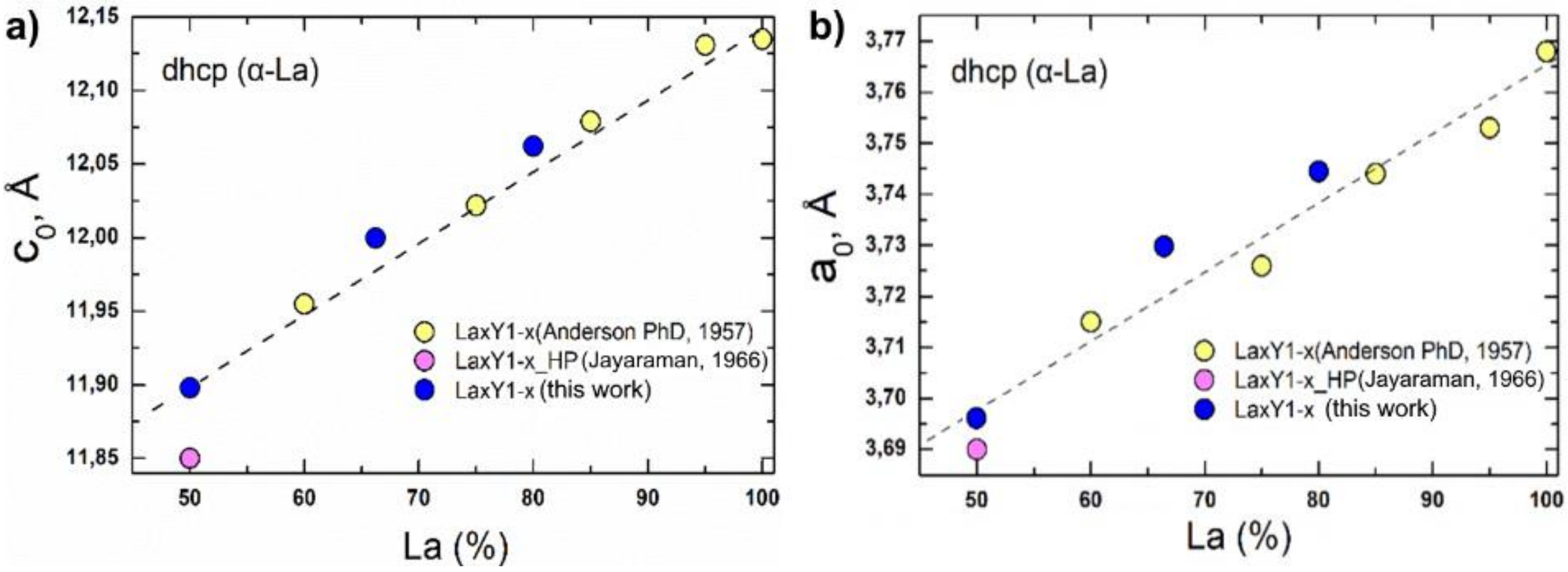

**Figure S15**. Dependencies of the unit cell parameters of La–Y alloys with the α-La structure on the lanthanum concentration (atom %). The dashed line indicates the linear fit of the experimental data.[42,45–47] Blue circles mark the data obtained in this work.

# X-ray diffraction data

**Table S4.** Experimental and predicted lattice parameters and unit cell volumes of $Fm\bar{3}m$-$LaH_{10}$ (Z=4), $P6_3mc$-$LaH_{10}$ (Z=2), $Fm\bar{3}m$-(La,Y)$H_{10}$ (Z = 4), *Imm*2-$YH_7$ (*Z* = 2), and *Cmcm*-$LaH_3$ (*Z* = 4). For $Im\bar{3}m$-(La,Y)$H_6$ and *I4/mmm*-(La,Y)$H_4$ the calculated unit cell volumes $V_{DFT}$ corresponds to $La_2YH_{18}$ and $La_2YH_{12}$, respectively.

| | | | | | | |
|---|---|---|---|---|---|---|
| | | | **$Fm\bar{3}m$-$LaH_{10}$** | | | |
| DAC | Pressure, GPa | *a*, Å | | | *V*, Å$^3$ | |
| SL3_S | 171 | 5.08(1) | | | 131.3(1) | |
| | 171 | 5.07(1) | | | 130.8(1) | |
| | | | **$P6_3mc$-$LaH_{10}$** | | | |
| DAC | Pressure, GPa | *a*, Å | | *c*, Å | *V*, Å$^3$ | |
| SL3_S | 171 | 3.60(1) | | 5.88(1) | 66.1(1) | |
| | | | **$Fm\bar{3}m$-(La,Y)$H_{10}$** | | | |
| DAC | Pressure, GPa | *a*, Å | | | *V*, Å$^3$ | |
| M1 | 180 | 5.038(1) | | | 127.86(2) | |
| M2 | 180 | 5.026(1) | | | 126.98(2) | |
| SL1 | 180 | 5.031(1) | | | 127.32(1) | |
| SL1_S* | 171 | a=5.10(1) Å, b=5.05(3) Å, c=5.08(1) Å, α = 89.9(2)°, β = 89.6(1)°, γ =89.6(4)° | | | 130.7(9) | |
| SL3 | 170 | 5.071(1) | | | 130.40(1) | |
| | | | **$Imm2$-$YH_7$** | | | |
| DAC | Pressure, GPa | *a*, Å | *b*, Å | *c*, Å | *V*, Å$^3$ | $V_{DFT}$, Å$^3$ |
| M1** | 166 | 3.29(4) | 3.33(6) | 4.68(7) | 51.50 | 50.85 |
| SL3 | 170 | 3.303(1) | 3.322(2) | 4.672(2) | 51.25(2) | 50.40 |
| M1 | 180 | 3.279(2) | 3.305(2) | 4.641(2) | 50.30(1) | 49.58 |
| | | | **$Cmcm$-$LaH_3$** | | | |
| DAC | Pressure, GPa | *a*, Å | *b*, Å | *c*, Å | *V*, Å$^3$ | $V_{DFT}$, Å$^3$ |
| M1 | 180 | 2.791(4) | 10.492(5) | 2.657(3) | 77.83(2) | 77.64 |
| SL3 | 170 | 2.737(2) | 10.507(3) | 2.727(2) | 78.44(3) | 78.96 |
| | | | **$Im\bar{3}m$-(La,Y)$H_6$** | | | |
| DAC | Pressure, GPa | *a*, Å | | | *V*, Å$^3$ | $V_{DFT}$, Å$^3$ |
| M2_S | 175 | 3.65(1) | | | 24.26 | 24.15 |
| M2_S | 180 | 3.64(1) | | | 24.16 | 23.96 |
| | | | ***I4/mmm*-(La,Y)$H_4$** | | | |
| DAC | Pressure, GPa | *a*, Å | | *c*, Å | *V*, Å$^3$ | $V_{DFT}$, Å$^3$ |
| M2_S | 180 | 2.74(1) | | 5.56(1) | 20.90 | 21.11 |

* Refined as *P*1 (pseudocubic)

** Data from Ref. [48]

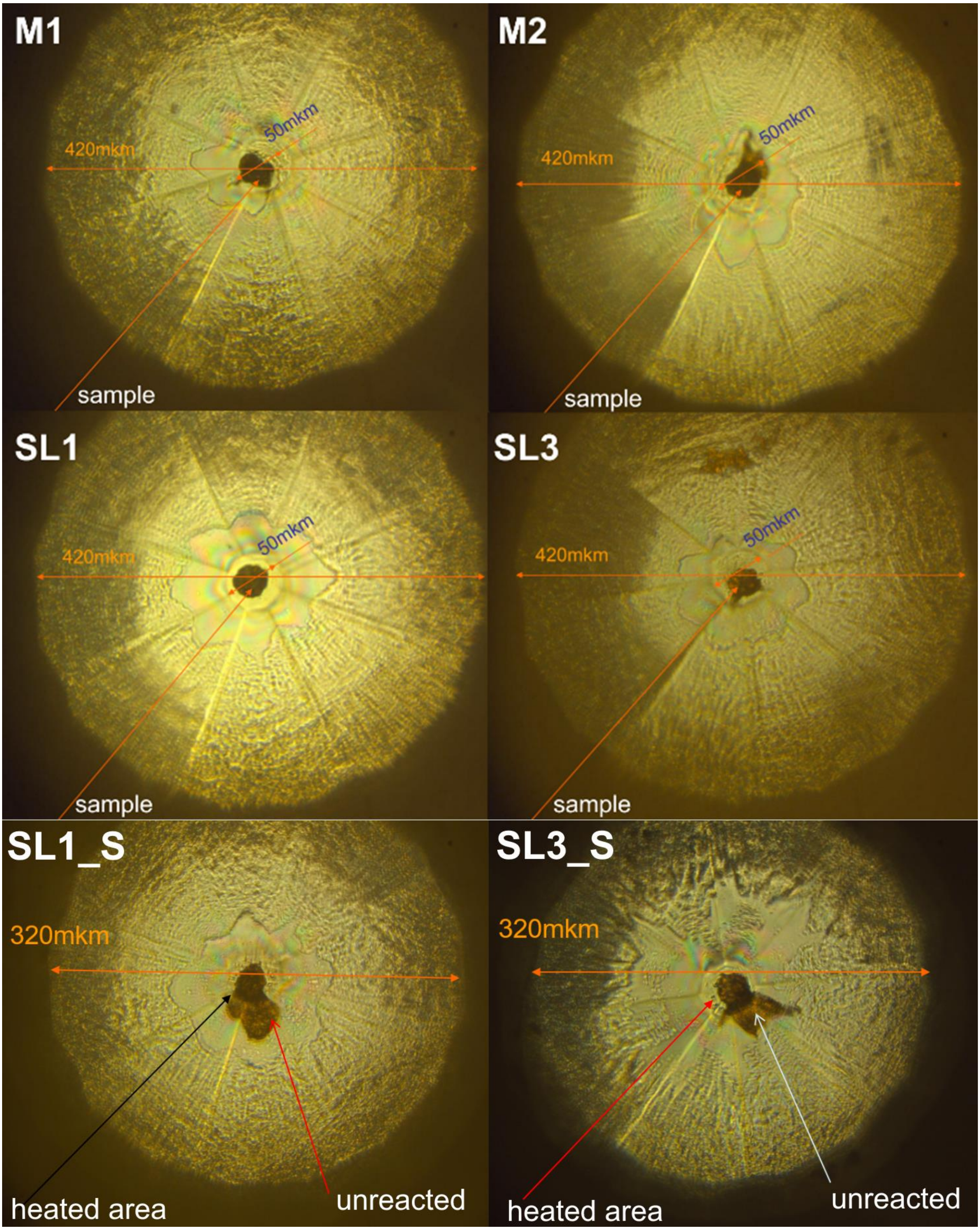

**Figure S16.** Diamond anvil culets of DACs M1, M2, SL1, SL1_S, SL3 and SL3_S, loaded with La–Y alloys and AB, after the laser heating.

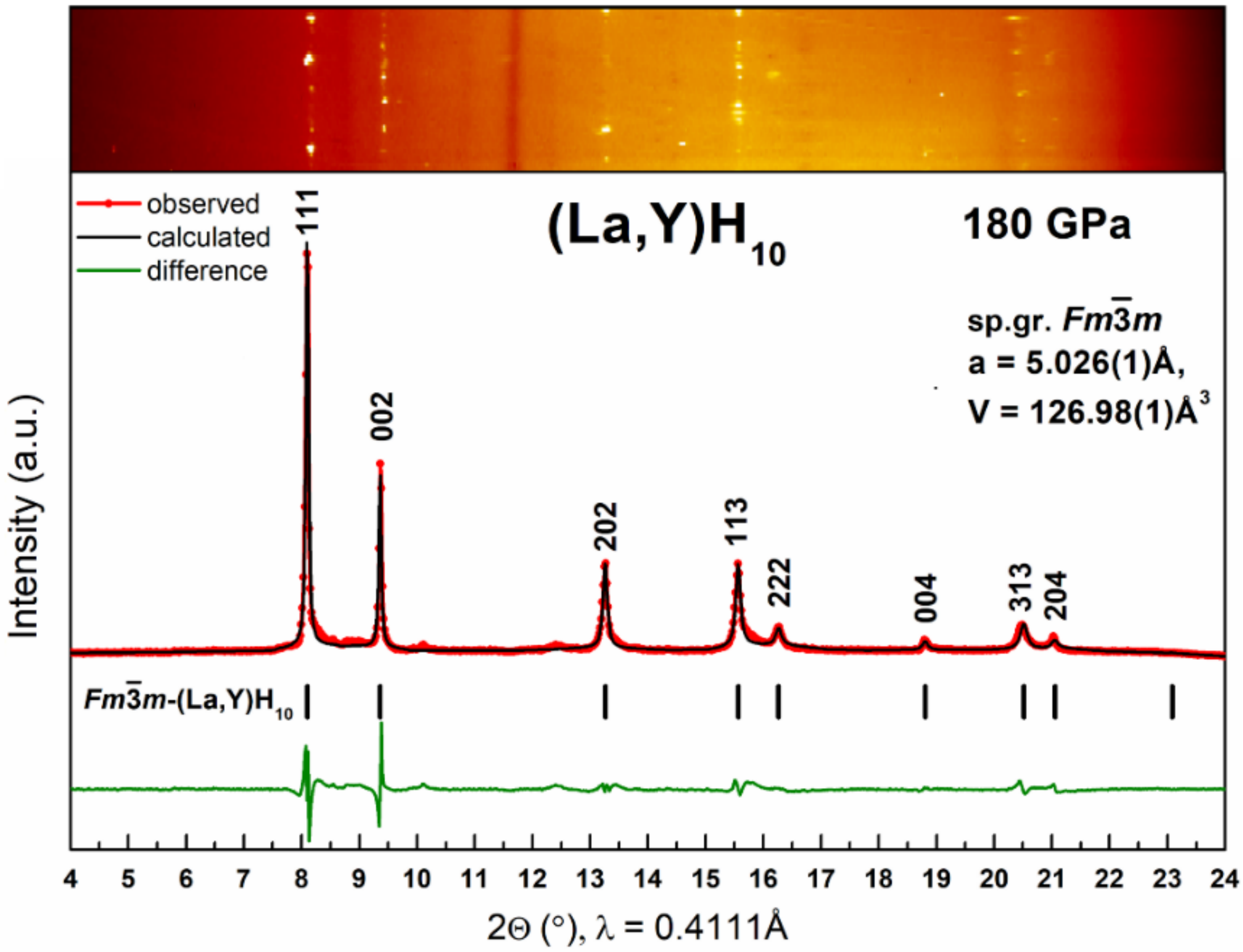

**Figure S17**. Le Bail refinement of *Fm*$\bar{3}$*m*-(La,Y)$H_{10}$ and the experimental XRD pattern at 180 GPa (DAC M2, $La_2Y$). The experimental data, model fit for the structure, and residues are shown in red, black, and green, respectively.

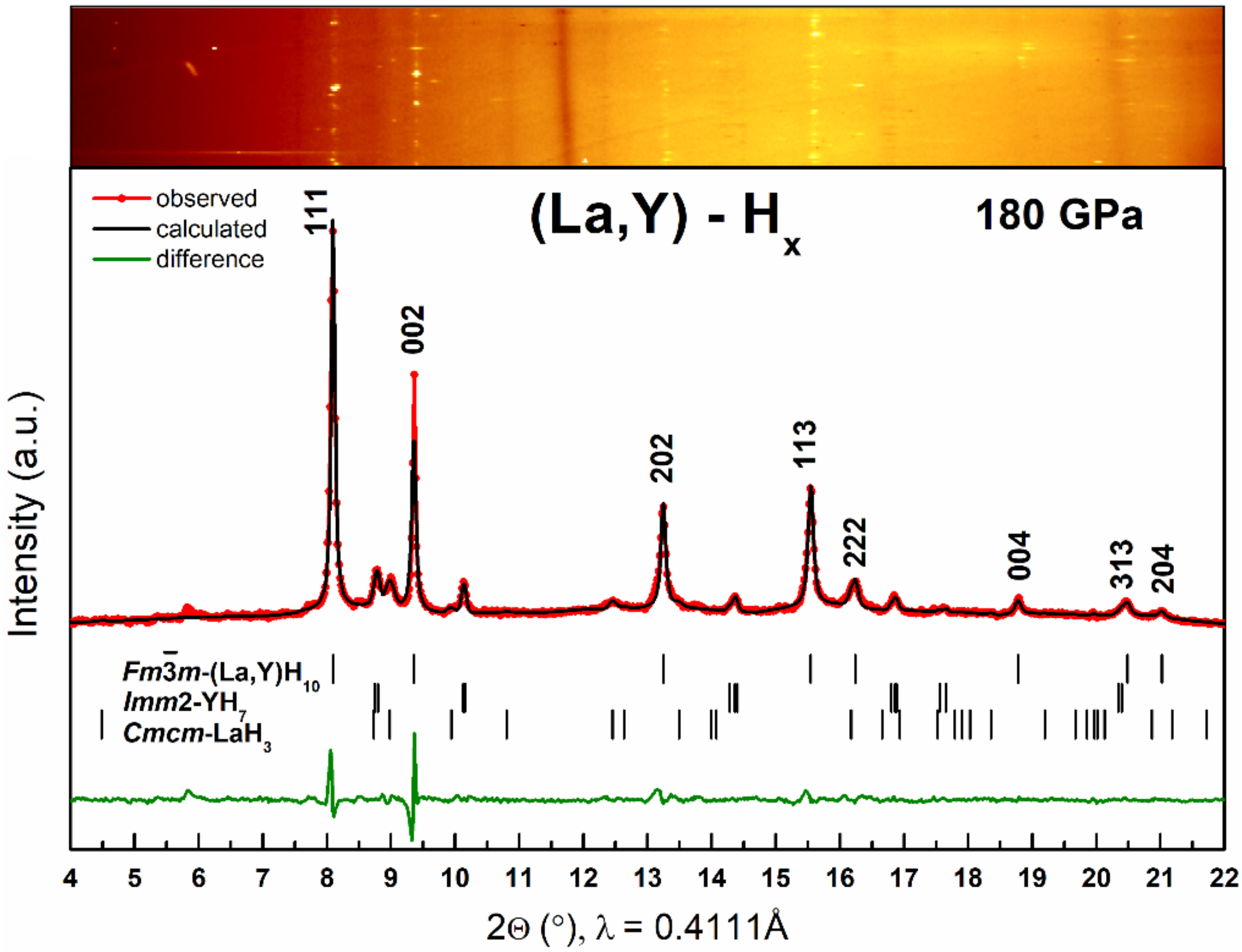

**Figure S18.** Le Bail refinement of *Fm*$\bar{3}$*m*-(La,Y)$H_{10}$, *Imm*2-$YH_7$, and *Cmcm*-$LaH_3$, and the experimental XRD pattern at 180 GPa (DAC M1, LaY). The experimental data, model fit for the structure, and residues are shown in red, black, and green, respectively.

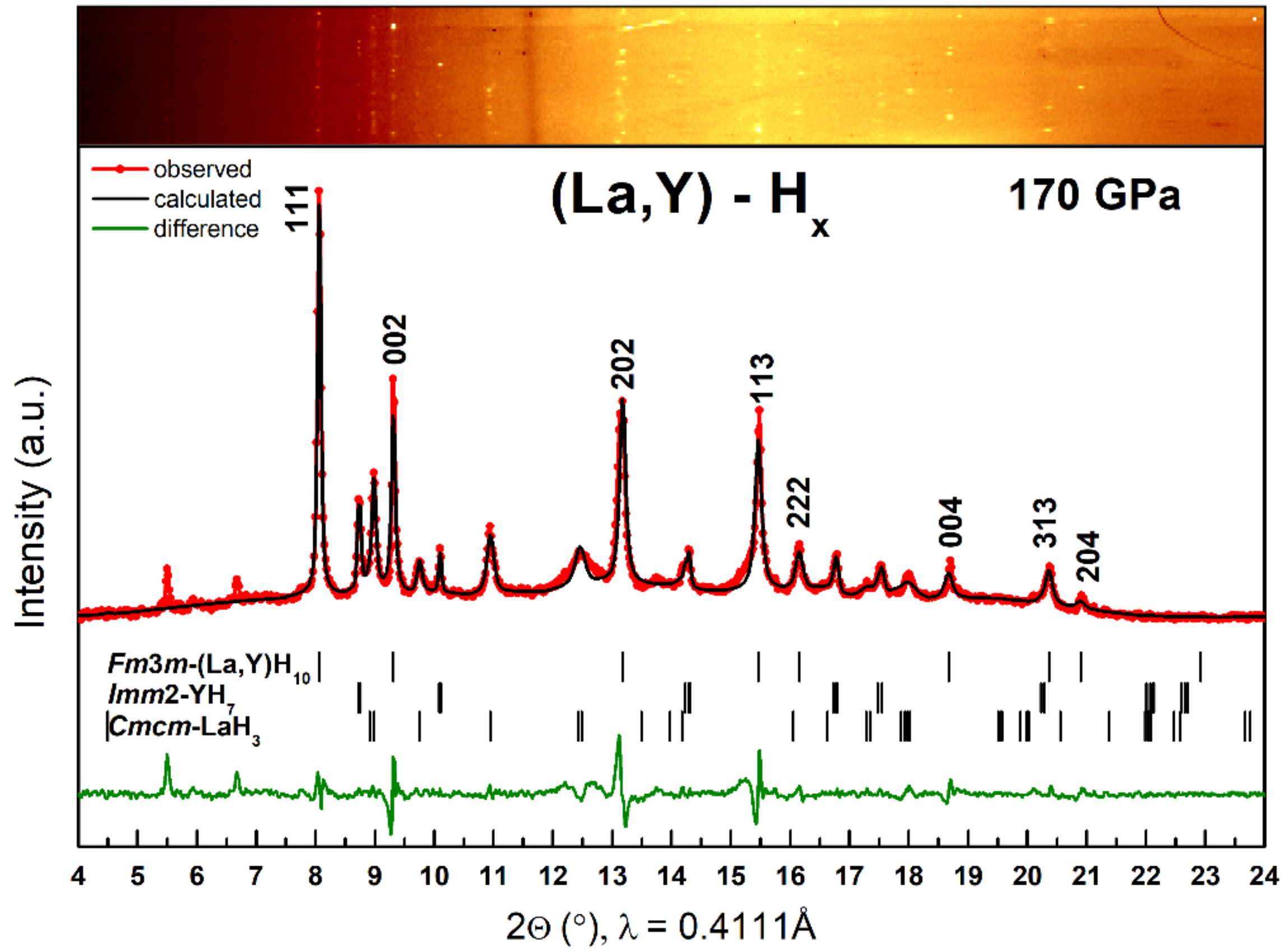

**Figure S19.** Le Bail refinement of *Fm*$\bar{3}$*m*-(La,Y)$H_{10}$, *Imm*2-$YH_7$, and *Cmcm*-$LaH_3$, and the experimental XRD pattern at 170 GPa (DAC SL3, $LaY_4$). The experimental data, model fit for the structure, and residues are shown in red, black, and green, respectively.

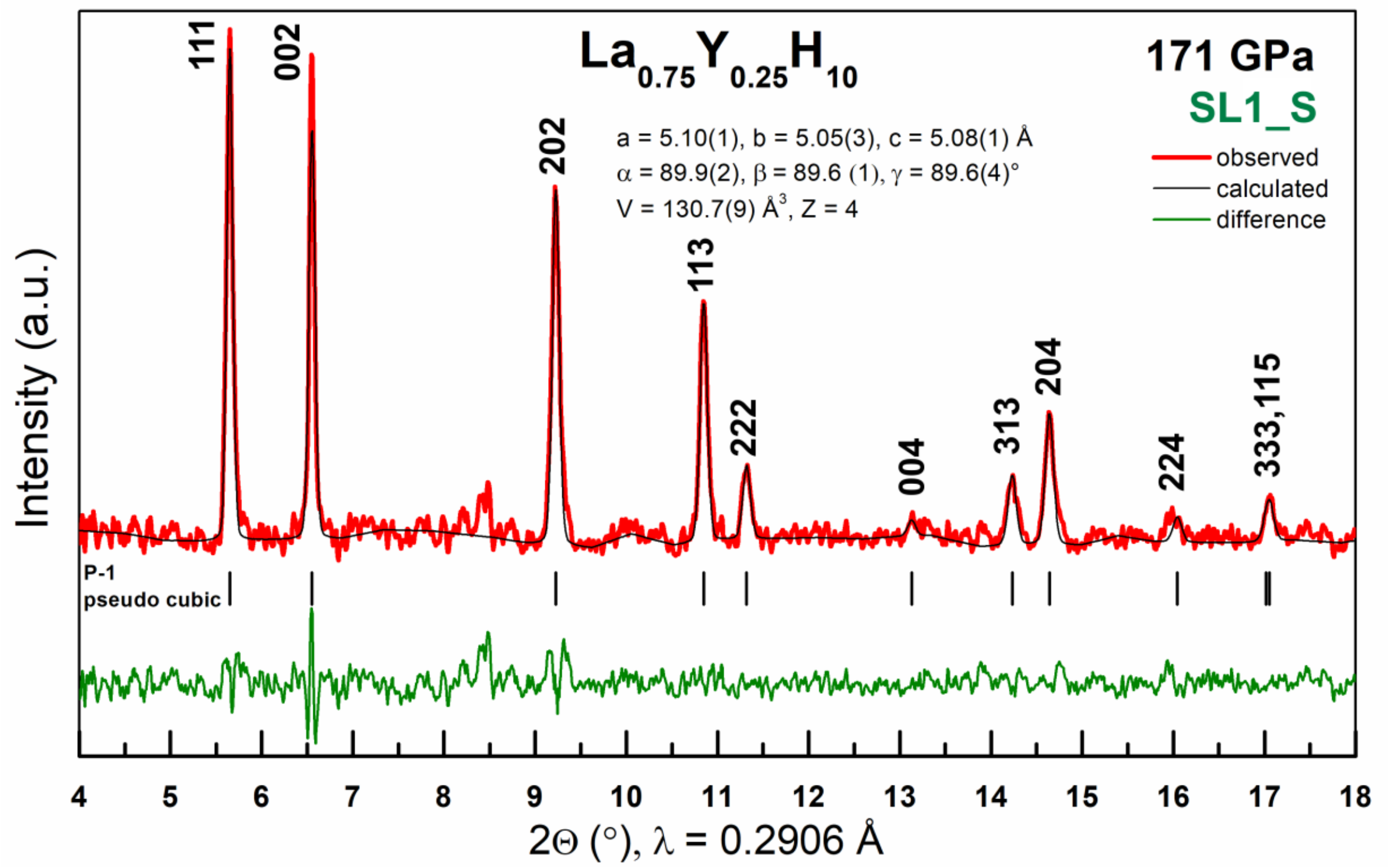

**Figure S20.** Le Bail refinement of pseudocubic *P*1 (distorted *Fm*$\bar{3}$*m)*-(La,Y)$H_{10}$, and the experimental XRD pattern at 171 GPa (DAC SL1_S, $La_3Y$). The experimental data, model fit for the structure, and residues are shown in red, black, and green, respectively. Reflections are marked as usual for cubic crystals.

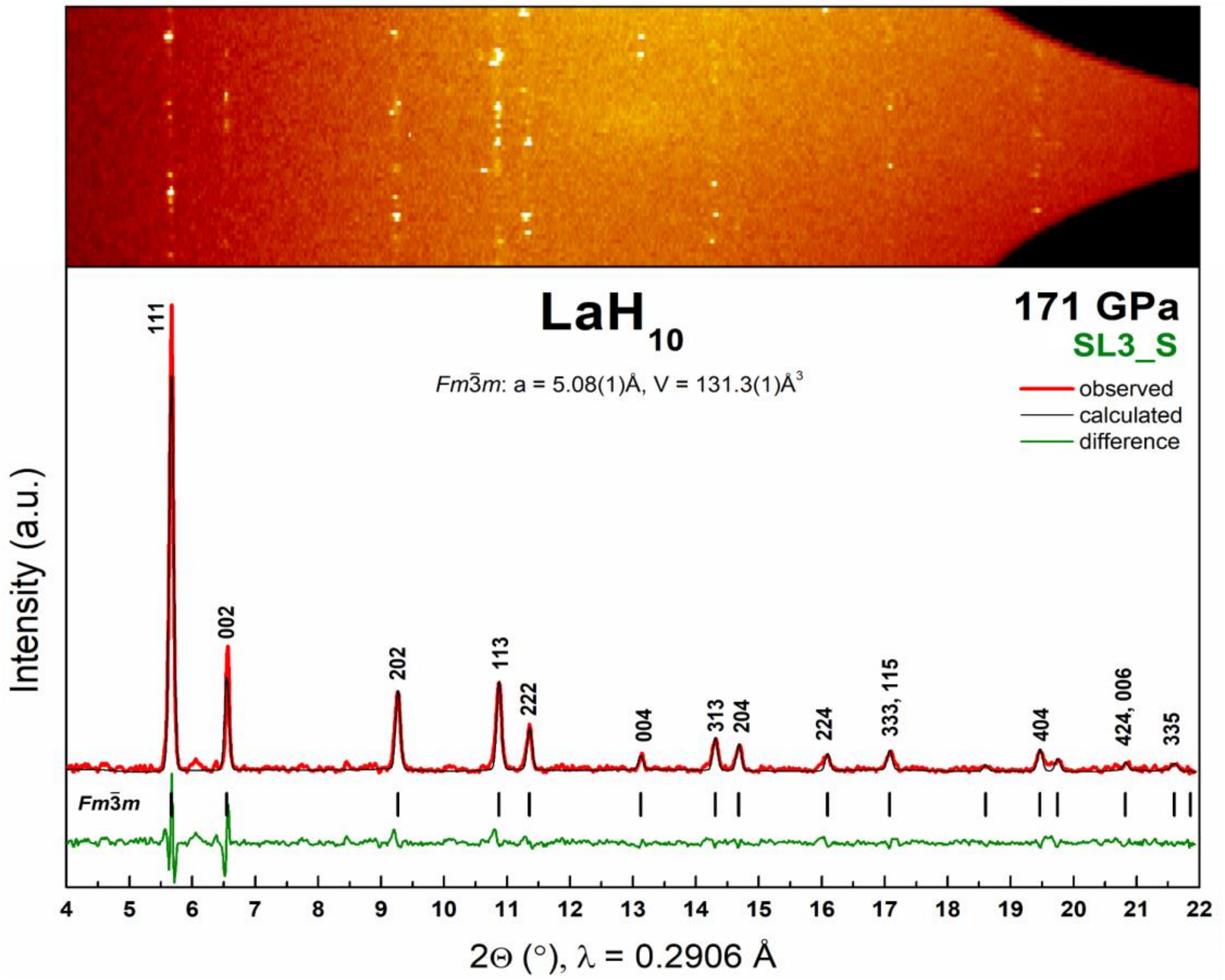

**Figure S21.** Le Bail refinement of *Fm*$\bar{3}$*m*-$LaH_{10}$ and the experimental XRD pattern at 171 GPa (DAC SL3_S, pure La). The experimental data, model fit for the structure, and residues are shown in red, black, and green, respectively.

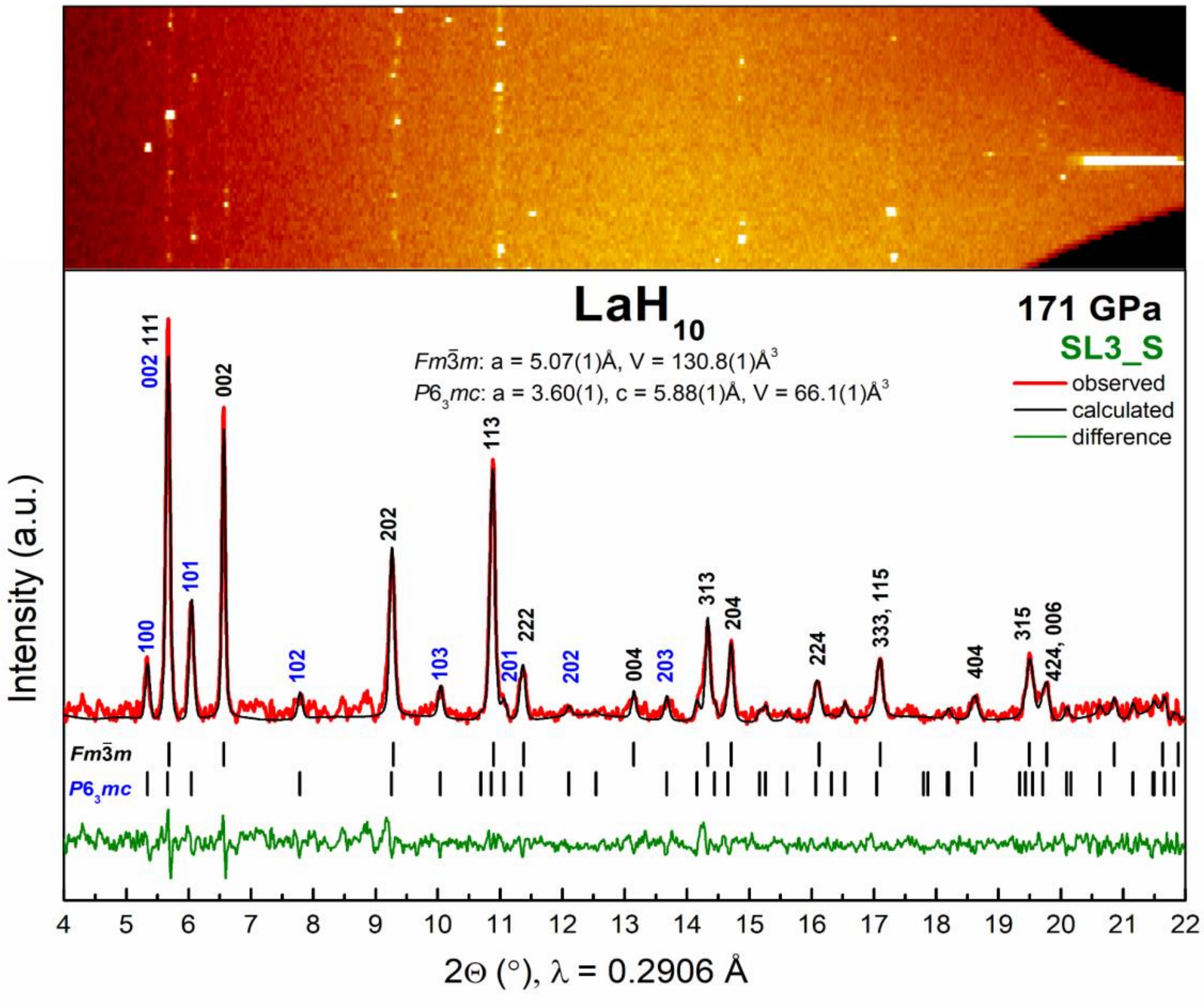

**Figure S22.** Le Bail refinement of *Fm*$\bar{3}$*m*-$LaH_{10}$, *P*$6_3$*mc*-$LaH_{10}$, and the experimental XRD pattern at 171 GPa (DAC SL3_S, pure La). The experimental data, model fit for the structure, and residues are shown in red, black, and green, respectively.

## Raman spectra

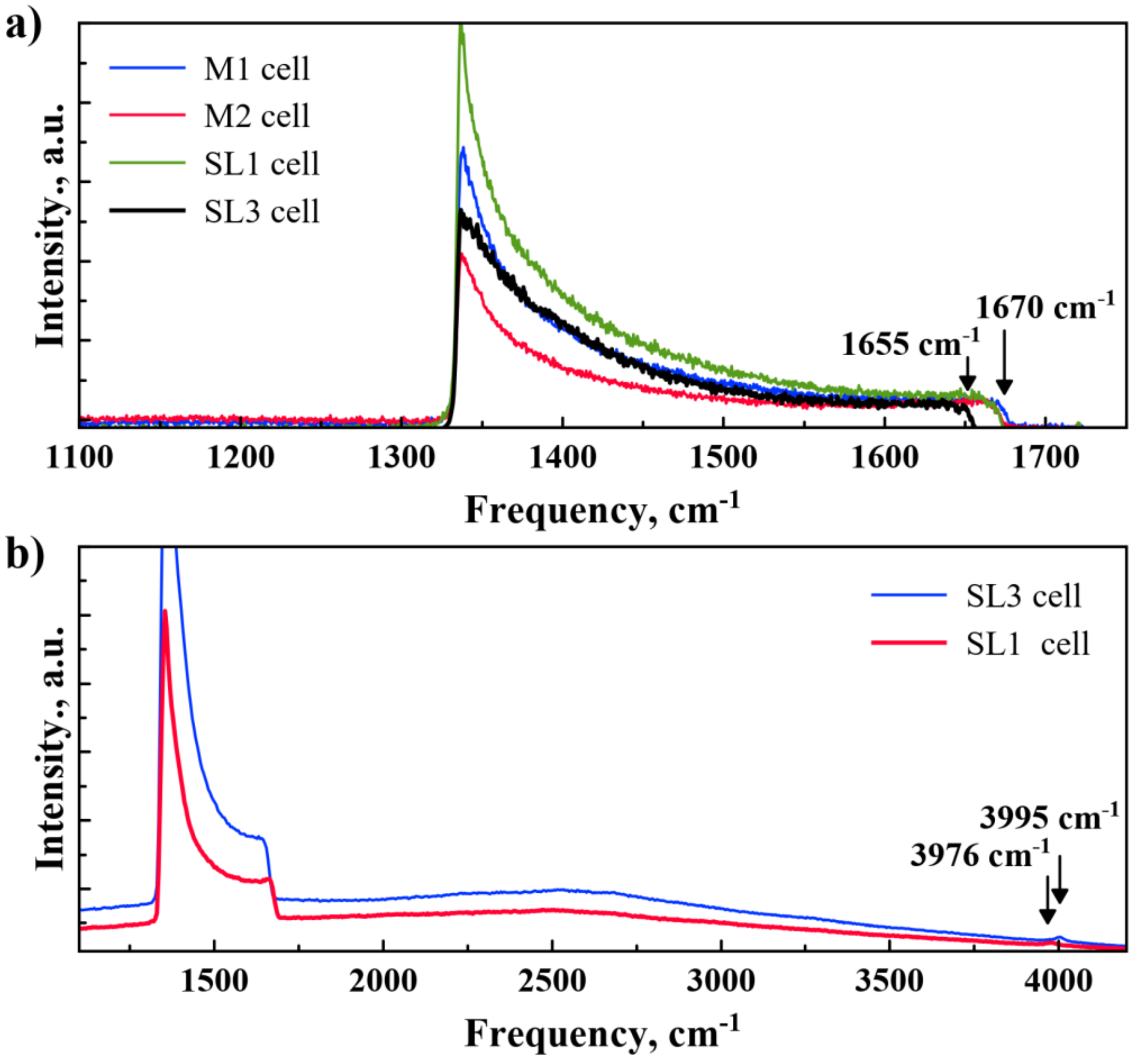

**Figure S23.** Raman spectra of the $Fm\bar{3}m$-(La,Y)$H_{10}$ samples in DACs (a) M1-2, SL1,3; (b) SL1 and SL3 at 170–180 GPa after the laser heating.

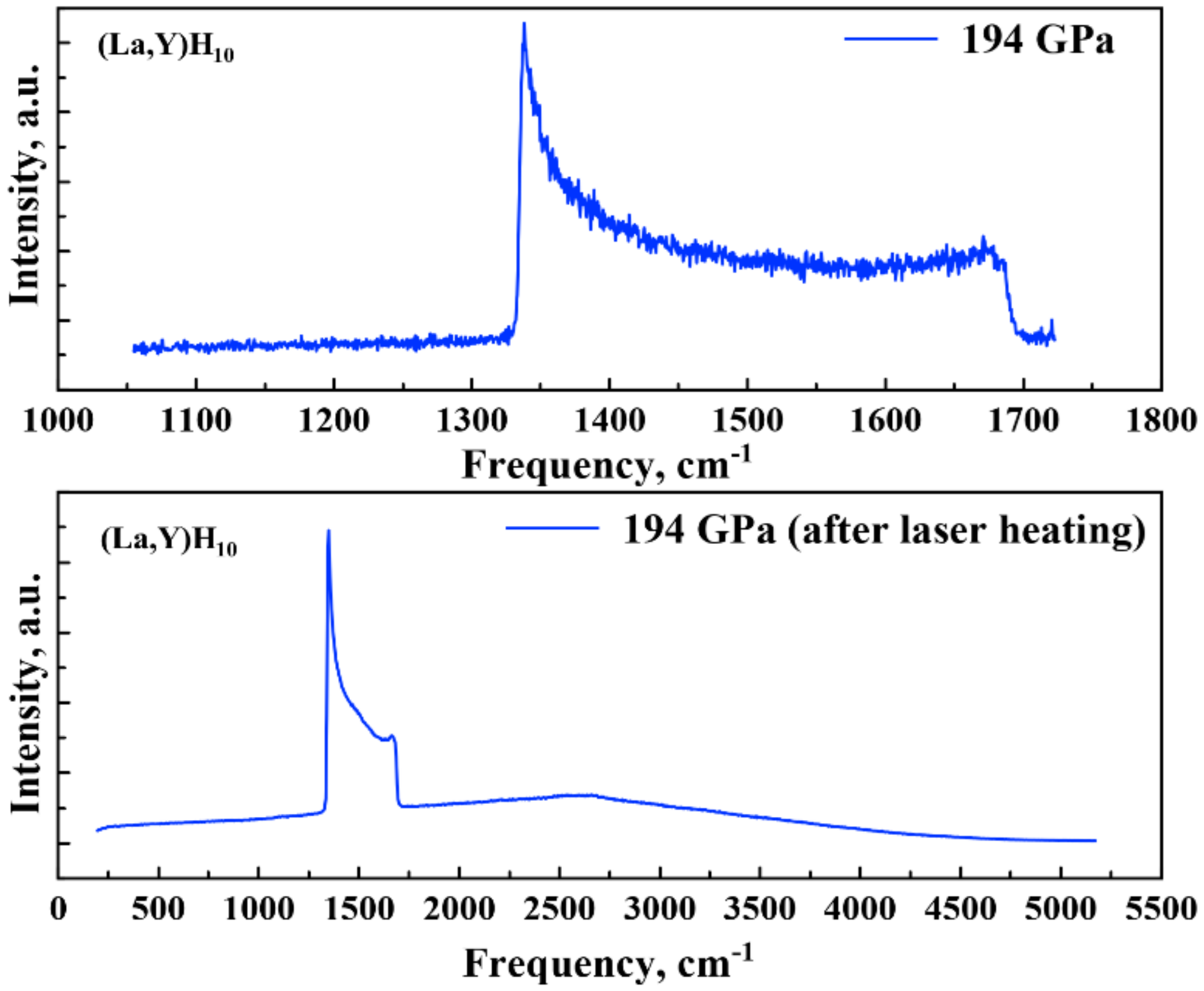

**Figure S24.** Raman spectra of the (La,Y)$H_{10}$ samples obtained from $La_2Y$ alloy for the electric transport measurements at 194 GPa. The absence of the $H_2$ vibron may be due to complete absorption of hydrogen by the sample.

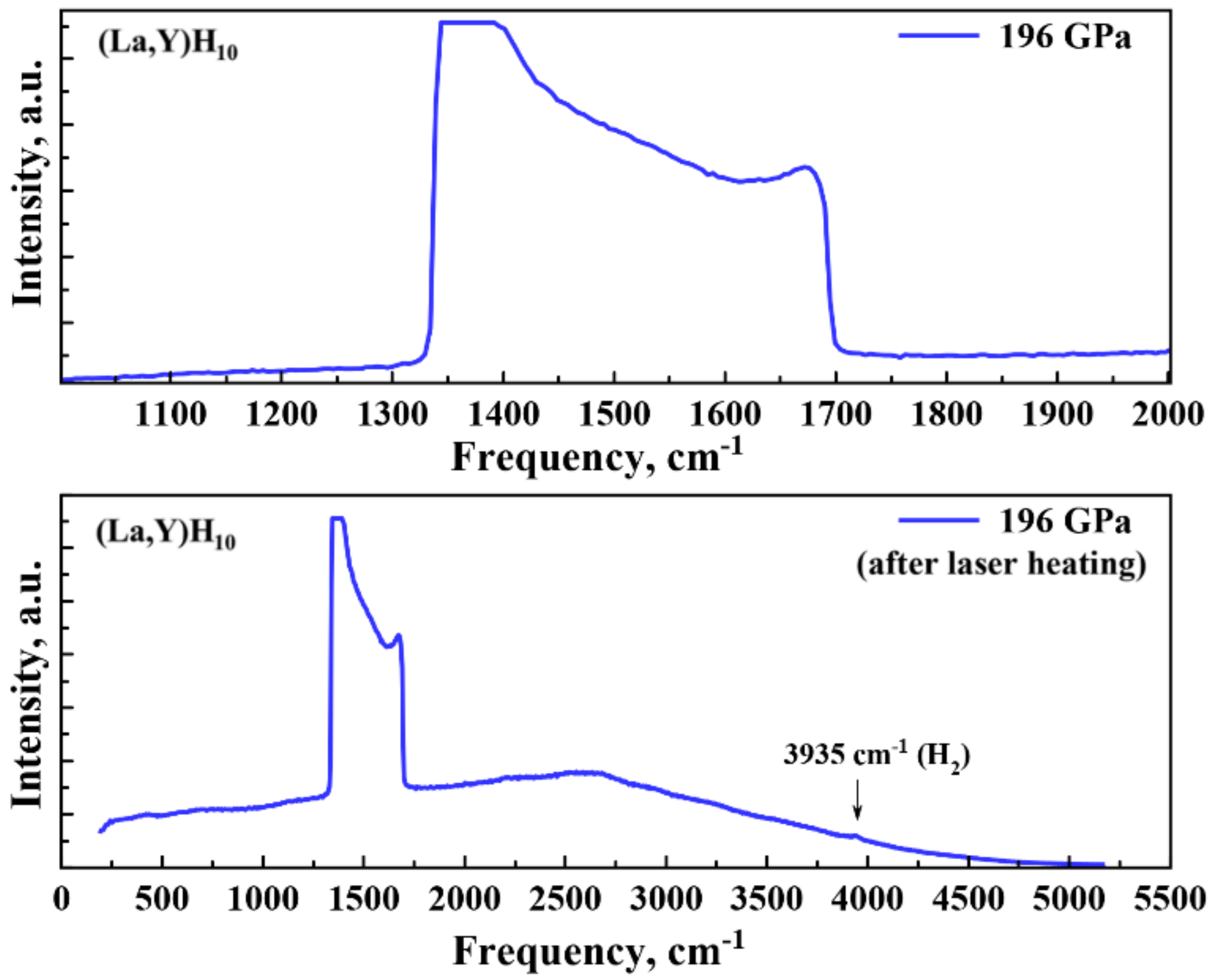

**Figure S25.** Raman spectra of the $(La,Y)H_{10}$ sample obtained from $La_2Y$ alloy for the electric transport measurements at 196 GPa after the laser heating.

## Transport measurements in magnetic fields

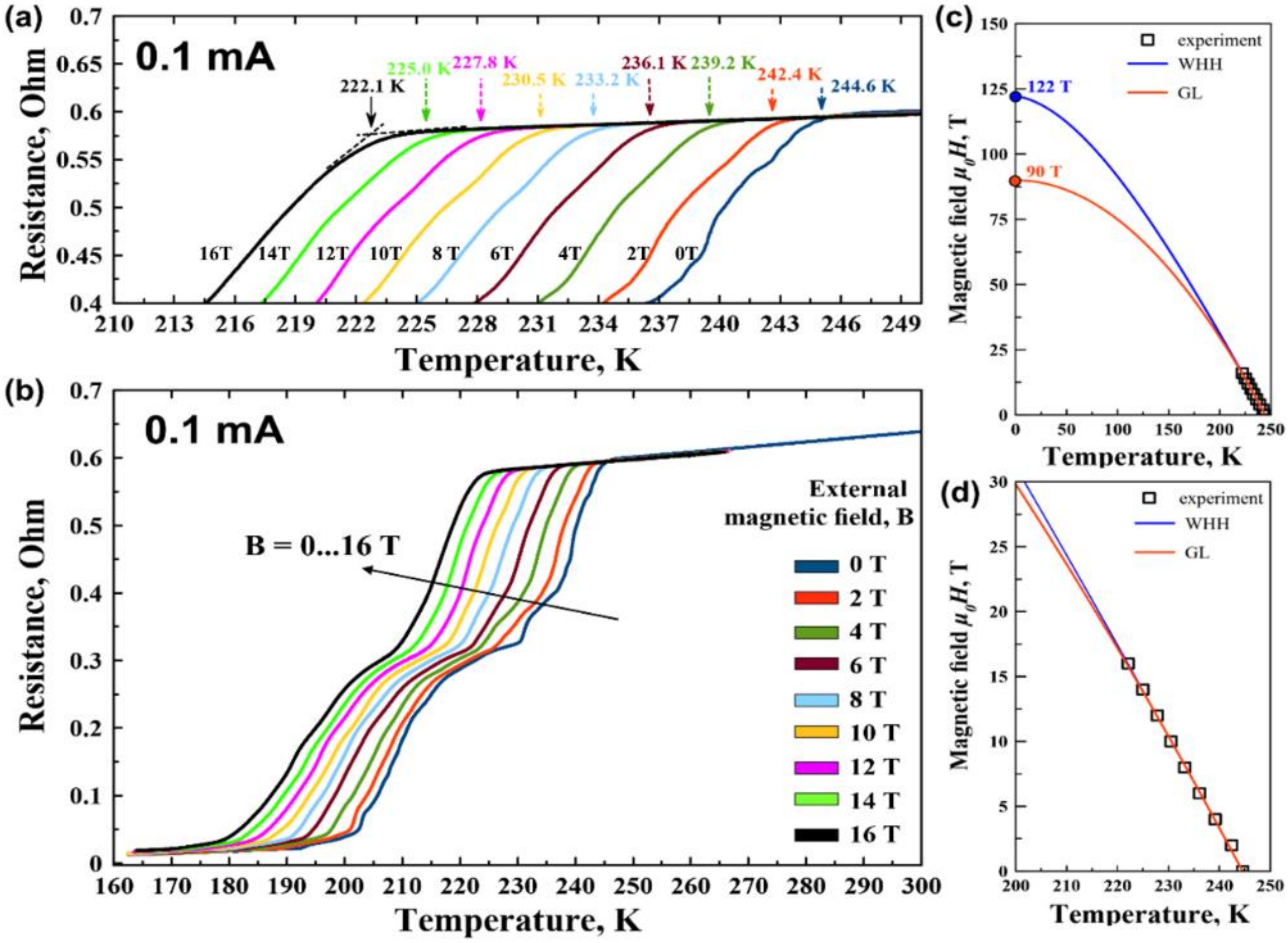

**Figure S26.** Electrical resistance and critical temperature $T_C$ of the (La, Y)$H_{10}$ sample, obtained from $La_2Y$ alloy, in an external magnetic field (0–16 T) at 183 GPa. (a) Temperature dependence of the electrical resistance at different external magnetic fields, $T$ = 210–250 K. Arrows indicate the onsets of the resistance drop where the critical temperatures were determined. (b) Same dependence at $T$ = 160–300 K. (c) Upper critical magnetic field extrapolated to 0 K. (d) Dependence of $T_C$ on the applied magnetic field.

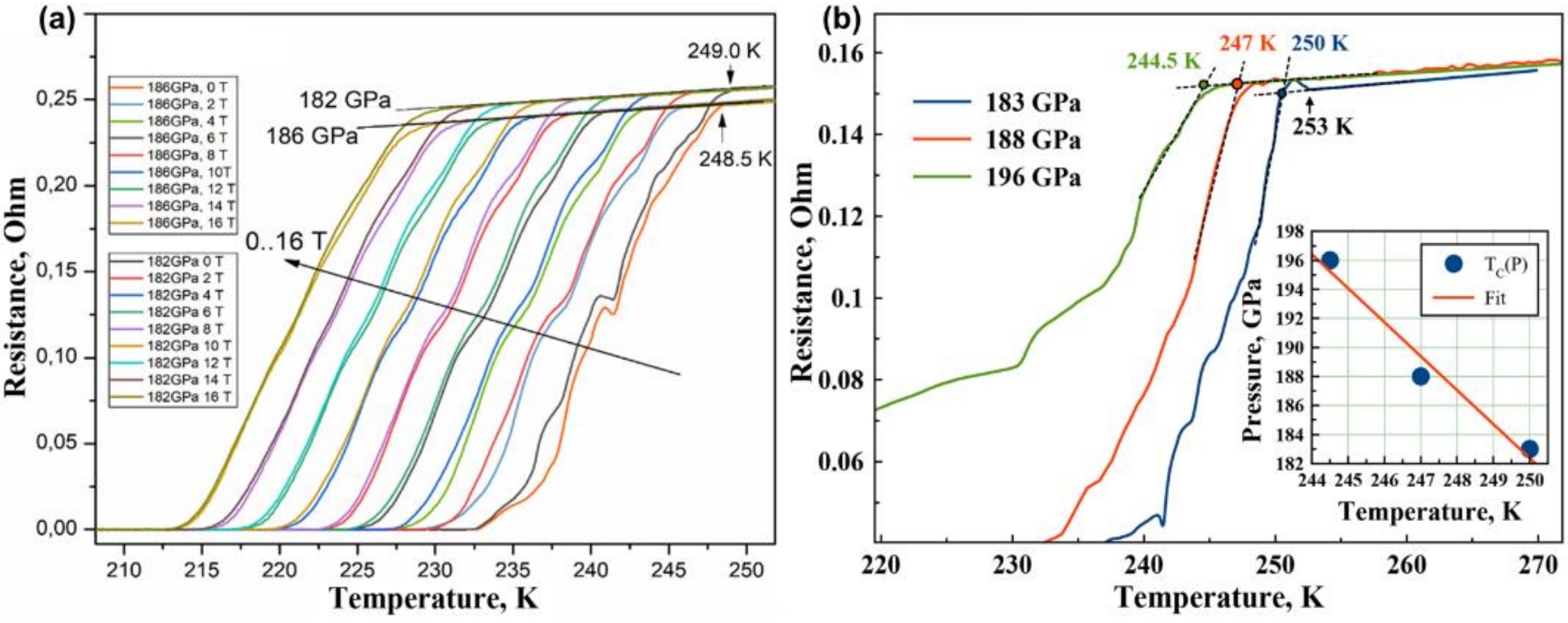

**Figure S27.** Electrical resistance and critical temperature $T_C$ of the (La, Y)$H_{10}$ sample in an external magnetic field (0–16 T): (a) at 182 and 186 GPa, for the sample obtained from $La_4Y$; (b) at 183, 188, and 196 GPa, with zero-field cooling, for the sample obtained from $La_2Y$. Inset: pressure dependence of $T_C$. The critical temperatures were determined at the onset of the resistance drop, $dT_C/dP$ = –0.43 K/GPa.

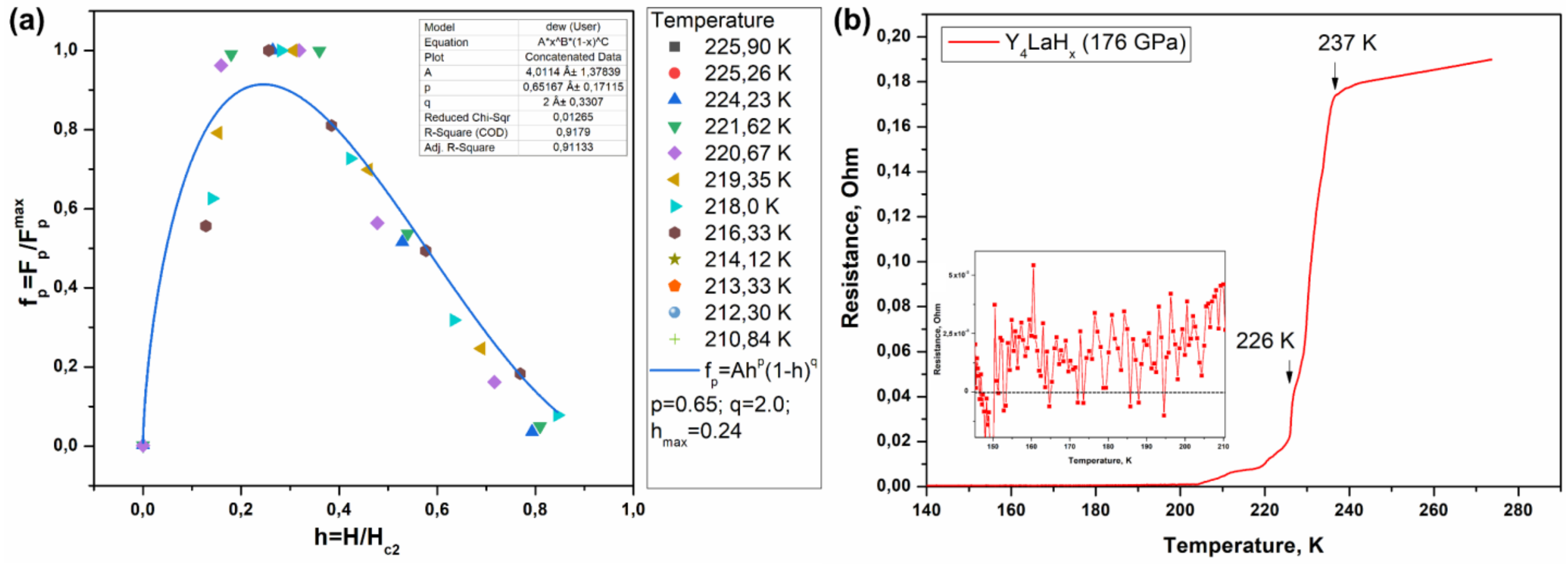

**Figure S28.** (a) Scaling of the normalized volume pinning forces ($F_p/F_p^{max}$) of the (La,Y)$H_{10}$ sample, obtained from $La_4Y$ alloy, at 186 GPa for several different temperatures versus the reduced field h = $H/H_{c2}$. Experimental data is fitted by the Dew-Hughes [9] model for surface type normal pinning centers: $f_p = h^p(1\text{-}h)^q$ . (b) Temperature dependence of the electrical resistance of the (La,Y)$H_x$ sample, obtained from $LaY_4$ alloy, at 176 GPa. The complex structure of the transition is probably due to the presence of impurities: $Im\bar{3}m$-$YH_6$ (~226 K), *Imm*2-$YH_7$, and, possibly, $La_xY_{1-x}H_6$.

# Eliashberg functions of La–Y–H phases

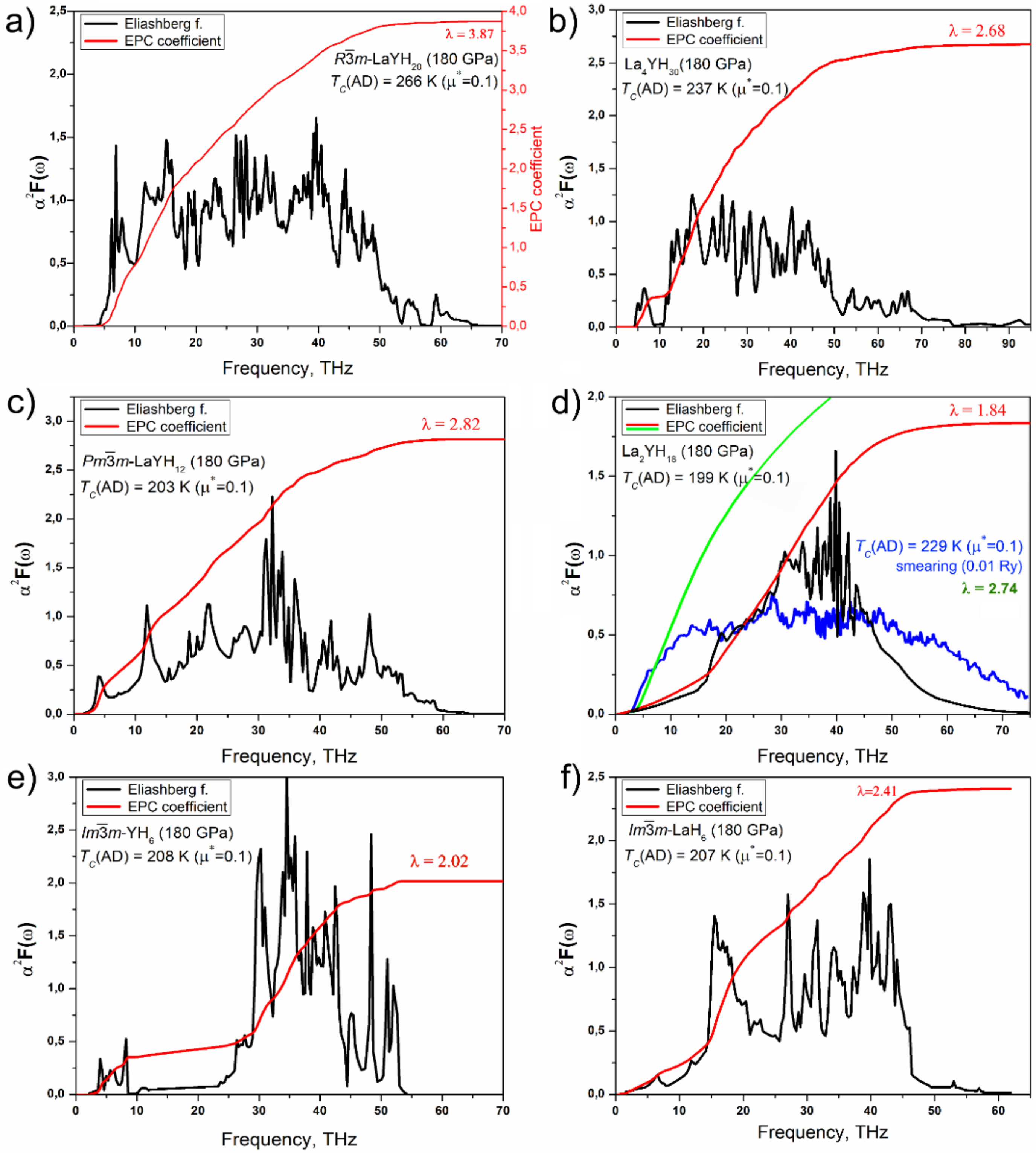

**Figure S29.** Ab initio calculated harmonic Eliashberg functions $\alpha^2F(\omega)$ and electron–phonon coupling (EPC) parameters at 180 GPa for (a) $R\bar{3}m$-$LaYH_{20}$, (b) cubic $La_4YH_{30}$, (c) $Pm\bar{3}m$-$LaYH_{12}$, (d) cubic $La_2YH_{18}$, (e) $Im\bar{3}m$-$YH_6$, and (f) $Im\bar{3}m$-$LaH_6$. The Eliashberg functions calculated within the tetrahedron method[49] and the interpolation method[50] (with $\sigma$ = 0.01 Ry) are shown in black and blue, respectively. $T_C$ (AD) is the critical temperature obtained using the Allen–Dynes formula.[31]

**Table S5.** Electron–phonon coupling (EPC) parameters for various La–Y hydrides calculated in Quantum ESPRESSO within different methods and $q$, $k$-meshes at 180 GPa. $T_C$ (E) corresponds to the solution of the isotropic ME equations, $T_C$ (AD) is obtained using the Allen–Dynes formula.[31]

| Compound | Cell | Method* | $q$, $k$-meshes | λ | $\omega_{log}$, K | $T_C$ (AD), K μ* = 0.1 | $T_C$ (E), K μ* = 0.1 |
|---|---|---|---|---|---|---|---|
| $Im\bar{3}m$-$YH_6$ | 14 atoms | T | $q$: 4×4×4, $k$: 12×12×12, 16×16×16 | 2.02 | 1190 | 208 | 250 |
| | | S | $q$: 2×2×2, $k$: 12×12×12, 16×16×16 | 2.24 | 929 | 184 | - |
| $Pm\bar{3}m$-$LaYH_{12}$ | 14 atoms | T | $q$: 4×4×4, $k$: 16×16×16, 24×24×24 | 2.82 | 847 | 203 | 241 |
| | | S | $q$: 3×3×3, $k$: 18×18×18, 24×24×24 | 2.47 | 875 | 192 | - |
| $Pm\bar{3}m$-$La_2YH_{18}$ | 42 atoms | T | $q$: 2×2×2, $k$: 6×6×6, 8×8×8 | 1.84 | 1257 | 199 | 223 |
| | | S | $q$: 1×2×4, $k$: 2×4×8, 4×8×16 | 2.74 | 964 | 229 | 270 |
| $Pm\bar{3}m$-$La_4YH_{30}$ | 35 atoms | T | $q$: 1×2×2, $k$: 2×8×8, 4×12×12 | 3.21 | 910 | 241 | - |
| | | S | $q$: 1×2×2, $k$: 2×8×8, 4×12×12 | 2.68 | 1046 | 237 | 265 |
| $Im\bar{3}m$-$LaH_6$ | 14 atoms | T | $q$: 4×4×4, $k$: 12×12×12, 16×16×16 | 2.41 | 1011 | 207 | 235 |
| | | S | $q$: 4×4×4, $k$: 12×12×12, 12×12×12 | 2.47 | 803 | 175 | - |
| $R\bar{3}m$-$LaYH_{20}$ | 22 atoms | T | $q$: 4×4×4, $k$: 12×12×12, 16×16×16 | 3.87 | 868 | 266 | 300 |

* Methods: T – the tetrahedron integration,[49] S – the interpolation method[50] (0.01–0.015 Ry).

**Table S6.** Parameters of the superconducting state of $R\bar{3}m$-$LaYH_{20}$, $Pm\bar{3}m$-$LaYH_{12}$, $Im\bar{3}m$-$LaH_6$, cubic $La_2YH_{18}$, and $La_4YH_{30}$ at 180 GPa calculated using the isotropic Migdal–Eliashberg equations (E)[29] and the Allen–Dynes formula (AD)[31] with μ* = 0.15–0.1.

| Parameter | $R\bar{3}m$-$LaYH_{20}$ | $Pm\bar{3}m$-$LaYH_{12}$ | $La_2YH_{18}$ | $La_4YH_{30}$ | $Im\bar{3}m$-$LaH_6$ |
|---|---|---|---|---|---|
| λ | 3.87 | 2.82 | 2.74 | 2.68 | 2.41 |
| $\omega_{log}$, K | 868 | 847 | 964 | 1046 | 1011 |
| $\omega_2$, K | 1208 | 1251 | 1525 | 1437 | 1312 |
| β | 0.48-0.49 | 0.48-0.49 | 0.47-0.49 | 0.48-0.49 | 0.47-0.49 |
| $T_C$ (AD), K | 232-266 | 176-203 | 197-229 | 206-237 | 180-207 |
| $T_C$ (E), K | 281-300* | 223-241 | 248-270 | 245-265 | 217-235 |
| $T_C$ (SCDFT), K | 252 | 191 | - | - | 176 |
| $N(E_F)$, states/eV/f.u. | 0.87 | 0.8 | 0.66 | 0.79 | 1.62 |
| $T_C$ (La–$YD_x$), K | 213 | 160-173 | 178-193 | 175-190 | 155-168 |
| Δ(0), meV | 67-71 | 53-57.5 | 58-64.4 | 57-62 | 49-54 |
| $\mu_0 H_C(0)$, T | 99-101 | 73-77 | 73-78 | 80-85 | 98-106 |
| $\Delta C/T_C$, mJ/mol·$K^2$ | 33.6-22.7 | 38-35 | 32-29 | 39-38 | 77-75.6 |
| γ, mJ/mol·$K^2$ | 19.9 | 14.3 | 11.6 | 13.7 | 26 |
| $R_\Delta = 2\Delta(0)/k_B T_C$ | 5.48-5.54 | 5.5 | 5.5 | 5.36-5.45 | 5.23-5.35 |

* $T_C$ (E) = 266 K at μ* = 0.2.

# Equations for calculating $T_C$ and related parameters

To calculate the isotopic coefficient β, the Allen–Dynes interpolation formulas were used:

$$\beta_{McM} = -\frac{d\ln T_C}{d\ln M} = \frac{1}{2}\left[1 - \frac{1.04(1+\lambda)(1+0.62\lambda)}{[\lambda - \mu^*(1+0.62\lambda)]^2}\mu^{*2}\right] \quad \text{(S2)}$$

$$\beta_{AD} = \beta_{McM} - \frac{2.34\mu^{*2}\lambda^{3/2}}{(2.46+9.25\mu^*)\cdot((2.46+9.25\mu^*)^{3/2}+\lambda^{3/2})} - \frac{130.4\cdot\mu^{*2}\lambda^2(1+6.3\mu^*)\left(1-\frac{\omega_{\log}}{\omega_2}\right)\frac{\omega_{\log}}{\omega_2}}{\left(8.28+104\mu^*+329\mu^{*2}+2.5\cdot\lambda^2\frac{\omega_{\log}}{\omega_2}\right)\cdot\left(8.28+104\mu^*+329\mu^{*2}+2.5\cdot\lambda^2\left(\frac{\omega_{\log}}{\omega_2}\right)^2\right)} \quad \text{(S3)}$$

where the last two correction terms are usually small (~0.01).

The Sommerfeld constant was found as

$$\gamma = \frac{2}{3}\pi^2 k_B^2 N(0)(1+\lambda) \quad \text{(S4)}$$

and was applied to estimate the upper critical magnetic field and the superconductive gap in yttrium hydrides using well-known semiempirical equations of the BCS theory (Ref. [51], eq 4.1 and 5.11), which works for $T_C/\omega_{\log} < 0.25$:

$$\frac{\gamma T_C^2}{B_{C2}^2(0)} = 0.168\left[1 - 12.2\left(\frac{T_C}{\omega_{\log}}\right)^2 \ln\left(\frac{\omega_{\log}}{3T_C}\right)\right] \quad \text{(S5)}$$

$$\frac{2\Delta(0)}{k_B T_C} = 3.53\left[1 + 12.5\left(\frac{T_C}{\omega_{\log}}\right)^2 \ln\left(\frac{\omega_{\log}}{2T_C}\right)\right] \quad \text{(S6)}$$

The lower critical magnetic field was calculated according to the Ginzburg–Landau theory:[52]

$$\frac{H_{C1}}{H_{C2}} = \frac{\ln k}{2\sqrt{2}k^2}, \qquad k = \lambda_L/\xi \quad \text{(S7)}$$

where $\lambda_L$ is the London penetration depth, found as

$$\lambda_L = 1.0541\cdot 10^{-5}\sqrt{\frac{m_e c^2}{4\pi n_e e^2}} \quad \text{(S8)}$$

where $c$ is the speed of light, $e$ is the electron charge, $m_e$ is the electron mass, and $n_e$ is the effective concentration of charge carriers, evaluated from the average Fermi velocity $V_F$ in the Fermi gas model:

$$n_e = \frac{1}{e\pi^2}\left(\frac{m_e V_F}{\hbar}\right)^3 \quad \text{(S9)}$$

The coherence length ξ was found as $\xi = \sqrt{\hbar/2e(\mu_0 H_{C2})}$ and was used to estimate the average Fermi velocity

$$V_F = \frac{\pi\Delta(0)}{\hbar} \quad \text{(S10)}$$

The critical temperature of superconducting transition was calculated using the Matsubara-type linearized Eliashberg equations:[29]

$$\hbar\omega_j = \pi(2j+1)k_B T, \qquad j = 0, \pm1, \pm2, \ldots \quad \text{(S11)}$$

$$\lambda(\omega_i - \omega_j) = 2\int_0^\infty \frac{\omega\cdot\alpha^2 F(\omega)}{\omega^2 + (\omega_i - \omega_j)^2}d\omega \quad \text{(S12)}$$

$$\Delta(\omega = \omega_i, T) = \Delta_i(T)$$

$$= \pi k_B T \sum_j \frac{[\lambda(\omega_i - \omega_j) - \mu^*]}{\rho + |\hbar\omega_j + \pi k_B T \sum_k (sign\ \omega_k) \cdot \lambda(\omega_i - \omega_j)|} \cdot \Delta_j(T) \quad \text{(S13)}$$

where $T$ is the temperature in kelvins, $\mu^*$ is the Coulomb pseudopotential, $\omega$ is the frequency in Hz, $\rho(T)$ is a pair-breaking parameter, the function $\lambda(\omega_i - \omega_j)$ is related to an effective electron–electron interaction via the exchange of phonons.[53] The transition temperature can be found as the solution of the equation $\rho(T_C) = 0$, where $\rho(T)$ is defined as max($\rho$), provided that $\Delta(\omega)$ is not a zero function of $\omega$ at a fixed temperature.

These equations can be rewritten in a matrix form as[30]

$$\rho(T)\psi_m = \sum_{n=0}^{N} K_{mn}\psi_n \Leftrightarrow \rho(T)\begin{pmatrix}\psi_1\\ \dots\\ \psi_N\end{pmatrix} = \begin{pmatrix}K_{11} & \dots & K_{1N}\\ \dots & K_{ii} & \dots\\ K_{N1} & \dots & K_{NN}\end{pmatrix} \times \begin{pmatrix}\psi_1\\ \dots\\ \psi_N\end{pmatrix}, \quad \text{(S14)}$$

where $\psi_n$ relates to $\Delta(\omega, T)$, and

$$K_{mn} = F(m-n) + F(m+n+1) - 2\mu$$

$$* -\delta_{mn}\left[2m + 1 + F(0) + 2\sum_{l=1}^{m} F(l)\right] \quad \text{(S15)}$$

$$F(x) = F(x,T) = 2\int_0^{\omega\max} \frac{\alpha^2 F(\omega)}{(\hbar\omega)^2 + (2\pi k_B T x)^2}\hbar\omega d\omega, \quad \text{(S16)}$$

where $\delta_{nn} = 1$ and $\delta_{nm} = 0$ ($n \neq m$) is a unit matrix. Now we can replace the equation $\rho(T_C) = 0$ with the vanishing of the maximum eigenvalue of the matrix $K_{nm}$: {$\rho$ = max_eigenvalue($K_{nm}$) = $f(T)$, $f(T_C) = 0$}.

# Electronic properties

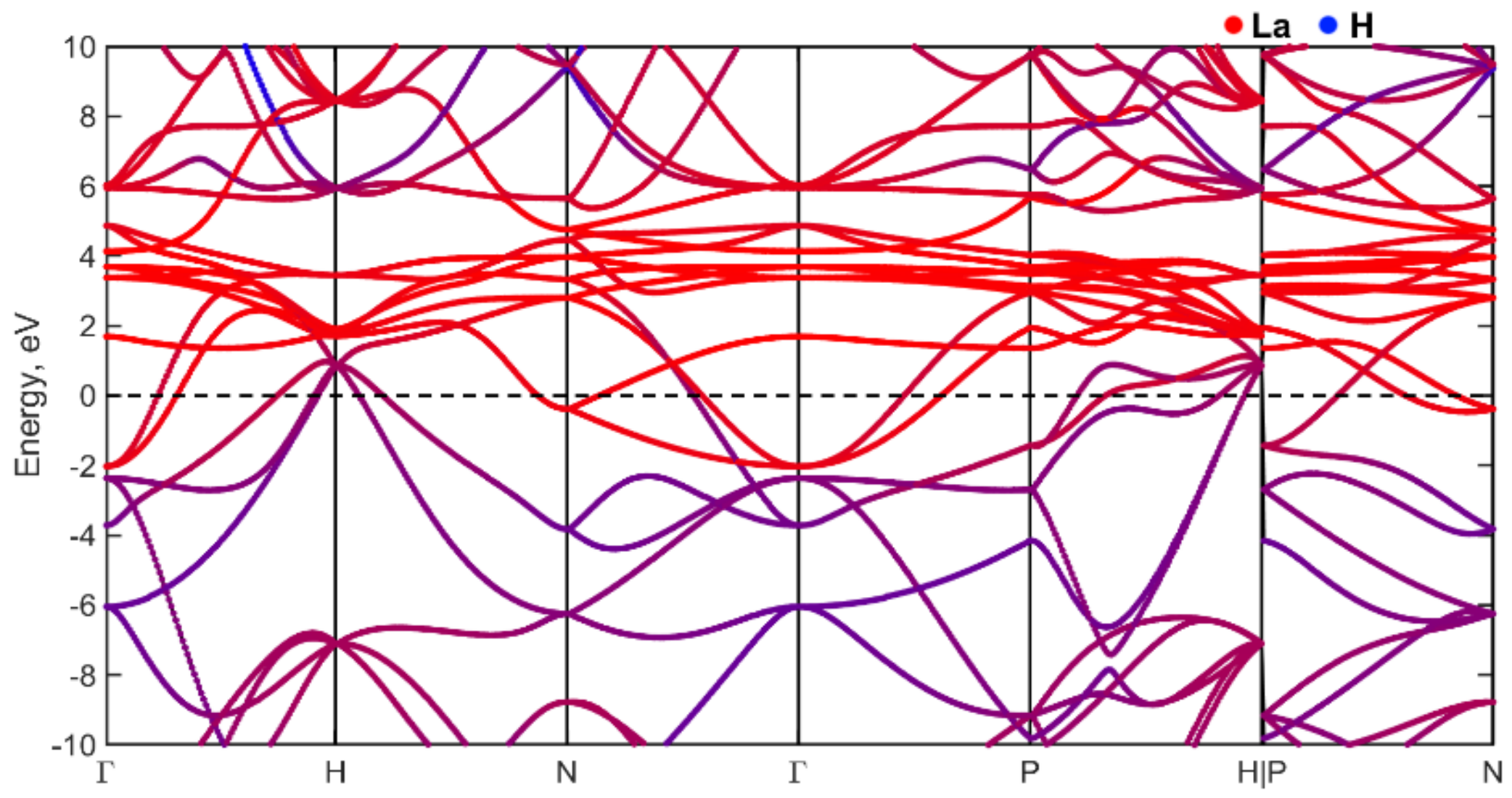

**Figure S30.** Electronic band structure of proposed $Im\bar{3}m$-$LaH_6$ at 180 GPa.

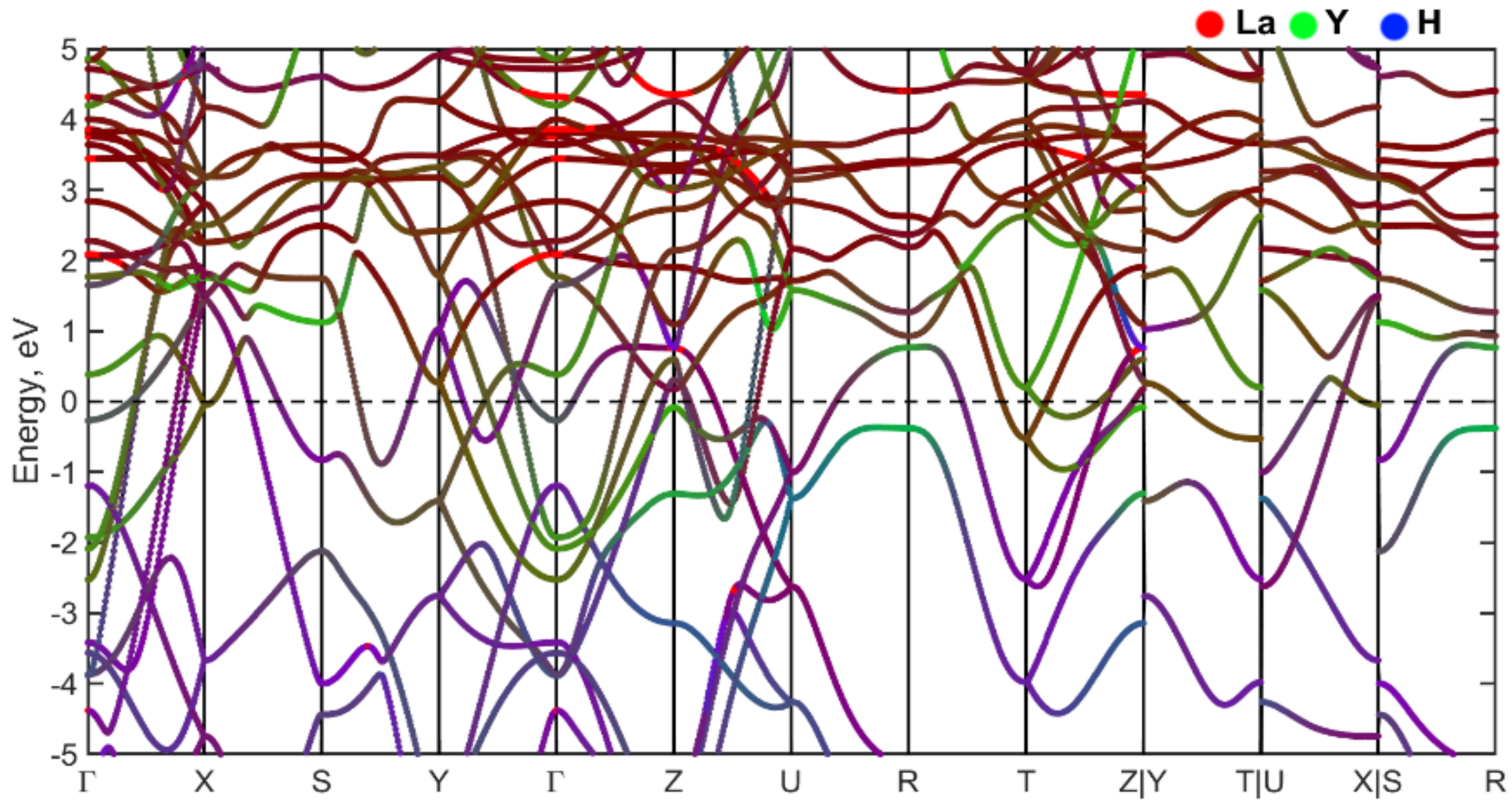

**Figure S31.** Electronic band structure of $Pm\bar{3}m$-$LaYH_{12}$ at 180 GPa.

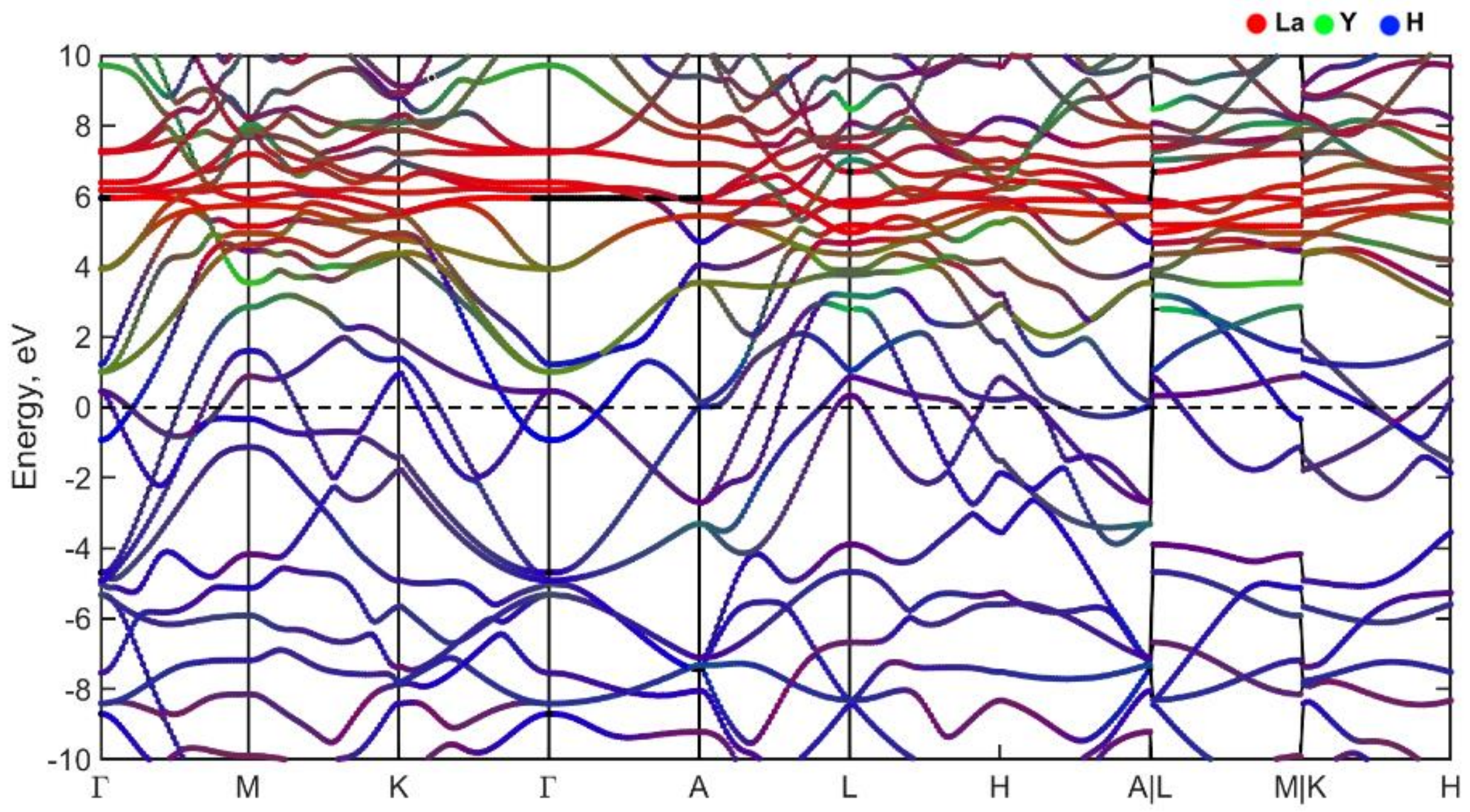

**Figure S32.** Electronic band structure of $R\bar{3}m$-$LaYH_{20}$ at 180 GPa.

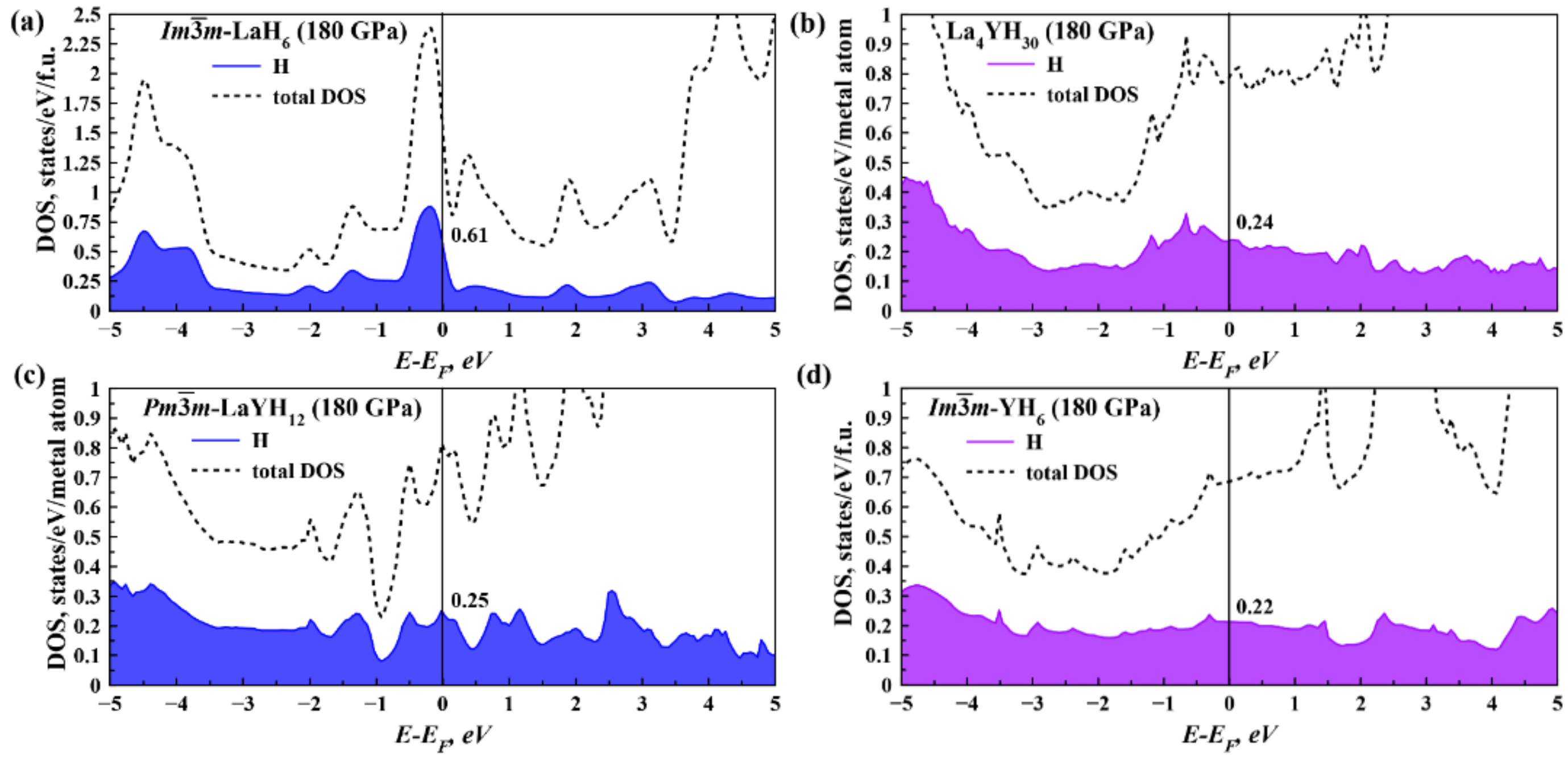

**Figure S33.** Electronic density of states of (a) proposed $Im\bar{3}m$-$LaH_6$, (b) cubic $La_4YH_{30}$, (c) $Pm\bar{3}m$-$LaYH_{12}$, and (d) $Im\bar{3}m$-$YH_6$ at 180 GPa.

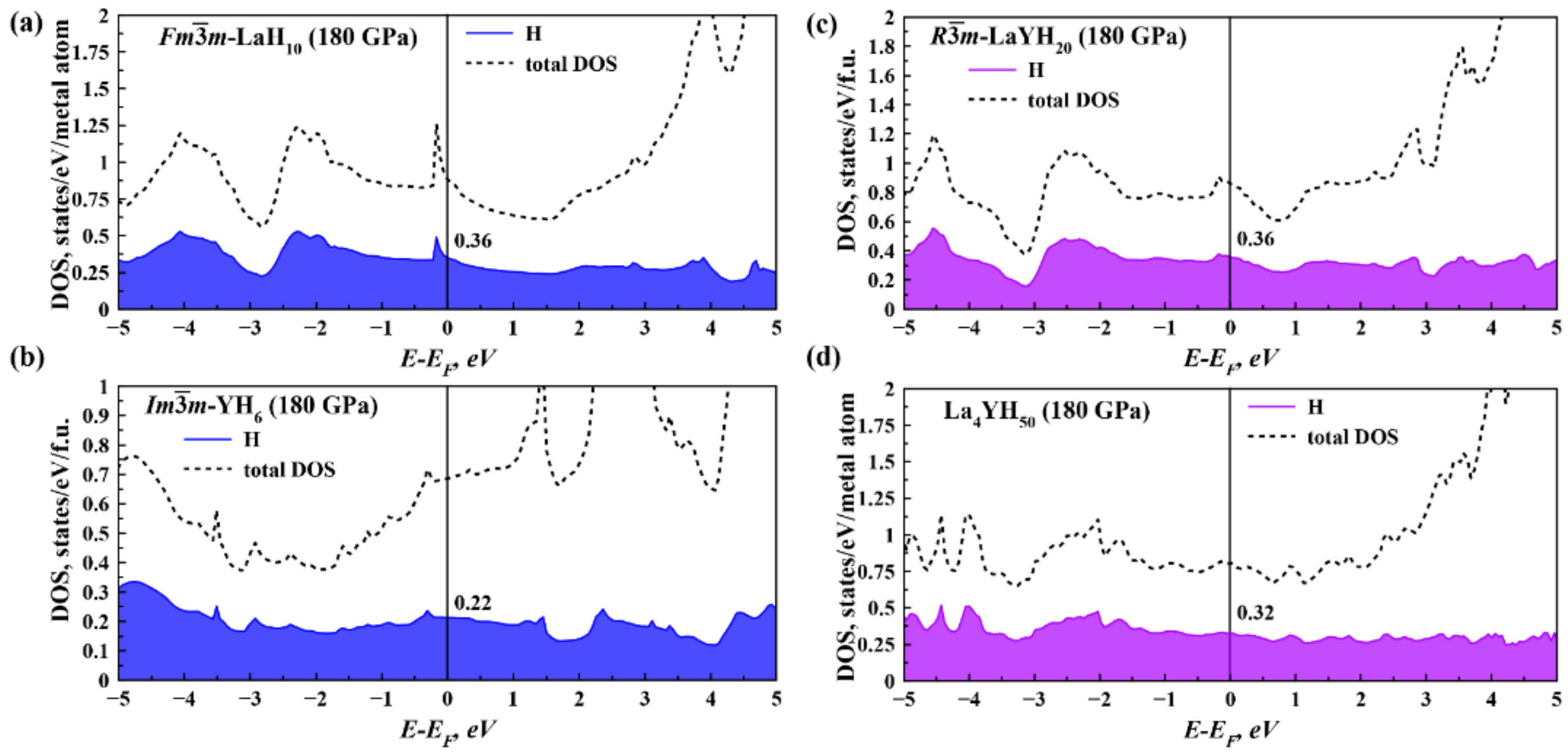

**Figure S34.** Electronic density of states of (a) $Fm\bar{3}m$-$LaH_{10}$, (b) $Im\bar{3}m$-$YH_6$, (c) $R\bar{3}m$-$LaYH_{20}$, and (d) cubic $La_4YH_{50}$ at 180 GPa.